\documentclass[twocolumn,showpacs,preprintnumbers, amsmath,amssymb,amsfonts, final, a4paper,aps,
citeautoscript,footinbib,prb,superscriptaddress]{revtex4}
\usepackage{times}
\usepackage{graphicx}
\usepackage[latin1]{inputenc}
\usepackage{mathrsfs}
\usepackage{epstopdf}

\newcommand{\ZZ}{\mathbb{Z}}

\begin{document}
\title{Competing orders in one-dimensional half-filled multicomponent
  fermionic cold atoms: The Haldane-charge conjecture}
  \author{H. Nonne}
\affiliation{Laboratoire de Physique Th\'eorique et Mod\'elisation,
  CNRS UMR 8089, Universit\'e de Cergy-Pontoise, Site de Saint-Martin,
  F-95300 Cergy-Pontoise Cedex, France}
\author{P. Lecheminant} 
\affiliation{Laboratoire de Physique Th\'eorique et Mod\'elisation,
  CNRS UMR 8089, Universit\'e de Cergy-Pontoise, Site de Saint-Martin,
  F-95300 Cergy-Pontoise Cedex, France}
 \author{S.\ Capponi} 
\affiliation{Laboratoire de Physique Th\'eorique, CNRS UMR 5152,
  Universit\'e Paul Sabatier, F-31062 Toulouse, France.}
\author{G. Roux}
\affiliation{Laboratoire de Physique Th\'eorique et Mod\`eles
  statistiques, Universit\'e Paris-Sud, CNRS UMR 8626, 91405 Orsay,
  France}
\author{E. Boulat}
\affiliation{Laboratoire Mat\'eriaux et Ph\'enom\`enes Quantiques,
  CNRS UMR 7162, Universit\'e Paris Diderot, 75013 Paris, France}

\date{\today}

\begin{abstract}
We investigate the nature of the Mott-insulating phases 
of  half-filled $2N$-component fermionic cold atoms loaded into a one-dimensional
optical lattice. By means of conformal field theory techniques and large-scale DMRG 
calculations, we show that the phase diagram  strongly depends on
the parity of $N$.
First, we single out charged, spin-singlet, degrees of freedom, that carry a pseudo-spin
${\cal S}=N/2$ allowing to formulate a Haldane conjecture: for attractive interactions,
we establish the emergence of Haldane insulating phases when $N$ is even,
whereas a metallic behavior is found when $N$ is odd. 
We point out that  the $N=1,2$ cases do \emph{not} have the generic properties of each family.
The metallic phase for $N$ odd and larger than 1 has a quasi-long range singlet pairing ordering
with an interesting edge-state structure.
Moreover, the properties of the Haldane insulating phases with even $N$ further depend on the parity
  of $N/2$. In this respect, within the low-energy approach, we argue that the Haldane phases with $N/2$ even 
  are not topologically protected but equivalent to a topologically trivial insulating phase and thus confirm
  the recent conjecture put forward by Pollmann {\it et al.} [Pollmann {\it et al.}, arXiv:0909.4059 (2009)]. 
  \end{abstract}

\maketitle

\section{Introduction}
Topological phases have attracted much interest in recent years
due to their robustness against perturbations and their relevance 
to quantum computation.
A topological ordered phase is a gapped phase which displays
a protected ground-state degeneracy dependent on the topology
of the manifold in which the model is embedded \cite{wenbook}.
This phase is not characterized by a local order parameter 
and falls beyond the usual symmetry breaking paradigm of condensed matter 
physics \cite{anderson}.

One of the simplest examples of topologically ordered
phases is the Haldane phase in quantum spin chains.
In 1983, Haldane argued that the spin-$S$ Heisenberg
chain displays striking different properties
depending on the parity of $2S$ \cite{haldane}.
While half-integer Heisenberg spin chains have a gapless
behavior, a finite gap  from the singlet ground-state (GS) to the first triplet excited states is found when
$2S$ is even.  On top of the existence a gap, 
the spin-1 phase (the so-called Haldane phase) has remarkable
exotic properties which may be regarded as manifestations of  the existence
of a topological ordered phase.
This phase is not characterized by a local order but 
displays non-local string long-range ordering which signals the presence of
a hidden N\'eel antiferromagnetic order \cite{dennijs}.
The latter can be revealed through a non-local
unitary transformation and the emergence of 
a complete breaking of a $\ZZ_2$ $\times$
$\ZZ_2$ symmetry \cite{kennedy}.
One remarkable resulting consequence of the Haldane 
phase is the liberation of fractional spin-1/2 degrees of freedom
at the edge of the sample when the chain is 
doped by non-magnetic impurities \cite{hagiwara}.

Haldane's conjecture is now well understood and has been confirmed
experimentally in quasi-1D compounds as well as numerically (see for instance
Refs. \onlinecite{phlereview, mikeska}).  The Haldane phase displays unusual and interesting physical properties so that it is important to experimentally stabilize
it in other contexts. In this
respect, it has been argued that the Haldane phase is relevant to
Josephson junction array systems \cite{auerbach98}. Furthermore,
it is likely that the Haldane physics will be explored experimentally
in the near future in trapped ultracold atomic systems thanks to the
tunability of interactions in these systems using optical lattices and
Feshbach resonances.  A first possible direction is to consider spin-1
bosons loaded into a one-dimensional (1D) optical lattice with one
atom per site so that the Haldane phase is one of the possible
insulating phases of this model \cite{cirac}.  A second route consists in preparing
1D ultracold quantum gases with dipolar
interactions, like $^{52}$Cr bosonic atoms, where a Haldane insulating
(HI) phase has been predicted
\cite{Berg2008,Lee2008,amico,dalmonte}. Finally, we have recently
shown that a similar phase can also be stabilized by considering 1D
spin-3/2 cold fermions at half-filling with \emph{contact interactions only}~\cite{Nonne2009}.

In this paper, we pursue our investigation of the HI phase in
the context of 1D ultracold fermionic alkaline atoms in the general half-integer (hyperfine) spin $F =
N -1/2$ case at half-filling ($N$ atoms per site). In this respect, we will use 
complementary analytical (renormalization group (RG)
analysis, conformal field theory (CFT) \cite{dms})  and density-matrix 
renormalization group (DMRG) \cite{DMRG}
techniques to fully determine the nature of the Mott-insulating phases at 
half-filling when $N\geq 2$. 
The starting point of the analysis is the lattice model of $2N$ components cold fermions with contact interactions.
Due to Pauli principle, low-energy
s-wave scattering processes of spin-$F$ fermionic atoms are allowed only in
the even total spin $J=0,2,\ldots, 2N-2$  channels, so that the general
effective Hamiltonian with contact interactions reads as follows in absence of a magnetic field:
\cite{ho}
\begin{eqnarray}
{\cal H}
&=& -t \sum_{i,\alpha} \left[c^{\dagger}_{\alpha,i}
c^{\phantom\dagger}_{\alpha, i+1} + {\rm H.c.} \right]
- \mu \sum_{i,\alpha} c^{\dagger}_{\alpha,i} c^{\phantom\dagger}_{\alpha,i}
\nonumber \\
&+& \sum_{i,J} U_J \sum_{M=-J}^{J}
P_{JM,i}^{\dagger} P_{JM,i},
\label{hubbardSgen}
\end{eqnarray}
where $c^{\dagger}_{\alpha,i}$ is the fermion creation operator
corresponding to the $2N$ hyperfine states ($\alpha = 1,\ldots, 2N$)
at the $i^{\text{th}}$ site of the optical lattice. The pairing
operators in Eq.~(\ref{hubbardSgen}) are defined through the
Clebsch-Gordan coefficients for spin-$F$ fermions: $P^{\dagger}_{JM,i}
= \sum_{\alpha \beta} \langle{JM|F,F;\alpha \beta}\rangle
c^{\dagger}_{\alpha,i} c^{\dagger}_{\beta,i}$. In the general spin-$F$
case, there are $N$ coupling constants $U_J$ in
model~(\ref{hubbardSgen}), which are related to the $N$
two-body scattering lengths of the problem. In the following, in order to simplify the analysis of the Mott-insulating phases when
$N > 2$, we perform a fine-tuning of the different scattering lengths in channels $J \ge 2$, i.e.
$U_2 = ... = U_{2N-2}$. Using the identity $\sum_{JM}P_{JM,i}^{\dag} \,P_{JM,i}^{\phantom\dag}=n_i^2-n_i$, model (\ref{hubbardSgen}) can then be mapped onto the following:
\begin{eqnarray}
{\cal H} &=& -t \sum_{i,\alpha} [c^{\dagger}_{\alpha,i} c^{\phantom\dagger}_{\alpha,i+1} +
{\rm H.c.} ]
- \mu \sum_i n_i
\nonumber \\
&+& \frac{U}{2} \sum_i n_i^2 + V \sum_i P^{\dagger}_{00,i}
P^{\phantom\dagger}_{00,i},
\label{hubbardS}
\end{eqnarray}
with $U= 2 U_2$, $V = U_0 - U_2$, and $n_i = \sum_{\alpha}
n_{\alpha,i} = \sum_{\alpha} c^{\dagger}_{\alpha,i} c^{\phantom\dagger}_{\alpha,i}$ is
the density at site $i$.  In Eq.~(\ref{hubbardS}), the singlet BCS
pairing operator for spin-F fermions is $ \sqrt{2N} P^{\dagger}_{00,i}
= \sum_{\alpha \beta} c^{\dagger}_{\alpha,i} {\cal J}_{\alpha \beta}
c^{\dagger}_{\beta,i} = - \sum_{\alpha} \left(-1\right)^{\alpha}
c^{\dagger}_{\alpha,i} c^{\dagger}_{2N+1-\alpha,i}$, the matrix
${\cal J}$ being a $2N \times 2N$ antisymmetric matrix with ${\cal J}^2 =
- I$. When $N=1$, $P^{\dagger}_{00,i}$ coincides with the Cooper pairing
$c^{\dagger}_{\uparrow,i} c^{\dagger}_{\downarrow,i}$
so that model (\ref{hubbardS}) is equivalent to the spin-1/2 Hubbard model. 

Model (\ref{hubbardS}) obviously conserves the total number of fermions -- no atoms
are dynamically created or annihilated. This conservation law is associated to a
U(1) continuous symmetry
$c_{\alpha,i} \rightarrow e^{i \theta} c_{\alpha,i}$, which, by analogy with
condensed matter dealing with electrons, we will refer to as a "U(1)$_c$ charge symmetry",
the charge being simply the number of fermions.
On top of this symmetry,
model (\ref{hubbardS}) displays an extended continuous symmetry for $N > 1$ in  spin space.
When $V=0$ ($U_0=U_2$) model (\ref{hubbardS}) is
the Hubbard model for $2N$-component fermions with a U(2$N$)$=$ U(1)$_c$
$\times$ SU(2$N$) invariance. 
The Hamiltonian (\ref{hubbardS}) for $V \ne
0$ still displays an extended symmetry since the BCS singlet-pairing operator $P^{\dagger}_{00,i}$ 
is invariant under the Sp($2N$) group which 
consists of  $2N \times 2N$ unitary matrices $U$ that satisfy
$U^* {\cal J} U^{\dagger} = {\cal J}$.
When $V\ne 0$, the continuous symmetry of model (\ref{hubbardS}) is thus
U(1)$_c$ $\times$ Sp(2$N$) \cite{sachdev,wuzhang}.
 In the $F=3/2$ case, i.e. $N=2$, there is no fine-tuning; models (\ref{hubbardSgen}) and (\ref{hubbardS}) are equivalent and share an exact
Sp(4)$\simeq$ SO(5) spin symmetry \cite{zhang}.
The zero-temperature
phase diagram of model (\ref{hubbardS}) away from half-filling has
been investigated by means of a low-energy approach
\cite{Lecheminant2005,phle,Wu2005} in the general $N$ case, and by Quantum Monte Carlo and DMRG calculations
for $N=2$ \cite{sylvainMS,wunum}. A rich exotic physics emerge when $N \ge 2$ with,
in particular, the stabilization of a superconducting instability with charge
$2Ne$ for attractive interactions and at sufficiently low density
\cite{Lecheminant2005,phle,Wu2005,sylvainMS}.

At half-filling (when $\mu = N U + V/N$), 
model (\ref{hubbardS}) enjoys a particle-hole symmetry $c_{\alpha,i} \rightarrow (-1)^{i}
\sum_{\beta} {\cal J}_{\alpha \beta} c^{\dagger}_{\beta,i}$ which
plays a crucial role in the following. In the $N=1$ case, it is well
known that the particle-hole symmetry enlarges the U(1)$_c$ charge
symmetry of the spin-1/2 Hubbard model to an SU(2)$_c$ symmetry at
half-filling~\cite{yang,yangzhang}. In addition, the physics of the
half-filled spin-1/2 Hubbard model for repulsive and attractive
interactions are related through a canonical transformation
$c_{\uparrow,i} \rightarrow (-1)^{i} c^{\dagger}_{\uparrow,i}$,
$c_{\downarrow,i} \rightarrow c_{\downarrow,i}$. While for $U>0$ a
Mott-insulating phase with one gapless spin modes is stabilized, there
is a spin gap for attractive interaction which marks the emergence of
a singlet-pairing phase \cite{bookboso,giamarchi}.  When $N >1$, all
these properties do not generalize; in particular, the symmetry
enlargement of the charge degrees of freedom at half-filling requires
an additional fine tuning $V = N U$ to display an SU(2)$_c$ $\times$
Sp(2$N$) global invariance \cite{Nonne2009}.  We have
shown in Ref.~\onlinecite{Nonne2009} that this SU(2)$_c$ symmetry is
central to the emergence of an even-odd scenario for attractive
interactions in close parallel to the famous Haldane conjecture in
spin-$S$ SU(2) Heisenberg chains. In this respect, we have identified a
spin-singlet pseudo-spin $N/2$ operator which governs the low-energy
properties of the model in the vicinity of the SU(2)$_c$ line for
attractive interactions.  This operator gives rise to a Haldane-charge
conjecture with the emergence of a HI phase when $N$ is even, while a
metallic phase is stabilized when $N$ is odd. Such a scenario has been
checked in Ref.~\onlinecite{Nonne2009} by a low-energy approach in
the $N=2$ case and DMRG calculations for $N=2,3$ in the vicinity of
the SU(2)$_c$ line. In the special $N=2$ case, these complementary
techniques reveals unambiguously the existence of a HI phase with
non-local string charge correlations and pseudo-spin-1/2 edge states.

In this paper, we extend the results of our letter
Ref.~\onlinecite{Nonne2009} by determining the zero temperature phase
diagram of model (\ref{hubbardS}) at half-filling by means of a
low-energy approach in the general $N$ case and DMRG calculations for
$N=2,3,4$.  On top of the confirmation of the Haldane-charge
conjecture, we show that the $N=1$ and $N=2$ cases are special and are
not the generic cases of each family. In particular, for $N>1$ odd,
the metallic phase with dominant singlet-pairing correlation has an
interesting edge-state structure when open-boundary conditions (OBC)
are used, similarly to the spin-3/2 Heisenberg chain
\cite{Ng1994,Ng1995}. For all $N$ even $>2$, a new gapless phase with
dominant singlet-pairing instability is stabilized between the HI
phase and the rung-singlet (RS) phase. In addition, we show, within
the low-energy approach, that the HI phase has striking different
properties
depending on the parity of $N/2$. When $N/2$ is even, the HI phase turns out
to be equivalent to the topologically trivial RS insulating phase
whereas it is a topologically ordered phase when $N/2$ is odd in full
agreement with the recent findings in the study of integer Heisenberg
spin chains \cite{pollman09,tonegawa}.

The rest of the paper is organized as follows.
In Sec. II, we discuss the strong-coupling analysis of
model (\ref{hubbardS}) along
special highly-symmetric lines which give some clues
about the nature of the Mott-insulating phases.
The low-energy approach of the general $N$ case is presented in 
Sec. III. In Sec. IV, we map out the phase diagram of model (\ref{hubbardS}) with $N=2$
by means of intensive DMRG calculations. 
Section V and VI describe our DMRG  results respectively for the $N=3,4$ cases
to complement the low-energy approach.
Finally, our concluding remarks are given in Sec. VII.

\section{Strong-coupling analysis}

Before investigating the zero-temperature phase diagram of model (\ref{hubbardS})
by means of the low-energy and DMRG approaches, a strong-coupling analysis 
along the highly-symmetric lines of the model might be useful to shed light on
the possible Mott-insulating phases. To this end, let us first consider
the energy-spectrum for the single-site problem. 

The Hubbard term of Eq.~(\ref{hubbardS}) distributes the different
states into energy levels with the same number of particles $n$,
with $n=0,\ldots,2N$; this is the one-site spectrum 
of the U($2N$) Hubbard model. The singlet-pairing term in Eq.
(\ref{hubbardS}) with coupling constant $V$ will split 
these levels into levels with different pairing schemes, denoted
$(n,k)$. The level ($n,k$) group states with $n$ particles, among
which $2k$ particles are in $k$ Sp($2N$) singlets. These states
transform in the ${\bar \omega}_{n-2k}$ representation of Sp($2N$). Note
that, for a given number of particles $n$, $0<k<E(n/2)$ if $n\leq
N$, and $n-N<k<E(n/2)$ if $n>N$, where $E(x)$ is the floor
function. In order to write down
the eigenstates in terms of fermionic operators, let us define the
pair operator that creates a pair of fermions with spins $\alpha$
and $2N+1-\alpha$, by:
\begin{equation}
	P^{\dagger}_{\alpha,i}=c_{\alpha,i}^\dagger c_{2N+1-\alpha,i}^\dagger .
\end{equation}
In terms of these operators, the
singlet pairing operator $P^{\dagger}_{00,i}$ is :
\begin{equation}	
P^{\dagger}_{00,i}=\frac{-2}{\sqrt{2N}}
\sum^{N}_{\alpha=1}\left(-1\right)^\alpha P_{\alpha,i}^{\dagger} .
\end{equation}
We now need to define a set of $N-1$ linear combinations of
$P^{\dagger}_{\alpha,i}$, ``orthogonal'' to $P^{\dagger}_{00,i}$
that we label as $\Pi^{\dagger}_{l,i}$ (with $l=1,\cdots,N-1$).
The $n-2k$ particles that are not Sp($2N$) singlets then divide
into two kinds: they can be either written as linear
combinations of pairs of particles with spin
$(\alpha,2N+1-\alpha)$ and thus as a combination of pair operators
$\Pi_{l,i}^\dag$, or they are unpaired and can be only written
with a single creation operator $c_{\alpha,i}^\dag$. In the end,
the eigenstates that belong to the energy level $(n,k)$ are
written as:
\begin{equation}
	|n;k,m\rangle=\frac{1}{{\cal M}_{n,k,m}}
	c_{\alpha_1,i}^\dagger\ldots c_{\alpha_p,i}^\dagger  \Pi^\dagger_{l_1}\ldots\Pi^\dagger_{l_{q}}
	(P^{\dagger}_{00,i})^k|0\rangle,
		\label{SCsetstates}
\end{equation}
where $m$ labels the state, ${\cal M}_{n,k,m}$ is a normalization
factor, $p$ is the number of ``single'' particles, and $2q$ is the
number of ``paired'' particle that cannot be penned down in terms
of $P^{\dagger}_{00,i}$. The energy of the eigenstates
(\ref{SCsetstates}) only depends on $(n,k)$ and reads:
\begin{equation}
	E(n,k)=\frac{n^2}{2}U +\left[2k\left(1+\frac{k+1}{N}-\frac{n}{N}\right)\right]V-n\mu.
	\label{SCenergylevel}
\end{equation}
The energy level $(n,k)$ is ${\cal D}(n,k)$-fold degenerate, with:
\begin{equation}
	{\cal D}(n,k)
	=\frac{2(N-n+2k+1)(2N+1)!}{(n-2k)!(2N-n+2k+2)!}.
	\label{SCspectrumdegeneracy}
\end{equation}

At half-filling, $\mu$ is set by the particle-hole symmetry: 
$\mu=NU+V/N$, and the energy levels read:
\begin{eqnarray}
	E(n,k)&=&\left(\frac{n^2}{2}-nN\right)U\nonumber\\
	&&+\left[2k\left(1+\frac{k+1}{N}
	-\frac{n}{N}\right)-\frac{n}{N}\right]V.
	\label{SCenergylevelhalff}
\end{eqnarray}

At this point, we can consider two important highly-symmetric lines for all $N$: 
$V=0$ (respectively $V = NU$) with the emergence of a U($2N$) (respectively 
SU(2)$_c$ $\times$ Sp(2$N$)) extended symmetry. We also mention that in the special $N=2$ case, there is an additional 
SO(7) symmetric line at half-filling  when $V = - 2 U$ \cite{zhang}.
However, despite the fact that we indeed find an additional degeneracy
for the one-site problem in the general $N$ case on the special line $V=-N^2U/2$, 
the latter seems not to correspond to an enlarged symmetry since
the kinetic term lifts it.

\subsection{Strong-coupling argument close to the $V=0$ line.}
\label{sec:strong-couplingV=0}
\begin{figure}[ht]
\centering
\includegraphics[width=0.95\columnwidth,clip]{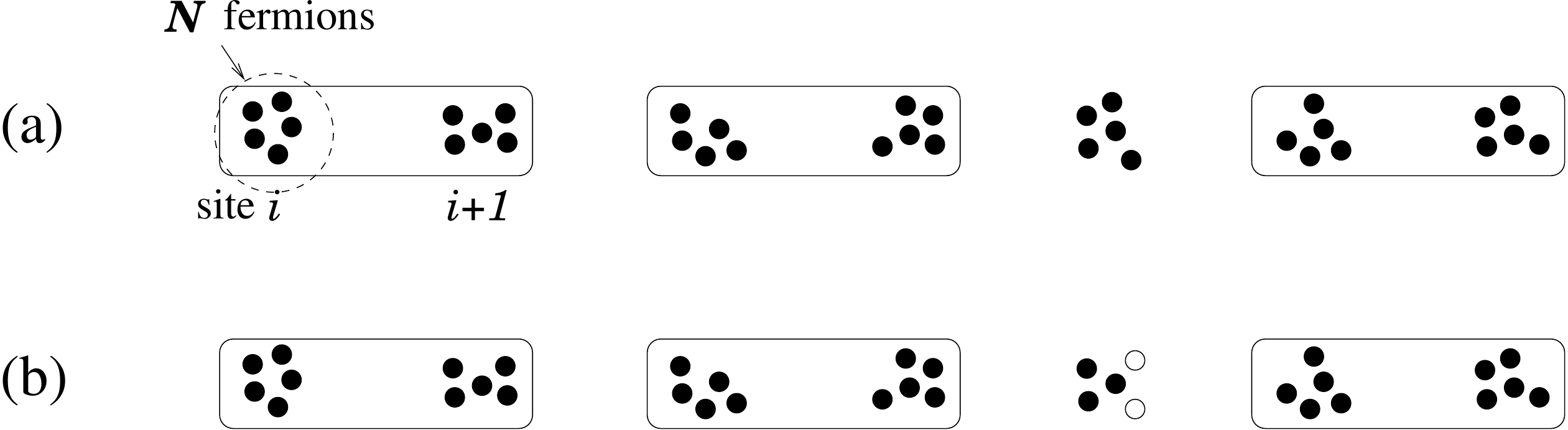}
\caption{Sketch of the kinks supported by the U($2N$) repulsive Hubbard model,
that interpolate between the two degenerate dimerized vacua (the boxes indicate an
SU($2N$) singlet made of $2N$ fermions). Here $N=5$. At large $U$, in the low-energy sector, 
all sites have exactly $N$ fermions. (a) Neutral kinks,
that are the only low-energy excitations at large $U$, and 
transform in the antisymmetric representation $\omega_N$. Note that they are their own antiparticle. 
(b) Charged kinks, that play a 
role at small $U$, depicted here with charge $Q=-2$.}
\label{figU2N_GS_SP}
\end{figure}

When $V=0$, as already stated in the introduction, model (\ref{hubbardS}) is equivalent
to the U($2N$) Hubbard model. The degeneracies of the 
energy-spectrum (\ref{SCenergylevelhalff}) with $V=0$ are related to the dimensions of representations of the 
SU($2N$) group. In particular, when $U>0$, we observe from Eq.~(\ref{SCenergylevelhalff}) that
the lowest-energy states correspond to $n=N$ and transform in  the antisymmetric
self-conjugate representations of SU($2N$) (representation described by a Young tableau with one column
of $N$ boxes). This case has been studied in Ref.~ \onlinecite{afflecksun} and in the strong coupling limit
the model is equivalent to an SU($2N$)  Heisenberg spin chain where the spin operators belong
to  the antisymmetric self-conjugate representation of SU($2N$). The latter model is expected to 
have a dimerized or Spin-Peierls (SP)  two-fold degenerate GS, where dimers are formed between two neighboring sites \cite{afflecksun,marston07}.  In the $N=2$ (i.e. SU(4)), 
such a SP phase has been ascertained by means of a low-energy approach, Quantum Monte Carlo
and DMRG calculations  \cite{assaraf,marston,nonne2010}. 
The strong $U$ limit gives the opportunity to get a simple physical picture of the GS
as well as of the low-lying excitations. The two-fold degenerate GS allows for \emph{kink}
configurations that interpolate between the two vacua. As depicted in  Fig.\ref{figU2N_GS_SP}, these kinks
have zero charge but carry a non-zero SU($2N$) spin since they transform in the 
antisymmetric self-conjugate representation of SU($2N$). Note that the system also allows
for \emph{charged} kinks, that carry charge $Q_k= k-N$ with $k=0,\ldots, 2N$. These states transform in 
$\omega_{k}$, the 
antisymmetric representation of SU($2N$) with Young tableau made
of a single column with $k$ boxes. 
Although at large $U$ they are expected to have a large gap of order
 $\Delta_k\sim U\,(N-k)^2/2$
as seen from (\ref{SCenergylevelhalff}), we nevertheless introduce them here
since they will play an important role
at small $U$ (see Sec. \ref{lowen_generalNcase}). Notice also that these kink 
excitations have a collective nature, i.e. their quantum number cannot be 
reproduced by states built by using a finite number of fermions.

In the attractive case ($U<0$), the lowest energy states are the empty and the fully occupied state, which is an SU($2N$) (and Sp($2N$) as well) singlet. At second order of perturbation theory, 
the effective model is thus: \cite{ueda}
\begin{equation}
	{\cal H}_{\textrm{eff}} = \frac{t^2}{N(2N-1)|U|} \sum_i \left(n_i n_{i+1} - N  n_i\right),
\end{equation}
The first term introduces an effective repulsion interaction between nearest neighbor sites. This leads to a fully-gapped charge-density wave (CDW) where empty and fully occupied states alternate. This phase has a long-range order and is two-fold degenerate.

\subsection{Strong-coupling argument close to the $V=NU$ line.} 
\label{sec:strong-coupling}
The second highly-symmetric line corresponds to the
additional SU(2)$_c$ symmetry  in the charge sector  for $V=NU$  that we have identified in Ref.~\onlinecite{Nonne2009}. On this line, one easily verifies from Eq.~(\ref{SCenergylevelhalff}), 
that all pure $(P^{\dagger}_{00,i})^k$ states (i.e., the states with $n=2k$, $k=0,\ldots,N$) 
are degenerate, with energy $E=0$. Let us give the 
proper normalization factor for these states:
\begin{eqnarray}
	&&|P^k_{00,i}\rangle=\frac{1}{{\cal M}(k)} (P^{\dagger}_{00,i})^k |0\rangle,\nonumber\\ 
	&&\textrm{with } 
	{\cal M}(k)=\sqrt{\left(\frac{2}{N}\right)^k \left(\prod^{k-1}_{q=0}(k-q)(N-q)\right)}.
\end{eqnarray}
They transform in the spin-$N/2$ representation of SU($2$) and we define the 
corresponding pseudo-spin operator acting on them as: 
\begin{eqnarray}\label{spinop.eq}
{\cal S}^{\dagger}_i &=& \sqrt{N/2} \; P^{\dagger}_{00,i}\nonumber\\
{\cal   S}^{z}_i &=& ( n_i - N)/2.
\end{eqnarray}
 This operator carries charge and is a Sp(2$N$) spin-singlet. It generalizes the $\eta$-pairing operator introduced
by Yang for the half-filled spin-1/2 (i.e. $N=1$) Hubbard
model~\cite{yang} or by Anderson in his study of the BCS superconductivity \cite{anderson58}. 
It is easy to observe that ${\vec {\cal S}}_i$
satisfies the SU(2) commutation relations; ${\cal S}^{\dagger}_i$ allows to construct the whole set of 
states with $E=0$ from $|0\rangle$ with:
\begin{eqnarray}
	&&{\cal S}_i^{\dagger}|P^{k}_{00,i}\rangle
	 =\sqrt{(N-k)(k+1)}|P^{k+1}_{00,i}\rangle,\nonumber\\
	&&{\cal S}_i^-|P^{k}_{00,i}\rangle 
	=\sqrt{k(N-k+1)}|P^{k-1}_{00,i}\rangle.
\end{eqnarray}

Let us check the commutation relation of $\vec{{\cal S}}$ with the
Hamiltonian. For the interacting part alone (${\cal H}_{\rm int}= {\cal H}\left[t=0\right]$),
we have (for a generic filling):
\begin{eqnarray}
	&&\left[{\cal H}_{\rm int},{\cal S}_i^{\dagger} \right] 
		=\left[-2\mu -2U \phantom{\frac{2V}{N}} \right.\nonumber\\
	&&\qquad\qquad\qquad\left.+\frac{2V}{N}(N+2)
	+2\left(U-\frac{V}{N}\right) n_i\right]
		{\cal S}_i^{\dagger},\nonumber\\
	&&\left[{\cal H}_{\rm int},{\cal S}_i^z\right] = 0,
\end{eqnarray}
so that they commute only at half-filling and when $V=NU$; as for the
hopping term, ${\cal H}_t=-t\sum_{i,\alpha} \left(c_{\alpha,i}^\dag
c_{\alpha,i+1}^{\phantom \dag} +\textrm{H.c.} \right)$, it commutes with 
the total charge pseudo-spin operator if we define it as:
\begin{eqnarray}
	&&{\cal S}^{\dagger} =\sum_i\left(-1\right)^i {\cal S}^{\dagger}_i,\nonumber\\
	&&{\cal S}^z=\sum_i {\cal S}^z_i .
\end{eqnarray}

The pseudo-spin operator thus generates a higher SU(2)$_c$
$\times$ Sp(2$N$) symmetry at half-filling along the line
$V = N U$ and we can recast the interacting (on-site) Hamiltonian 
as	${\cal H}_{{\rm int}}=2U\sum_i [\vec{{\cal S}}_i^2-N(N+2)/4]$; 
the pseudo-spin ${\vec {\cal S}}_i$ is a spin-$N$/2 operator. 
The existence of such an extended SU(2) symmetry in the
charge sector for $N=2$ has been first noticed in
Ref.~ \onlinecite{wucapponi}.

For a strong attractive $U$, one can derive an effective Hamiltonian 
in the strong-coupling regime $|U| \gg t$ using the standard strategy \cite{auerbach}. To 
second order of perturbation 
theory, one obtains the effective model:
\begin{eqnarray}
	{\cal H}_{\textrm{eff}} = \sum_i \left(J\vec{\cal S}_i \cdot \vec{\cal S}_{i+1} 
		+ D  ({\cal S}_i^z)^2 \right),
\label{HaldaneSU2Hamiltonian}
\end{eqnarray}
with		
\begin{eqnarray}		
   		J&=&\frac{4t^2}{N(2N+1)|U|},\nonumber\\
			D&=&2\left(U-\frac{V}{N}\right).
\label{Haldanecouplings}
\end{eqnarray}
On the SU(2)$_c$ symmetric line ($V = N U$), $D=0$, and model (\ref{HaldaneSU2Hamiltonian}) is the spin-$N$/2 antiferromagnetic SU(2) Heisenberg chain.
From this strong-coupling approach, we thus expect the emergence of an even-odd dichotomy for attractive
interactions along the SU(2) line. For even $N$, i.e. integer pseudo spin,
the HI phase is formed while a metallic phase is stabilized when $N$ is odd, i.e. half-integer pseudo spin.
This is the same as Haldane's conjecture for model (\ref{HaldaneSU2Hamiltonian}) except that the underlying
spin $\vec{{\cal S}}$ is non-magnetic and carries charge. In this respect, we coin it 
Haldane-charge conjecture. 
When we deviate from this SU(2)$_c$  line, the
SU(2)$_c$  charge symmetry is broken down to U(1)$_c$ and the single-ion
anisotropy appears. The phase diagram of the resulting model for general $N$ is
known from the bosonization work of Schulz~\cite{schulz}. For even $N$, on top of
the Haldane phase, N\'eel and large-D singlet gapful phases appear.
Using the expression of the pseudo-spin operator (\ref{spinop.eq}),
the N\'eel and large-D singlet phases correspond respectively to the CDW and RS phases.
When $N$ is odd, gapless (XY) and gapful (Ising) phases are stabilized
in the vicinity of the SU(2) line. The gapless XY phase can be viewed as singlet-pairing phase
since ${\cal S}_i^{\dagger} \sim P^{\dagger}_{00,i}$.

\section{Low-energy approach}
\label{sec:lowenergy}

In this section, we present the low-energy description of the model (\ref{hubbardS})
in the general $N$ case. This will lead us to map out the phase diagram 
at zero temperature and to show the emergence
of the Haldane-charge conjecture in the weak-coupling limit. As it will be shown, the $N=2$ case, which was already
presented in Ref.~\onlinecite{Nonne2009}, turns out to be very particular and is not
the generic case of the even $N$ family.

\subsection{Continuum limit}

The low-energy effective field theory of the lattice model
(\ref{hubbardS}) is derived by taking the standard continuum limit of 
the lattice fermionic operators $c_{\alpha\,i}$, written in terms of 
left- and right-moving $L_\alpha,\,R_\alpha$ Dirac fermions: \cite{bookboso,giamarchi}
\begin{equation}
  \frac{c_{\alpha\,i}}{\sqrt{a_0}}\to R_\alpha e^{ik_F x}+ L_\alpha e^{-i k_F x},
\end{equation}
with $x=ia_0$ ($a_0$ being the lattice spacing) and $k_F = \pi/(2 a_0)$ is the Fermi momentum. 
In the continuum limit, the non-interacting part of the Hamiltonian (\ref{hubbardS}) 
corresponds to the Hamiltonian density of $2N$ free relativistic massless fermions:
\begin{equation}
  {\cal H}_0=-i v_F\left(R_\alpha^\dag \partial_x R_\alpha^{\phantom \dag} - L_\alpha^\dag \partial_x L_\alpha^{\phantom \dag}\right) ,
\end{equation}
where $v_F =  2t a_0$ is the Fermi velocity and we assume in the following a summation over 
repeated indices.
The continuous symmetry of the non-interacting part of the model is enlarged 
to SO($4N$)$|_L\times$SO($4N$)$|_R$ since  $2N$ complex (Dirac) fermions 
are equivalent to $4N$ real (Majorana) fermions. This SO($4N$) symmetry is the maximal continuous 
symmetry of $2N$ Dirac fermions. The corresponding CFT is the SO($4N)_1$ with 
central charge $c= 2N$ \cite{dms}.

The crucial point is now to find a good basis describing the low-energy properties of the model.
Some simple considerations on its symmetries guide us to choose the relevant conformal embedding of 
the problem. No spin-charge separation at half-filling is expected
for $N >1$, and since the global symmetry invariance  of model (\ref{hubbardS})
is Sp($2N$),  we need to understand how the non-interacting  conformal symmetry SO($4N)_1$
decomposes into Sp($2N$)$_1$ CFT. The general list of conformal embeddings can be found in 
Ref.~\onlinecite{cftembedding} and the one which is directly relevant to our problem is: \cite{altschuler}
\begin{equation}
  \mbox{SO}(4N)_1 \sim \mbox{SU}(2)_N \times \mbox{Sp}(2N)_1,
\label{embedding}
\end{equation}
 where the SU(2)$_N$ (respectively 
Sp($2N$)$_1$) CFT has central charge $c = 3N/(N+2)$ (respectively $c = N(2N+1)/(N+2)$).

The next step of the approach is to express the $2N(4N-1)$ SO($4N$)$_1$
currents, which are made from all Dirac fermionic bilinears, in terms of
the currents
of the SU(2)$_N$ and Sp($2N$)$_1$ CFTs, in order to write down the effective
interacting Hamiltonian in the new basis.  To this end, let us consider the following
left currents which appear in the low-energy description of the model
away from half-filling: \cite{Lecheminant2005,phle}
\begin{eqnarray}
  J_L^{A}&=& L_{\alpha}^\dag T^{A}_{\alpha \beta} L_{\beta}, \;\textrm{the SU($2N$)$_1$ spin currents,}\nonumber\\
  J_L^a&=& L_{\alpha}^\dag T^a_{\alpha \beta} L_{\beta}, \;\textrm{the Sp($2N$)$_1$ spin currents,}\nonumber\\
  J_L^i&=& L_{\alpha}^\dag T^i_{\alpha \beta} L_{\beta}, \;\textrm{the SU($2N$)$_1$/Sp($2N$)$_1$ currents,}
  \label{LEcurrents}
\end{eqnarray}
where $T^A$ are the $4N^2-1$ SU($2N$) generators in the fundamental representation, 
$T^a$ being the $N(2N+1)$ Sp($2N$) generators in the fundamental representation, 
and $T^i$ are the $2N^2-N-1$ remaining generators of SU($2N$). 
All these generators are normalized in such a way that they satisfy $\text{Tr}(T^A T^B)=\delta^{AB}/2$. 
We need then to introduce the SU(2)$_N$ currents to complete the conformal embedding.
In this respect, one may use the recognition of the SU(2) pseudo-spin operator (\ref{spinop.eq})
to consider the following left current:
\begin{eqnarray}
   \mathcal{J}_L^{\dagger}
    &=& \frac{1}{2} \; L_\alpha^\dag J_{\alpha\beta} L_\beta^\dag\nonumber\\
    \mathcal{J}_L^z &=& \frac{1}{2}  \; :L_{\alpha}^\dagger L_{\alpha}: ,
    \label{su2Ncurrents}
\end{eqnarray}
where $:\,:$ stands for the normal ordering with respect to the Fermi sea of the non-interacting theory.
Note the  unusual definition of the charge current $\mathcal{J}_L^z$; the $1/2$ factor  in Eq.~(\ref{su2Ncurrents}) 
is there to realize the SU$(2)_N$ algebra: \cite{knizhnik}
\begin{equation}
\mathcal{J}_L^a\left(z\right) \mathcal{J}_L^b\left(0\right) \sim \frac{N \delta^{ab}}{8 \pi^2 
z^2} + 
\frac{i \epsilon^{a b c}}{2 \pi z} \mathcal{J}_L^c\left(0\right),
\label{su2Ncurralg}
\end{equation}
with $a,b = 1,2,3$ and $z = v_F \tau + i x$ ($\tau$ being the imaginary time).

At this point, we have only defined $4N^2 + 2$ currents, and we now have to introduce the 
$4N^2-2N -2$ other pseudo-currents in order to take into account the 
umklapp terms of the form $K_\alpha^\dag K_\beta^\dag, K_\alpha K_\beta$ ($K=L,\,R$). In this respect,
let us consider 
\begin{equation}
       J_L^{i+}
    = L_\alpha^\dag \tilde{T}^i_{\alpha\beta} L_\beta^\dag,
    \label{LEpseudocurrents}
\end{equation}
where the generators $\tilde{T}^i_{\alpha\beta}$ are such that, together with $J_{\alpha\beta}$, they form the set of antisymmetric generators of SU$(2N)$; there are $N(2N-1)-1$ of them, so that all the left 
$2N(4N-1)$ SO($4N$)$_1$ currents are described by: 
$J_L^a, J_L^i, \mathcal{J}_L^{\pm}, \mathcal{J}_L^z, J_L^{i \pm}$, 
with $a = 1, \ldots, N(2N+1)$, and $i = 1, \ldots, 2N^2-N-1$. 

With these currents at hands, we can now derive the low-energy
effective expression of the interacting part of model (\ref{hubbardS}) at half-filling.
The interacting part of this low-energy Hamiltonian can then be deduced by symmetry, simply by requiring the Sp($2N$) invariance:
\begin{eqnarray}
  \mathcal{H}_{\textrm{int}}&=&
  g_1 J_R^aJ_L^a + g_2 J_R^iJ_L^i+g_3 \mathcal{J}_R^z\mathcal{J}_L^z\nonumber\\
  &+&\frac{g_4}{2}(J_R^{i+} J_L^{i-}+\textrm{H.c.}) 
  + \frac{g_5}{2} (\mathcal{J}_R^{+}\mathcal{J}_L^{-}+\textrm{H.c.}),
  \label{LEeffectiveinteracting}
\end{eqnarray}
where we have neglected four-fermion chiral interactions which would only introduce a velocity anisotropy.
A direct continuum limit leads to the identification:
\begin{eqnarray}
  &&g_1=-a_0\left(2U+\frac{4V}{N}\right)\nonumber\\
  &&g_2=-a_0\left(2U-\frac{4V}{N}\right)\nonumber\\
  &&g_3=\frac{2}{N}a_0\left(U(2N-1)+\frac{2V}{N}\right)\nonumber\\
  &&g_4=2U a_0\nonumber\\
  &&g_5=\frac{2}{N}a_0(U+2V) .
\end{eqnarray}
Along the U($2N$) line with $V=0$, we have $g_1 = g_2 = - \frac{N}{2N-1} g_3 = -g_4 = - N g_5$.
When $V=NU$, model (\ref{LEeffectiveinteracting}) displays a manifest 
SU(2)$_c$ $\times$ Sp(2$N$) invariance with $g_2 = g_4$ and $g_3 = g_5$.
The interaction of model (\ref{LEeffectiveinteracting}) is marginal, so that a one-loop 
RG analysis can be performed to deduce its low-energy properties.

\subsection{Phase diagram for $N=2$}
\label{lowenergy_N2}
The RG analysis for $N=2$ has been presented in details in
Refs. ~\onlinecite{Nonne2009, nonne2010}.  For completeness and
especially for the comparison with the DMRG calculations of Sec. IV,
we give here a brief account of the main results of this case.

\begin{figure}[t]
\centering
\includegraphics[width=0.75\columnwidth,clip]{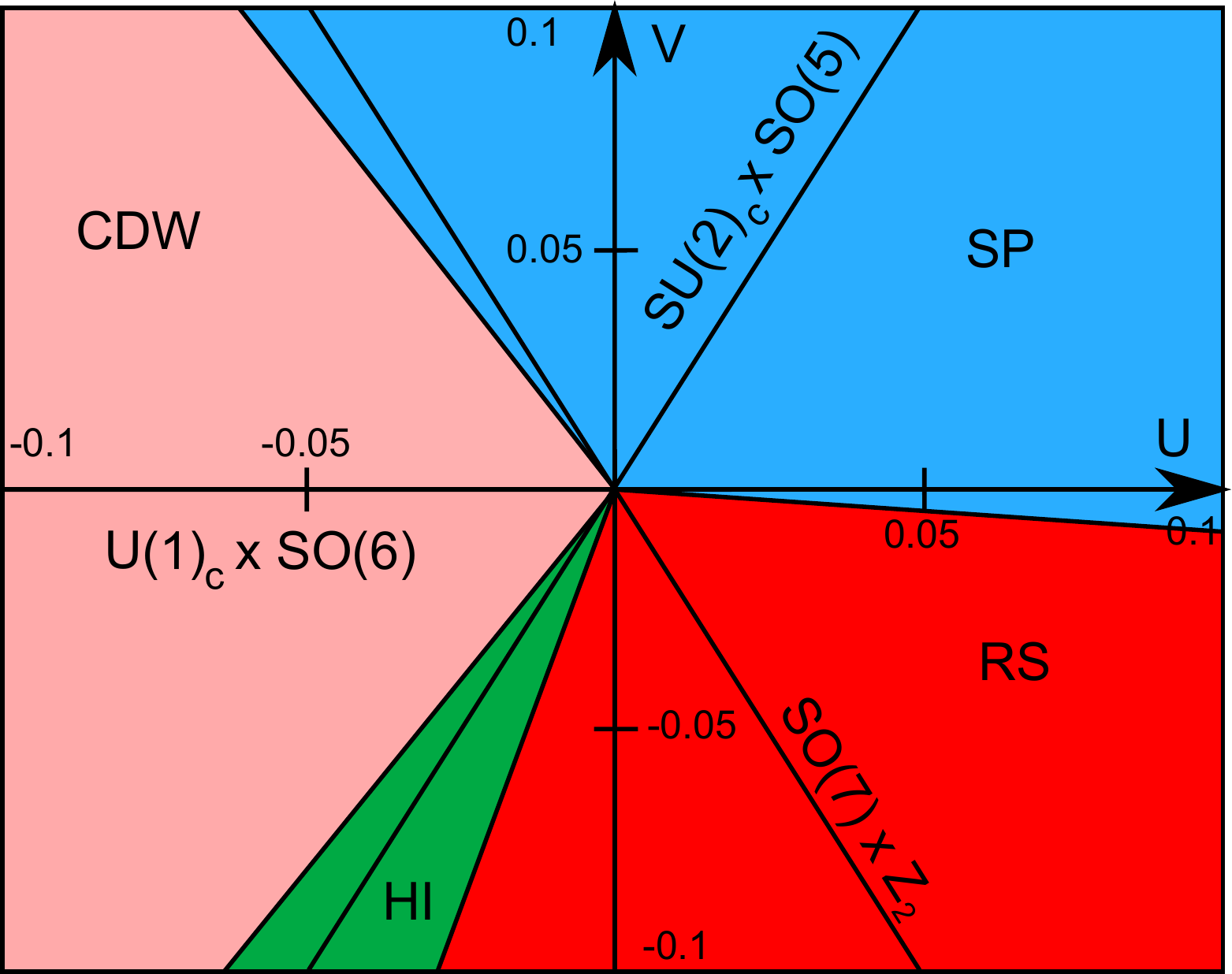}
\caption{Phase diagram obtained by the low-energy approach in the $N=2$ case.}
\label{figRGN2}
\end{figure}

Since Sp(4)$_1$ $\sim$ SO(5)$_1$ and SU(2)$_2$ $\sim$ SO(3)$_1$, the
currents in Eq.~(\ref{LEeffectiveinteracting}) admit a free-field
representation in terms of eight Majorana fermions: $\xi_{R,L}^{a},
a=1,\ldots, 5$ describe the fluctuations of the Sp(4) spin degrees of
freedom whereas $\xi_{R,L}^{6,7,8}$ account for the remaining ones. The
interacting Hamiltonian (\ref{LEeffectiveinteracting}) reads as
follows in terms of these Majorana fermions:
\begin{eqnarray}
{\cal H}_{\rm int} &=&
\frac{g_1}{2} \; \left(\sum_{a=1}^{5} \xi_R^{a} \xi_L^{a} \right)^2
+ g_2 \; \xi_R^{6} \xi_L^{6}
\sum_{a=1}^{5} \xi_R^{a} \xi_L^{a}
\nonumber \\
&+& \frac{g_3}{2} \; \left(\xi_R^{7} \xi_L^{7} 
+ \xi_R^{8} \xi_L^{8}  \right)^2
 \label{majofthalfilling} \\
&+& \left( \xi_R^{7} \xi_L^{7} + \xi_R^{8} \xi_L^{8} \right)
\left( g_4 \;
\sum_{a=1}^{5} \xi_R^{a} \xi_L^{a}
+ 
g_5 \; \xi_R^{6} \xi_L^{6} \right).
\nonumber
\end{eqnarray}
One particularity of the Majorana fermion basis
is that it allows for a very simple representation
 of  non-perturbative hidden duality symmetries in
the low-energy effective Hamiltonian (\ref{majofthalfilling}). These discrete 
symmetries are very useful to determine the zero-temperature phase diagram \cite{boulat}.
For model (\ref{majofthalfilling}), we can define two independent duality symmetries:
\begin{eqnarray}
 \Omega_1&:&\quad \xi^{7,8}_L \to -\xi^{7,8}_L\nonumber\\
 \Omega_2&:&\quad \xi^{6}_L \to -\xi^{6}_L, 
 \label{dualitiesN2}
 \end{eqnarray}
 while the right-moving fermions remain invariant. The transformations (\ref{dualitiesN2})
are exact symmetries of Eq.~(\ref{majofthalfilling}) if the couplings are simultaneously
changed according to
$g_{4,5} \rightarrow  - g_{4,5}$  for $\Omega_1$, and $g_{2,5} \rightarrow  - g_{2,5}$ 
for the second duality $\Omega_2$.
These duality symmetries along with the one-loop RG equations enable us 
to map out the phase diagram of the $N=2$ case.
This analysis has been done in Refs.  ~\onlinecite{Nonne2009, nonne2010} 
and we find four insulating phases in the phase diagram, depicted in Fig.~\ref{figRGN2} .

A first two-fold degenerate phase, which contains the SU(4) line with repulsive $U$,
is a SP phase with  a non-zero order parameter ${\cal O}_{\rm SP} = \sum_{
  \alpha} (-1)^i c^{\dagger}_{\alpha,i+1} c_{\alpha,i}$. The duality symmetry $\Omega_1$
gives a second gapful two-fold degenerate phase which is a CDW phase
with order parameter ${\cal O}_{\rm CDW} = \sum_{\alpha} (-1)^i c^{\dagger}_{\alpha,i}
c_{\alpha,i}$.  This phase contains the SU(4) line with negative $U$, in full 
agreement with the numerical result of Ref.  ~\onlinecite{ueda}.
On top of these two-fold degenerate phases, there are two non-degenerate insulating
phases which are stabilized with help of the $\Omega_2$ duality symmetry.
Starting from the CDW phase and applying the $\Omega_2$ transformation, 
one obtains a  HI phase which includes the SU(2)$_c$ line 
$V = 2 U$ with attractive $U$. This gapful non-degenerate phase is equivalent to the Haldane
phase of the spin-1 Heisenberg chain and displays a hidden ordering
which can be revealed through a non-local string order parameter. This order parameter
is built from the pseudo-spin operator (\ref{spinop.eq}) and the HI phase is characterized
by the long-range ordering:
\begin{equation}
\lim_{|i-j| \rightarrow \infty}
\langle {\cal   S}^{z}_i
e^{ i \pi \sum_{k=i+1}^{j-1}
{\cal   S}^{z}_k}
{\cal   S}^{z}_j 
\rangle \ne 0.
\label{stringcharge}
\end{equation}
As a consequence of this ordering, this phase displays  pseudo-spin-1/2 edge states which 
carry charge but are spin-singlet states (holon edge states) \cite{Nonne2009}. 
Finally, the last insulating phase is obtained from this HI phase by applying the $\Omega_1$ duality. 
One obtains a gapful non-degenerate RS phase, equivalent to the RS phase of the two-leg
spin ladder with antiferromagnetic interchain coupling \cite{bookboso,giamarchi}.
This RS phase has no
edge states and is characterized by the string-ordering: \cite{Nonne2009}:
\begin{equation}
\lim_{|i-j| \rightarrow \infty}
\langle 
e^{ i \pi \sum_{k=i+1}^{j-1}
{\cal   S}^{z}_k} 
\rangle \ne 0.
\label{stringRS}
\end{equation}
Finally, the different quantum phase transitions of Fig.~\ref{figRGN2} can also be determined
by means of the duality symmetries (\ref{dualitiesN2}). 
The transitions SP/CDW and HI/RS are Berezenskii-Kosterlitz-Thouless (BKT)  transitions with central charge $c=1$
whereas SP/RS and CDW/HI transitions belong to the 2D Ising universality class
with central charge $c=1/2$.  

\subsection{Renormalization Group analysis - General $N$ case}
\label{lowen_generalNcase}
\begin{figure}
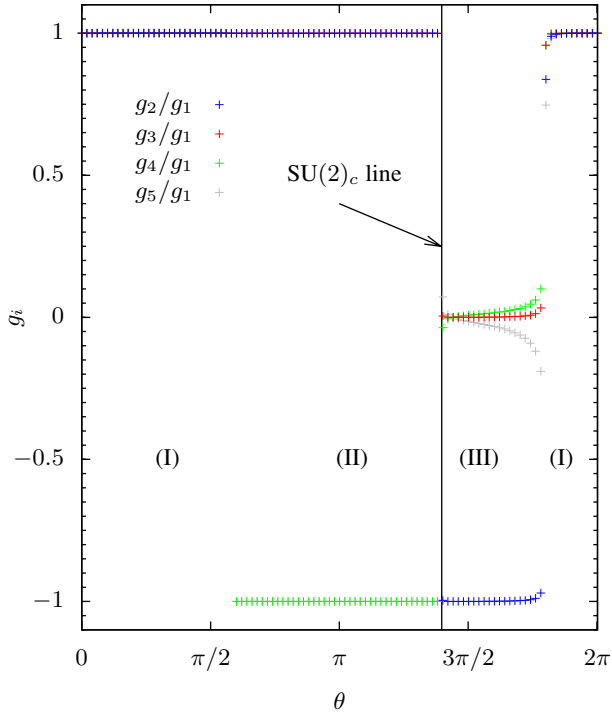

\begin{center}
\include{figringdiagramN3}
\caption{Ratio of the coupling constants, to which the RG flow leads in the IR limit, for $N=3$; $\theta$ spans a ring with fixed radius $R = 0.1t$ in the phase diagram: $U = R \cos\theta$, 
$V = R \sin \theta$. Three different
regions can be defined as function of $\theta$.}
\label{figRGN3}
\end{center}
\end{figure}

We now turn to the general $N >2$ case which is much more involved.
The leading effects of the current-current interaction  of 
model (\ref{LEeffectiveinteracting}) can be inferred from a one-loop RG approach.
In this respect, it is useful to rescale the coupling constants as:
$g_{1,2,4} \to  2\pi v_F  N g_{1,2,4}$ and $g_{3,5} \to  2\pi v_F g_{3,5}$, to obtain the one-loop RG equations:
\begin{eqnarray}
  \dot{g_1}&=&\frac{N}{2}\left[ (N+1) g_1^2 + (N-1) g_2^2 +2 (N-1)g_4^2\right]\nonumber\\
  \dot{g_2}&=&N^2 g_1g_2 +(N^2-N-2) g_4^2 + 2 g_4 g_5\nonumber\\
  \dot{g_3}&=&(2N^2-N-1)g_4^2+g_5^2\nonumber\\
  \dot{g_4}&=&N^2 g_1g_4 +(N^2-N-2) g_2 g_4 + g_3 g_4 + g_2 g_5\nonumber\\
  \dot{g_5}&=&(2N^2-N-1) g_2 g_4+ g_3 g_5, 
  \label{RGbetafunctionredef}
\end{eqnarray}
where ${\dot g}_i = \partial g_i/ \partial l (i =1,\ldots, 5)$, $l$ being the RG time.

Here, we remark the particular character of the case $N=2$, where the term in $N^2-N-2$ cancel 
out in the equation for $g_{2}$ and $g_4$. It is this very term that makes 
the general $N$ case tricky. Indeed, because of it, the duality symmetry 
$\Omega_2$ of the $N=2$ case (\ref{dualitiesN2}) disappears for $N>2$
and, with it, fades out a very satisfactory way to precisely identify  and 
characterize the different phases in the phase diagram.
The only duality symmetry $\Omega_1$ which remains when $N>2$,
corresponds to the transformation $ L_\alpha \to i L_\alpha$ on the left-moving Dirac fermions, so that 
\begin{eqnarray}
 \Omega_1: \mathcal{J}_L^{\pm} \to - \mathcal{J}_L^{\pm} , \;
J_L^{i \pm} \to - J_L^{i \pm} ,
 \label{dualityomega1}
\end{eqnarray}
while the other currents are invariant. 
Thus, the duality transformation $\Omega_1$ 
is an exact symmetry of model (\ref{LEeffectiveinteracting}) 
when $g_{4,5}  \to -  g_{4,5}$. In particular, the RG Eqs. (\ref{RGbetafunctionredef})
are indeed symmetric with respect to $g_{4,5}  \to -  g_{4,5}$.

We have solved numerically the RG equations by a standard Runge-Kutta method. In order to have a picture of the RG flow, it is useful to draw diagrams on a ring defined by $R^2=U^2+V^2$, $U=R\cos{\theta}$, $V=R\sin{\theta}$
with $\theta=0\dots 2\pi$; in our numerical calculations, we set $R=0.1 t$. The procedure is the following: we initiate the couplings $g_i$ for a given value of $\theta$ and run the Runge-Kutta algorithm on the RG Eqs. (\ref{RGbetafunctionredef}). The coupling constants $g_i$ flow to the strong coupling regime under RG time so we need to stop the procedure at one point. 
To this end, we stop the RG iterations as soon 
as one of the coupling reaches a limit value $G$; this happens at RG time $l_0$, and defines a mass scale $\Lambda=a_0^{-1}\, e^{-l_0}$ that gives an estimate of the largest gap in the model ($a_0^{-1}$ is a UV cutoff).
At this point, we extract the values of all the $g_i(l_0)$ and draw them, renormalized by $g_1(l_0)$ ($g_i/g_1(l_0)$) on the ring diagram.
We re-initiate the couplings for a new $\theta$ and restart the procedure for all values of $\theta$. 
 The resulting diagram looks very similar for all $N>2$ and different regimes can 
be defined (see Fig.~\ref{figRGN3} for $N=3$) as a function of the lattice coupling constants $U$ and $V$.
Two qualitatively different behaviors of the flow can be identified in the asymptotic limit of weak coupling, where $l_0$ is large (small $R a_0$).
In regions (I) and (II) of Fig.~\ref{figRGN3}, at large $l_0$, the ratios $g_i/g_1$ do not evolve anymore with the RG time and have already reached fixed values when the RG iterations are stopped.
On the other hand, in region (III), $g_1$ is always the first coupling to reach the limit value $G$ at which we stop the RG ; the ratios $g_i/g_1(l_0)$ for $i=3,4,5$ vanish in the weak coupling limit: 
\begin{equation}
\lim_{R\to 0} \big[g_i/g_1(l_0) \big]=0, 
\label{flowIII}
\end{equation}
whereas the ratio $g_2/g_1(l_0)$ remains finite. This property will be important when we will derive effective models to describe this last region.
We now turn to the description of the physical properties of the different regimes.

In the region (I) of Fig.~\ref{figRGN3}, all coupling constants of the low-energy effective
Hamiltonian (\ref{LEeffectiveinteracting}) flow to strong coupling in the infrared (IR) limit
at fixed ratio: $g_i/g_1 = 1$ ($ i=1,\ldots, 5$). Along this special isotropic ray, model  (\ref{LEeffectiveinteracting}) 
displays an extended global SO($4N$) symmetry and becomes equivalent to 
the SO($4N$) Gross-Neveu (GN) model \cite{GN}, in the sense that the low-energy spectrum model of 
(\ref{LEeffectiveinteracting}) is adiabatically connected to that of the SO($4N$) GN model -- for a precise discussion see Ref. ~\onlinecite{boulat}. 
This phenomenon is an example of a dynamical symmetry enlargement by the interactions
as found in half-filled two-leg Hubbard ladders or in the half-filled U(4) Hubbard chain,
with the emergence of an SO(8) symmetry that becomes
asymptotically exact in the weak-coupling limit \cite{lin,assaraf,saleur}. 

The  SO($4N$) GN  model is a massive integrable field theory
whose mass spectrum is known exactly \cite{zamolo,karowski}. It consists of the elementary
fermions with mass $m$, their bound states, and of kinks.
The bound states have masses ($N >1$):
\begin{equation}
m_n = m\; \frac{\sin \left( \frac{\pi n}{2\left(2 N - 1\right)} \right)}{\sin \left( \frac{\pi }{2  \left(2N - 1\right)} \right)},
 \label{masspecO4N}
\end{equation}
with $n=2, \ldots, 2 N - 2$, while the kinks' mass reads:
\begin{equation}
m_{\rm kinks} = \frac{m}{2 \sin \left( \frac{\pi }{2  \left(2N - 1\right)} \right)} .
 \label{kinkO4N}
\end{equation}
The $N=2$ case is special since the kinks mass is equal to that of the fermions.
The SO(8) GN model enjoys a triality symmetry which has been exploited in the 
study of the half-filled two-leg Hubbard ladder \cite{lin,konik}. In the $N>2$ case, 
the lowest excitations are the fermions which transform into the vectorial representation
of the SO(4$N$) group: they have the same quantum numbers as the original fermions $R_\alpha,L_\alpha$
and $R_\alpha^\dagger,L_\alpha^\dagger$. 

The kinks transform in the spinorial representations
of SO($4N$). It is more transparent to characterize these states by giving their charge and spin quantum numbers
under U(1)$_c$ and SU($2N$) respectively: the
kinks are those $2^{2N}$ states that carry charge $Q_k = k-N$ and transform in the $\omega_k$ 
representation 
of SU($2N$) where $k$ varies
from $0$ to $2N$. One can distinguish even and odd kinks, that transform
in the even (odd respectively) spinorial representation and corresponds to even (odd respectively) $k$'s.
 In particular, the low-energy spectrum of the SO(4$N$) GN model
contains Sp(2$N$) spin-singlet states with charge $Q = \pm N$ which 
can be viewed as the generalization of the Cooperon excitations of the $N=2$ case \cite{lin,konik}.
These kink states identify with those discussed in Sec. \ref{sec:strong-couplingV=0}.

The development of the strong-coupling regime in the SO($4N$) GN model
leads to the generation of a spectral gap and the formation of a SP phase for all $N>1$
with order-parameter: 
\begin{equation}
 {\cal O}_{\rm SP}  = 
 i \left( L_\alpha^\dagger R_\alpha - H.c.\right).
\label{SP}
\end{equation}
This order parameter is the continuum limit of the Spin Peierls operator 
${\cal O}_{\rm SP}= (-1)^i\sum_{\alpha}c^\dagger_{\alpha,i+1}c_{\alpha,i}$ 
and it has a non-zero expectation value in the GS
as can be seen by a direct semi-classical approach of the SO($4N$) GN model.
The phase is two-fold degenerate and breaks spontaneously the one-step
translation symmetry ($T_{a_0}$): $L_\alpha \to -i L_\alpha,  R_\alpha \to i R_\alpha$ 
since ${\cal O}_{\rm SP} \to - {\cal O}_{\rm SP}$ under  $T_{a_0}$. This SP phase contains
the U$(2N)$ line ($V=0$) with $U >0$, i.e. the repulsive U($2N$) Hubbard model.
 
The second region of Fig.~\ref{figRGN3} can be easily determined 
with help of the duality symmetry $\Omega_1$. The transformation of the SO($4N$) 
isotropic line under $\Omega_1$ is  $1 = g_2/g_1 = g_3/g_1 = - g_4/g_1= - g_5/g_1$, which turns
out to be the  asymptote of the RG flow in the region (II) of Fig.~\ref{figRGN3}.
We thus deduce a second Mott-insulating phase which is obtained from the SP phase
by applying the duality symmetry $\Omega_1$. Since $ L_\alpha \to i L_\alpha$ under $\Omega_1$,
its order parameter can be obtained from Eq.~(\ref{SP}):
\begin{equation}
 {\cal O}_{\rm CDW}  = 
  L_\alpha^\dagger R_\alpha + H.c.,
\label{CDW}
\end{equation}
which is nothing but the continuum limit of the $2k_F$ 
CDW operator: ${\cal O}_{\rm CDW} = \sum_{\alpha} (-1)^i c^{\dagger}_{\alpha,i}
c_{\alpha,i}$.  Region (II) is thus a fully gapped CDW phase. This CDW phase contains
the U$(2N)$ line ($V=0$) with $U <0$. 

What happens in region (III) of Fig.~\ref{figRGN3} is clearly of a different nature:
the RG flow displays no symmetry enlargement, and we will have to develop
other tools to tackle the physics in this interesting region. This will be done in the next
section, where we will reveal striking differences according to the parity of $N$.
Before that, we would like to give
hand-waving arguments, based on the spectrum of the SO($4N$) GN model, 
supporting this even-odd scenario. To understand what happens
to the system when one leaves regions (I) and (II) where  symmetry enlargement occurs, one should
recast   the whole particle content of the SO($4N$) GN
model in multiplets of the internal continuous symmetry group of our problem, 
namely Sp($2N$)$\times$U(1)$_c$. One already knows
how the SO($4N$) multiplets split into U(1)$_c$ $\times$ SU($2N$) representations, that we write
 $(Q,\lambda)$ with $\lambda$ an SU($2N$) weight and $Q$ the U(1)$_c$ charge (the number
 of fermions measured with respect to the GS). Denoting   the vectorial representation 
(to which the GN "fundamental fermions" belongs)
by ${\cal V}$,  the even spinorial
(to which even kinks belong) by $S^{(+)}$, and the odd spinorial
(to which odd kinks belong) by $S^{(-)}$, one has:
\begin{eqnarray}
{\cal V}&=&(1, \omega_1)\oplus(-1,\omega_{2N-1})\nonumber\\
S^{(+)}&=&\oplus_{k=0}^{N}(2k-N,\omega_{2k})\nonumber\\
S^{(-)}&=&\oplus_{k=0}^{N-1}(2k+1-N,\omega_{2k+1}) .
\end{eqnarray}
A quick way to check those quantum numbers is to note that they
must be compatible with the fundamental fermions being a boundstate
of two kinks.  
 The only missing piece of information
 is the splitting of the SU($2N$) representations $\omega_k$ into
Sp$(2N$) representation. Denoting by $\bar \omega_k$ the $k^{\mbox{\scriptsize th}}$ Sp($2N$) 
fundamental representation, one has: $\omega_{2n}=\oplus_{k=0}^n\bar\omega_{2k}$
and $\omega_{2n+1}=\oplus_{k=0}^n\bar\omega_{2k+1}$, so that any 
SU($2N$) representation $\omega_{2n}$ contains one and only one Sp($2N$) singlet,
whereas all states of $\omega_{2n+1}$ carry non-zero Sp($2N$) spin. 

Now one notices that 
region (III) of Fig.~\ref{figRGN3} occurs at negative
$V$, where the system tends to favor Sp($2N$) singlets. Let us
assume that there is adiabatic continuity in the low-energy part
of the spectrum. Then, the quantum numbers of the lowest energy modes can be obtained
by looking at those states in the SO($4N$) GN spectrum that are 
Sp($2N$) singlets.  It results that  when $N$ is even, one expects
the "elementary" charged particle (with the smallest U(1)$_c$ charge) to carry
charge $Q= \pm 2$. On the other hand, when $N$ is odd, there is a kink state
that is a Sp($2N$) singlet \emph{and} carries charge $Q=\pm 1$. We will shortly see that
this even-odd dichotomy does indeed occur, and that the elementary
charged particles have the aforementioned charges.

\subsection{Even-odd scenario}

The last region of the RG flow of Fig.~\ref{figRGN3}, i.e. region (III), is difficult to analyze
due to the absence of the second duality symmetry $\Omega_2$ when  $N>2$.
In this region, which includes the SU(2)$_c$ line with $V = NU$ and $U<0$,
the operator with coupling constant $g_1$ in the low-energy effective
Hamiltonian (\ref{LEeffectiveinteracting}) reaches the strong-coupling regime
before the others. 
In the limit of weak coupling, one has a separation of energy scales, due to the property (\ref{flowIII}) of the RG flow.
Neglecting all other couplings for the moment, the corresponding perturbation is an integrable massive field
theory for $g_1 >0$ \cite{ahn90,babichenko}.
A spin gap $\Delta_s$ thus opens  for the Sp($2N$) spin sector in region (III).
The next step of the approach is to integrate out these spin degrees of freedom to derive
an effective Hamiltonian in the low-energy limit $E \ll \Delta_s$ from which the physical
properties of region (III) will be deduced.

\subsubsection{Parafermionization}

The resulting low-energy effective Hamiltonian involves the remaining degrees of freedom
of the initial conformal embedding (\ref{embedding}), i.e., the SU(2)$_N$ sector. Since
the global continuous symmetry of model (\ref{LEeffectiveinteracting})  is, in general,
U(1)$_c$ $\times$  Sp($2N$), we need to understand how we go from the SU(2)$_N$ CFT
to the U(1)$_c$ one.  Such a mapping is realized by the conformal embedding:
$\ZZ_N$ $\sim$  SU(2)$_N$ / U(1)$_c$, which defines the $\ZZ_N$ parafermionic CFT series
with central charge $c= 2 (N -1)/(N+2)$ \cite{para,gepner}. This CFT describes the critical properties of 
two-dimensional $\ZZ_N$ generalizations of the Ising model.
The $\ZZ_N$ CFT is generated by the parafermionic
currents $\Psi_{k L,R}$  ($k = 1, \ldots, N$) with scaling dimensions $h_k = k (N - k)/N$.

The different operators of Eq.~(\ref{LEeffectiveinteracting}) can be written in this parafermionic basis. 
First of all, the SU(2)$_N$ currents (\ref{su2Ncurrents}) can be directly expressed in terms of 
$\Psi_{1 L,R}$ and a bosonic field $\Phi_{c}$ which accounts for charge fluctuations: \cite{para}
\begin{eqnarray}
 \mathcal{J}_{L,R}^{\dagger} 
&\simeq& \frac{\sqrt{N}}{2\pi}
:\exp \left(\pm i \sqrt{8 \pi/N} \; \Phi_{c L,R} \right): \Psi_{1 L,R}
\nonumber \\
 \mathcal{J}_{L,R}^{z} 
&\simeq& \sqrt{\frac{N}{2\pi}} \partial_x \Phi_{cL,R} ,
\label{parasu2Ncurrent}
\end{eqnarray}
where the charge field  $\Phi_{c} = \Phi_{c L} + \Phi_{c R}$ is a compactified bosonic field with radius 
$R_c =  \sqrt{N/2\pi} $: $ \Phi_{c}  \sim \Phi_{c} +   \sqrt{2 \pi N}$.
The remaining currents of Eq.~(\ref{LEeffectiveinteracting}) can also be expressed in terms of the parafermionic
degrees of freedom using the results of Ref. ~\onlinecite{phle}: 
\begin{eqnarray}
 J_{L}^{i} J_{R}^{i} 
&\sim& \epsilon_1 {\rm Tr} {\phi}^{(2)} 
\nonumber \\
 J_L^{i+} J_R^{i-}
&\sim& \mu_2 {\rm Tr} {\phi}^{(2)} \exp \left(i \sqrt{8 \pi/N} \; \Phi_{c} \right),
\label{pararepcurr}
\end{eqnarray}
where ${\phi}^{(2)}$ is the second primary operator of the Sp($2N$)$_1$ CFT with scaling dimension
$2N/(N+2)$. In Eq.~(\ref{pararepcurr}), $\epsilon_1$ is the first thermal operator
of the $\ZZ_N$ CFT with scaling dimension $4/(N+2)$, and $\mu_2$ is the second disorder
operator with scaling dimension $2(N-2)/N(N+2)$ which orders when 
the $\ZZ_N$ symmetry is not spontaneously broken \cite{para}.

Before investigating the low-energy limit $E \ll \Delta_s$, it is crucial to analyze 
the hidden discrete symmetries of model (\ref{LEeffectiveinteracting}) which become explicit
thanks to the conformal embedding.
It is well known that the $\ZZ_N$ CFT has a global  $\ZZ_N$ $\times$ $\widetilde{\ZZ}_N$ 
discrete symmetry under which the parafermionic currents $\Psi_{k L}$ (respectively $\Psi_{k R}$) carry
a $(k,k)$ (respectively $(k,-k)$) charge: \cite{para}
\begin{eqnarray}
\Psi_{k L,R} &\rightarrow& e^{i 2 \pi m k/N} \Psi_{k L,R} \; \; {\rm under} \; \; \ZZ_N 
\nonumber \\ 
\Psi_{k L,R} &\rightarrow& e^{\pm i 2 \pi m k/N} \Psi_{k L,R} 
\; \; {\rm under} \; \; \widetilde{\ZZ}_N  ,
\label{chargepara}
\end{eqnarray}
with $m=0,\ldots, N-1$.
The thermal operator $\epsilon_1$ transforms as a singlet under these discrete symmetries
while the order and disorder operators $\sigma_k$, $\mu_{k}$ carry respectively  
a $(k,0)$ and $(0,k)$ charge:
\begin{eqnarray}
\sigma_k &\rightarrow& e^{i 2 \pi m k/N} 
\sigma_k \; \; {\rm under} \; \; \ZZ_N
\nonumber \\
\mu_{k} &\rightarrow& e^{i 2 \pi m k/N} \mu_{k}
\; \; {\rm under} \; \; \widetilde{\ZZ}_N  ,
\label{chargeorderdisorder}
\end{eqnarray}
and $\sigma_k$ (respectively $\mu_{k}$) remains unchanged
under the $\widetilde{\ZZ}_N$ (respectively $\ZZ_N$) symmetry.
The $\ZZ_N$ symmetry of the parafermions 
has a simple interpretation in terms of the original lattice fermions or the Dirac fermions 
of the continuum limit. It is nothing but a special phase transformation $ c_{\alpha,i} \to  e^{- i \pi m/N} 
c_{\alpha,i}$ or, in the continuum description:
\begin{equation}
 L_\alpha \to  e^{- i \pi m/N} L_\alpha, \;  \;  \;  R_\alpha \to  e^{- i \pi m/N} R_\alpha,
\label{ZN}
\end{equation}
with $m=0, \ldots, N -1$. This $\ZZ_N$ symmetry leaves invariant model (\ref{LEeffectiveinteracting}), and
the correspondences (\ref{parasu2Ncurrent}, \ref{pararepcurr}) are also compatible with the definition (\ref{ZN}).
In contrast, the $\widetilde{\ZZ}_N$ symmetry of the parafermions does not exist on the lattice. Away from half-filling,
it becomes an independent emergent symmetry of the model in 
the continuum limit and takes the form: \cite{Lecheminant2005,phle}  
\begin{equation}
L_{\alpha} \rightarrow e^{-i \pi m /N} L_{\alpha}, \; \;
R_{\alpha} \rightarrow e^{i \pi m /N} R_{\alpha} .
\label{tildeznfer}
\end{equation}
At half-filling, this transformation is no longer a symmetry of model (\ref{LEeffectiveinteracting})
due to the umklapp operators. The $\widetilde{\ZZ}_N$ symmetry has a more subtle role here: its combination with the following identification on the charge bosonic field:
\begin{equation}
\Phi_c \sim \Phi_c - m \sqrt{\frac{2 \pi}{N}} + p \sqrt{\frac{N \pi}{2}}, 
\; m=0, \ldots, N -1,
\label{tildeznchargebose}
\end{equation}
becomes a symmetry of model (\ref{LEeffectiveinteracting}), as it can be seen from 
Eq.~(\ref{pararepcurr}). In fact, this symmetry is a gauge redundancy since it corresponds
to the identity in terms of the Dirac fermions.
The last important discrete symmetries of the problem 
are the $\Omega_1$ duality transformation (\ref{dualityomega1})
and the one-step translation invariance ($T_{a_0}$), which only affect the charge field:
\begin{eqnarray}
 {\Omega_1:}\; \Phi_c  &\to& \Phi_c + \frac{1}{2} \sqrt{\frac{N \pi}{2}} \nonumber \\
  T_{a_0}:  \Phi_c  &\to& \Phi_c + \sqrt{\frac{N \pi}{2}} .
 \label{dualitytrans}
\end{eqnarray}

\subsubsection{Low-energy Hamiltonian}

We are now in position to derive the low-energy limit $E \ll \Delta_s$ 
by integrating out the gapful Sp($2N$) degrees of freedom. 
Using the parafermionization formulae (\ref{parasu2Ncurrent}, \ref{pararepcurr}),  one finds:
\begin{eqnarray}
\mathcal{H}_{\textrm{int}}&=&
\lambda_2  \epsilon_1+ \lambda_3 \partial_x \Phi_{cL} \partial_x \Phi_{cR}  
\nonumber \\
 &+& \frac{\lambda_4}{2} \left( \mu_2 \exp \left(i \sqrt{8 \pi/N} \; \Phi_{c} \right)
 + \textrm{H.c.} \right) \label{effectiveHam} \\
 &+&  \frac{\lambda_5}{2}  \left(  \Psi_{1 L}  \Psi^{\dagger}_{1 R}
 \exp \left(i \sqrt{8 \pi/N} \; \Phi_{c} \right) + \textrm{H.c.} \right),
\nonumber
\end{eqnarray}
where $\lambda_{2,4,5} \simeq \langle {\rm Tr} {\phi}^{(2)}  \rangle g_{2,4,5}$,
and $\lambda_{3} = g_3 N/2 \pi$.
The low-energy Hamiltonian (\ref{effectiveHam}) enables us to explore the 
whole phase diagram of the model for all $N$. Along the SU(2)$_c$ line with $V=NU$,
model (\ref{effectiveHam}) can be written in terms of the SU(2)$_N$ fields:
\begin{equation}
\mathcal{H}_{\textrm{int}} = g_3 {\vec{\cal{J}}}_R \cdot {\vec{\cal{J}}}_L
+ \lambda_2 {\rm Tr} {\Phi}^{(1)},
\label{heffsu2}
\end{equation}
where ${\Phi}^{(1)}$ is the spin-1 primary field of the SU(2)$_N$ CFT with scaling
dimension $4/(N+2)$. The effective Hamiltonian (\ref{heffsu2}) is the low-energy
theory of the spin-$N/2$ SU(2) Heisenberg chain derived by Affleck and Haldane
in Ref. ~\onlinecite{affleckhaldane}.  As shown by these authors, model (\ref{heffsu2}) has
a spectral gap, when $N$ is even,  while it describes a massless flow to the SU(2)$_1$ CFT when
$N$ is odd, in full agreement with Haldane's conjecture  \cite{affleckhaldane}. The latter result
has also be found by means of a parafermionic approach similar to Eq.~(\ref{effectiveHam}) in presence of 
an SU(2) symmetry  \cite{cabra}.

The crucial point to map out the general phase diagram of the low-energy Hamiltonian (\ref{effectiveHam})
for all $N$ stems from the status of the $\ZZ_N$ symmetry (\ref{ZN}). 
The first term in Eq. (\ref{effectiveHam}) describes an integrable deformation of the $\ZZ_N$ 
CFT which is always a massive field theory for all sign of $\lambda_2$ \cite{fateev}.
In our conventions, if $\lambda_2 >0$ (respectively $\lambda_2 < 0$) the $\ZZ_N$ symmetry
is unbroken (respectively spontaneously broken) and the disorder fields (respectively 
order fields) condense: $\langle \mu_k \rangle \ne 0$ (respectively $\langle \sigma_k \rangle \ne 0$)
for all $k=1,\ldots, N$.

Let us first re-investigate the emergence of the CDW, SP phases
in regions (I, II)
within this parafermionization approach. 
When $\lambda_2 >0$ (i.e. $g_2 >0$),  the $\ZZ_N$ symmetry remains
unbroken and one may integrate out the gapful parafermionic degrees of freedom to 
derive an effective field theory on the charge bosonic field. Since we have $\langle \mu_2 \rangle \ne 0$
and $ \langle \Psi^{\phantom \dagger}_{1 L}  \Psi^{\dagger}_{1 R} \rangle \ne 0$ in the $\ZZ_N$ high-temperature phase,
we obtain from Eq. (\ref{effectiveHam}):
\begin{eqnarray}
\mathcal{H}_{c} &=& \frac{v_c}{2} \left( \frac{1}{K_c} \left(\partial_x \Phi_c\right)^{2}
+ K_c \left(\partial_x \Theta_c\right)^{2} \right) \nonumber \\
&+& g_c  \cos \left(\sqrt{8 \pi/N} \; \Phi_{c} \right),
\label{hcZN}
\end{eqnarray}
where the Luttinger parameter is given by:
\begin{equation}
K_c =  \frac{1}{\sqrt{1 + g_3N/(2\pi v_F)}}.
\label{Luttingerpara}
\end{equation}
The low-energy Hamiltonian for the charge degrees of freedom (\ref{hcZN}) is the well-known 
sine-Gordon model at $\beta^2  =  8 \pi K_c /N$. We thus deduce the existence of a
charge gap when $K_c < N$ which is always the case at weak coupling as seen from
Eq. (\ref{Luttingerpara}). The nature of the Mott-insulating phase depends on the 
sign of $g_c$ which is changed by the duality transformation $ \Omega_1$ (\ref{dualitytrans}).
When $g_c <0$ (i.e. $g_4 <0$), the development of the strong-coupling regime of the sine-Gordon 
model (\ref{hcZN}) is accompanied by the pinning of the charged field on the minima:
$\langle \Phi_{c} \rangle = p \sqrt{N \pi/2}$, $p$ being an integer.
Since we have the identification $ \Phi_{c}  \sim \Phi_{c} +   \sqrt{2 \pi N}$ due to the periodicity
of the charge field, we deduce that the insulating phase is two-fold degenerate with
minima: $\langle \Phi_{c} \rangle = 0$ and  $\langle \Phi_{c} \rangle = \sqrt{N \pi/2}$, i.e.
the one-step translation symmetry (\ref{dualitytrans}) is spontaneously broken.
The low-lying excitations are massive kinks and antikinks which interpolate between
the two GS. The charges associated to these excitations are
 \begin{equation}
Q = \pm  \sqrt{2N/ \pi} \int dx \;  \partial_x \Phi_c =  \pm N.
\label{chargekink}
\end{equation}
For $N=2$, the excitations correspond to 
the Cooperon excitations of the half-filled two-leg Hubbard ladder \cite{lin}.  
The charge excitations (\ref{chargekink}) correspond to the generalization of these
Cooperons. That they are the charge excitations with the minimal charge can be deduced
from considerations on symmetry: amongst the  spectrum of the SO($4N$) GN model,
they are the only charged states that are both Sp($2N$) singlets and  neutral under $\ZZ_N$ (indeed,
the $\ZZ_N$ charge of any state can be simply read off from the way it transforms under SU($2N$): 
states in $\omega_k$ carry a $\ZZ_N$ charge given by $k\mbox{ mod }N$).

The physical nature of the two-fold degenerate Mott-insulating phases can be determined
by expressing the SP and CDW order parameters (\ref{SP}, \ref{CDW}) in terms of the charge
and the $\ZZ_N$ fields: \cite{phle}
\begin{equation}
L^{\dagger}_{\alpha} R_{\alpha}
   \sim 
\exp \left(i\sqrt{2 \pi/N} \; \Phi_c\right) \mu_1 {\rm Tr} \; \phi^{(1)} ,
\label{masstermrep}
\end{equation}
where $ \phi^{(1)}$ is the first Sp($2N$)$_1$ primary field with scaling dimension $(2N+1)/2 (N+2)$.
Averaging over the Sp($2N$) and $\ZZ_N$ degrees of freedom, we obtain:
\begin{eqnarray}
{\cal O}_{\rm CDW}   &\sim& \cos \left(\sqrt{2 \pi/N} \; \Phi_c\right) \nonumber \\
{\cal O}_{\rm SP}   &\sim& \sin \left(\sqrt{2 \pi/N} \; \Phi_c\right) .
\label{CDWSP}
\end{eqnarray}
The phase with $g_c <0$ (i.e. $g_4 <0$) is thus a CDW phase ($\langle {\cal O}_{\rm CDW}  \rangle
\ne 0$) and corresponds to the 
region (II) of Fig.~\ref{figRGN3}. The second phase with $g_c > 0$ is obtained
from the CDW phase by the application of the duality transformation $ \Omega_1$.
The pinnings of the charge field are then: 
$\langle \Phi_{c} \rangle = \sqrt{N \pi/8}$ and  $\langle \Phi_{c} \rangle = 3 \sqrt{N \pi/8}$,
which signals the formation of a SP phase in the region (I) of Fig.~\ref{figRGN3}
since from Eq. (\ref{CDWSP}), $\langle {\cal O}_{\rm SP}  \rangle
\ne 0$.
The quantum phase transition between the CDW/SP phases belongs to the self-dual
manifold of the duality symmetry $ \Omega_1$. Using the definition (\ref{dualitytrans}), 
one finds that the low-energy Hamiltonian of the transition is given by
\begin{eqnarray}
\mathcal{H}_{\rm selfdual} &=& \frac{v_c}{2} \left( \frac{1}{K_c} \left(\partial_x \Phi_c\right)^{2}
+ K_c \left(\partial_x \Theta_c\right)^{2} \right) \nonumber \\
&+& g_c  \cos \left(\sqrt{32 \pi/N} \; \Phi_{c} \right).
\label{hcselfdual}
\end{eqnarray}
The resulting quantum phase transition is of BKT type.  The transition displays 
a quantum-critical behavior with one gapless bosonic mode if $K_c > N/4$. At this point,
we need complementary numerical techniques to extract 
the value of $K_c$ in order to conclude on the nature of the transition. 

Finally, the case with $\lambda_2 < 0$  (or $g_2 < 0$)  corresponds to region (III) of  Fig.~\ref{figRGN3}
where  the $\ZZ_N$ symmetry is now spontaneously broken. In this $\ZZ_N$ low-temperature
phase, the $\ZZ_N$ degrees of freedom are still fully gapped and the disorder operators now average  to zero:
$\langle \mu_k \rangle = 0$. Similarly to the $\lambda_2 > 0$ case, we can integrate out the parafermionic fields
to obtain an effective field theory on the charge bosonic field. However, due to the presence
of the $\mu_2$ operator in Eq. (\ref{effectiveHam}), the resulting integration strongly depends
on the parity of $N$.
 
\subsubsection{Phase diagram in the $N$ odd case ($N>1$)}\label{part:odd_phase_diagram}

\begin{figure}
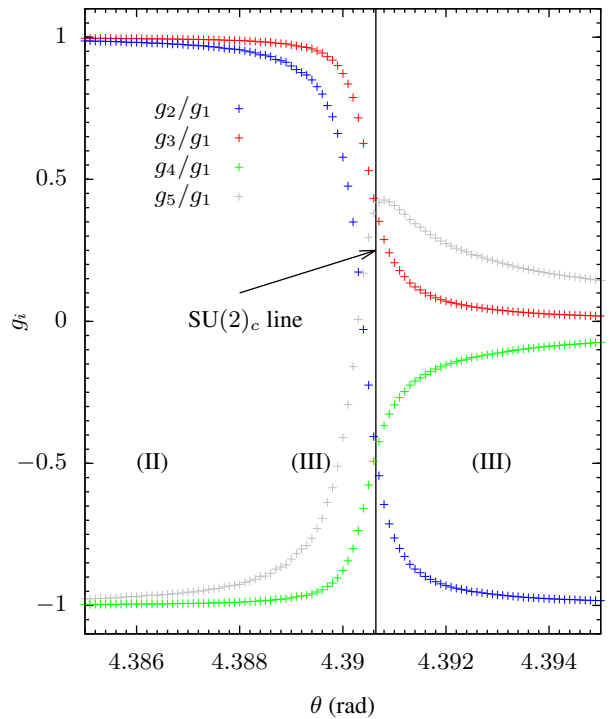

\begin{center}
\include{figringdiagramN3SU2}
\caption{Values of the coupling constants $g_i$ in the infrared limit, in the vicinity of the transition from CDW phase (II) to BCS critical phase (III), close to the 
SU(2)$_c$ symmetric line ($V=NU$). Notations are the same as in Fig.~\ref{figRGN3}.}
\label{figRGN3SU2}
\end{center}
\end{figure}
Let us first consider the case where $N$ is odd. 
Since all the  parafermionic operators in Eq.   (\ref{effectiveHam}) average to zero in 
the $\ZZ_N$ broken phase, one has to consider higher orders in perturbation theory
to derive an effective theory for the charge field.  The $\ZZ_N$ fields of model (\ref{effectiveHam})  carry a charge 2
under the $\widetilde{\ZZ}_N$ symmetry (see Eq. (\ref{chargeorderdisorder}) 
with $k=2$ for $\mu_2$). When $N$ is odd, one has to use the $N$th order
of perturbation theory to cancel out the $\widetilde{\ZZ}_N$  charge of $\mu_2$
so that we find:
\begin{eqnarray}
\mathcal{H}^{\rm odd}_{c} &=& \frac{v_c}{2} \left( \frac{1}{K_c} \left(\partial_x \Phi_c\right)^{2}
+ K_c \left(\partial_x \Theta_c\right)^{2} \right) \nonumber \\
&+& g_c  \cos \left(\sqrt{8 \pi N} \; \Phi_{c} \right),
\label{hcNodd}
\end{eqnarray}
with $g_c  \sim g^N_4$, while we do not have any estimate of the Luttinger
parameter except the bare one (\ref{Luttingerpara}). 
On symmetry grounds, the effective Hamiltonian (\ref{hcNodd}) can also be derived 
by finding the vertex operator in the charge sector with the smallest
scaling dimension which is compatible with translational invariance (\ref{dualitytrans})
and the redundancy  (\ref{tildeznchargebose}).
The resulting low-energy Hamiltonian  (\ref{hcNodd})
takes the form of a  sine-Gordon model at $\beta^2  =  8  \pi N K_c$ so that
a charge gap opens when $K_c < 1/N$. For $K_c =1/N$, this sine-Gordon model  
displays a hidden SU(2) symmetry which should correspond to the SU(2)$_c$ line $V=NU$ 
with $U<0$ that belongs to region (III). 
Close to this SU(2)$_c$ line, the RG flow of Fig.~\ref{figRGN3SU2} shows that the coupling
constant $g_4$ is negative so that $g_c<0$.  When $K_c < 1/N$,  the charge bosonic field is thus pinned
on the minima:
$\langle \Phi_{c} \rangle = p \sqrt{\pi/2N}$, $p$ being an integer.
Taking into account the gauge redundancy (\ref{tildeznchargebose}), we find 
that the strong-coupling phase of 
the sine-Gordon model  (\ref{hcNodd}) is two-fold degenerate with $\langle \Phi_{c} \rangle = 0$
and $\langle \Phi_{c} \rangle = \sqrt{\pi/2N}$.
The charges of the  massive kinks and antikinks excitations are now:
 \begin{equation}
Q = \pm \sqrt{2N/ \pi} \int dx \;  \partial_x \Phi_c =  \pm 1,
\label{chargekinkCDW}
\end{equation}
in sharp contrast to the  charge $Q=\pm N$ of excitations (\ref{chargekink}) of the CDW phase of region (II).
At this point, we need to find a local order parameter to fully characterize the two-fold
degenerate Mott insulating phase in region (III).
When the Sp($2N$) and $\ZZ_N$ degrees of freedom are 
integrated out, the expression of the bilinear Dirac fermions (\ref{masstermrep}) is naively 
short-ranged in region (III), since it contains the first disorder parameter.
However, by fusing this operator with the Hamiltonian (\ref{effectiveHam}) at the $(N-1)/2$ th order of 
perturbation theory, the disorder operator cancels out and one obtains the following low-energy
description:
\begin{equation}
L^{\dagger}_{\alpha} R_{\alpha}
   \sim 
\exp \left(i\sqrt{2 \pi N} \; \Phi_c\right).
\label{masstermrepIII}
\end{equation}
In region (III), where the $\ZZ_N$ is spontaneously broken, the CDW and SP operators then read as follows: 
\begin{eqnarray}
{\cal O}_{\rm CDW}   &\sim& \cos \left(\sqrt{2 \pi N} \; \Phi_c\right) \nonumber \\
{\cal O}_{\rm SP}   &\sim& \sin \left(\sqrt{2 \pi N} \; \Phi_c\right) ,
\label{CDWSPregionIII}
\end{eqnarray}
so that $\langle {\cal O}_{\rm CDW}  \rangle \ne 0$.
The insulating phase in region (III) when $K_c < 1/N$ is thus the continuation of the CDW 
phase of region (II). However, there is a striking difference at the level of the low-lying excitations:
the generalization of the Cooperon excitations with charge $Q= N$ is no longer
a stable excitation in region (III) but becomes a diffusive state made of the kinks
 (\ref{chargekinkCDW}) which are massive holons. The situation is very similar 
 to the SP phase of the half-filled U(4) Hubbard model between the weak and strong coupling regimes \cite{assaraf}.
As far as the GS properties are concerned, 
there is a smooth crossover when 
the $\ZZ_N$ symmetry changes its status at $\lambda_2 =0$ and not a $\ZZ_N$ quantum
phase transition as it is the case away from half filling \cite{Lecheminant2005,phle}.
\begin{figure}
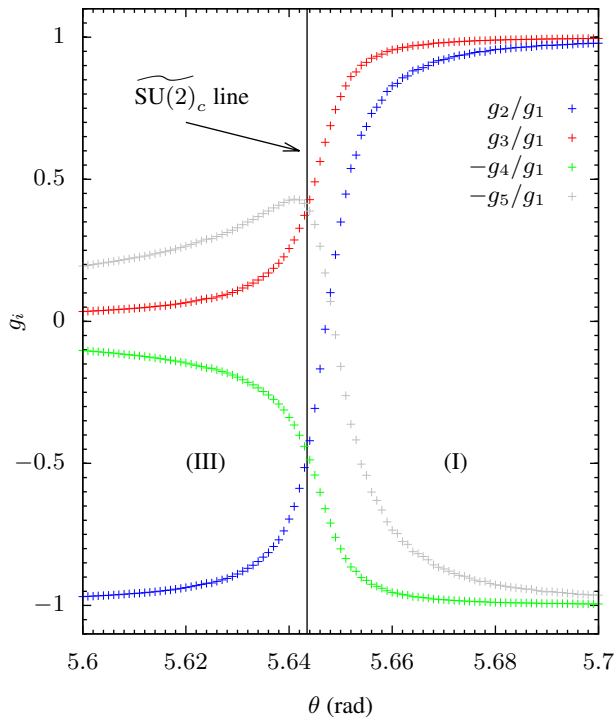

\begin{center}
\include{figringdiagramN3SU2dual}
\caption{Values of the coupling constants $g_i$ in the IR limit, in the vicinity of the transition between the BCS (III) and SP (I) phases, displaying the 
restoration of an SU$(2)$ symmetry, denoted $\widetilde{{\rm SU}(2)}_c$, which is the dual of the 
lattice SU(2)$_c$ symmetry 
($V=NU$): $g_2=-g_4$ and $g_3=-g_5$; we used the same notations as in Fig.~\ref{figRGN3}.}
\label{figRGN3SU2dual}
\end{center}
\end{figure}

When $K_c > 1/N$, the charge degrees of freedom become  gapless.
We then need to determine the leading instability of this phase, 
i.e. the one that has the slowest decaying correlation functions.
The singlet-pairing operator can be expressed in terms of the charge and the $\ZZ_N$ fields as: \cite{phle}
\begin{equation}
P^{\dagger}_{00}
   \sim 
\exp \left(i\sqrt{2 \pi/N} \; \Theta_c\right) \sigma_1 {\rm Tr} \; \phi^{(1)} .
\label{Pairingrep}
\end{equation}
Since the $\ZZ_N$ symmetry is broken, we have $\langle \sigma_1\rangle \ne 0$ and 
the low-energy representation of the singlet-pairing operator is thus:
$P^{\dagger}_{00} \sim \exp \left(i\sqrt{2 \pi/N} \; \Theta_c\right)$.
The gapless phase stems from the competition of this singlet-pairing operator, which
cannot condense, and the CDW operator (\ref{CDWSPregionIII}).
The leading asymptotics of their equal-time correlation functions
can then be straightforwardly determined:
\begin{eqnarray}
&&\langle  P^{\dagger}_{00} \left(x\right)   P_{00} \left(0\right) \rangle
\sim  A \; x^{- 1/NK_c} \nonumber \\
&&\langle   n \left(x\right)  n \left(0\right) \rangle \sim
-\frac{NK_c}{\pi^2 x^2} +  \left(-1\right)^{x/a_0} B \; x^{-NK_c},
\label{correlBCS}
\end{eqnarray}
where $n(x)$ is the continuum limit of the lattice density operator $n_i$, and 
$A,B$ are non-universal amplitudes. Since $K_c > 1/N$, the leading instability of this gapless
phase is the BCS singlet-pairing.

The quantum phase transition between the gapful CDW phase and the gapless BCS phase
occurs at $K_c =1/N$ which corresponds to the SU(2)$_c$ ($V=NU<0$) line. 
On this line, we observe that the exponents of the correlation functions
of Eq. (\ref{correlBCS}) are identical. Using the pseudo-spin operator (\ref{spinop.eq}),
we deduce the following leading asymptotics from Eq. (\ref{correlBCS}):
\begin{eqnarray}
\langle  {\cal S}^{\dagger} \left(x\right)   {\cal S}^{-} \left(0\right) \rangle
&\sim&   x^{- 1} \nonumber \\
\langle   {\cal S}^{z}\left(x\right)  {\cal S}^{z} \left(0\right) \rangle &\sim&
\left(-1\right)^{x/a_0} \; x^{-1}.
\label{correlsu2}
\end{eqnarray}
The model with $K_c =1/N$ displays  a quantum critical behavior with central charge
$c=1$ for all odd $N$ and corresponds to the SU(2)$_1$ universality class.
This result is in perfect agreement with the strong-coupling analysis of Sec. II
along the SU(2)$_c$ line, where the pseudo-spin operator (\ref{spinop.eq}) is a spin-$N/2$, i.e. half-integer,
operator. The low-energy properties of SU(2) half-integer Heisenberg spin chains are indeed
known to be governed by the SU(2)$_1$ CFT \cite{affleckhaldane}.
In the spin language, the CDW and BCS phases are respectively the analog of the Ising and XY phases
and the quantum phase transition occurs at the SU(2) Heisenberg point.

When we deviate from the SU(2)$_c$ line in the gapless BCS phase,
Figs. (\ref{figRGN3SU2}, \ref{figRGN3SU2dual}) show that $g_3$
decreases and then increases as a function of the interaction.  Using
the naive estimate of the Luttinger parameter (\ref{Luttingerpara}),
we deduce that $K_c$ increases from $K_c=1/N$ at the SU(2)$_c$ line and
then decreases until one reaches the SP phase of region (I).  The
resulting transition and its properties can be deduced from the
CDW/BCS transition by the duality symmetry $\Omega_1$. Indeed, under
the transformation (\ref{dualitytrans}), the sign of $g_c$ of model
(\ref{hcNodd}) is changed and the gapful insulating phase when $K_c <
1/N$ is two-fold degenerate with $\langle \Phi_{c} \rangle =
\sqrt{\pi/8N}$ and $\langle \Phi_{c} \rangle = 3\sqrt{\pi/8N}$. The SP
order parameter (\ref{CDWSPregionIII}) acquires a non-zero expectation
value in this phase: $\langle {\cal O}_{\rm SP} \rangle \ne 0$.  As far as
the GS properties are concerned, this phase is the
continuation of the SP phase of region (I). The quantum phase
transition between the BCS and SP phases occurs at $K_c=1/N$. Its
position corresponds to an SU(2) line with $g_2 = - g_4$, and $g_3 = -
g_5$ which is obtained from the lattice SU(2)$_c$ line $V=NU<0$ ($g_2 =
g_4$, $g_3 = g_5$) by the application of the duality symmetry
$\Omega_1$. The resulting SU(2) line, noted $\widetilde{\mbox{SU(2)}}_c$ in
Fig.~\ref{figRGN3SU2dual}, does not exist on the lattice: it is an
emergent SU(2) symmetry of the continuum limit.

\begin{figure}
\centering
\includegraphics[width=0.75\columnwidth,clip]{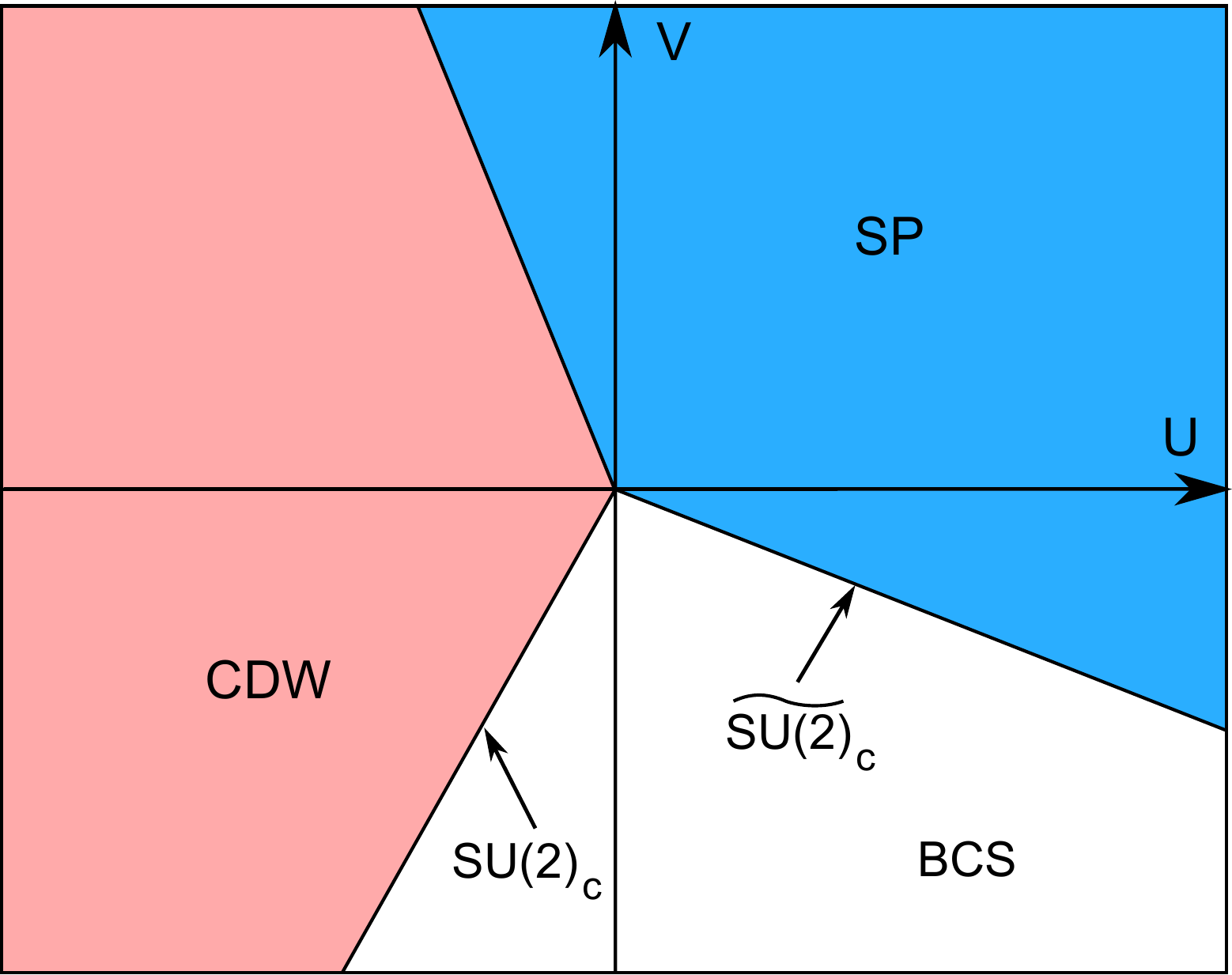}
\caption{Phase diagram obtained by the low-energy approach in the $N$ odd case ($N>1$).}
\label{PhDiagNodd}
\end{figure}

As a summary, Fig.~\ref{PhDiagNodd} shows the zero-temperature phase diagram of 
model (\ref{hubbardS}) 
in terms of the lattice parameters $U,V$ in the $N$ odd case ($N>1$), which results
from the low-energy approach.

\subsubsection{Phase diagram in the $N$ even case ($N>2$)}\label{part:even_phase_diagram}

As in the $N$ odd case, one has to consider higher orders in perturbation theory
to derive an effective theory for the charge field $\Phi_c$  since all the 
parafermionic operators in Eq.   (\ref{effectiveHam}) average to zero in the $\ZZ_N$ broken phase.
When $N$ is even, one needs the $N/2$ th order
of perturbation theory to cancel out the $\widetilde{\ZZ}_N$  charge of $\mu_2$.
The resulting low-energy Hamiltonian then reads as follows:
\begin{eqnarray}
\mathcal{H}^{\rm even}_{c} &=& \frac{v_c}{2} \left( \frac{1}{K_c} \left(\partial_x \Phi_c\right)^{2}
+ K_c \left(\partial_x \Theta_c\right)^{2} \right) \nonumber \\
&+& g_c  \cos \left(\sqrt{2 \pi N} \; \Phi_{c} \right) .
\label{hcNeven}
\end{eqnarray}
Alternatively, the effective Hamiltonian (\ref{hcNeven}) can also be 
obtained by considering the vertex operator in the charge sector with the smallest
scaling dimension which is compatible with translational invariance (\ref{dualitytrans})
and the gauge redundancy  (\ref{tildeznchargebose}).
The resulting low-energy Hamiltonian  (\ref{hcNeven})
takes the form of a  sine-Gordon model at $\beta^2  =  2 \pi N K_c$ so that
a charge gap opens when $K_c < 4/N$. 
One checks that, right on the
SU(2)$_c$ line where the Luttinger exponent is constrained ($K_c=1/N$), the
 sine-Gordon parameter takes the special value $\sqrt{2\pi}$, at which it is known
 that a hidden SU(2) symmetry emerges. \cite{afflecksu2} The lowest energy modes are a massive triplet,
 the magnon of the integer spin Heisenberg model.
Turning back to the generic situation  $K_c < 4/N$ where a charge gap opens,
the charge bosonic field is pinned into the following configurations 
\begin{eqnarray}
\langle \Phi_{c} \rangle &=& p \sqrt{\frac{2\pi}{N}}, \;  \;  \;  \;  \;  \;  \;  \;  \;  \;   \;  \;  \;  \;  \;  {\rm if}   \;  g_c < 0 \nonumber \\
\langle \Phi_{c} \rangle &=& \sqrt{\frac{\pi}{2N}} + p \sqrt{\frac{2\pi}{N}}, \; {\rm if}   \;  g_c > 0 ,
\label{pinningNeven}
\end{eqnarray}
$p$ being an integer. The lowest massive excitations are
the soliton and antisoliton of the sine-Gordon model; they carry charge
 \begin{equation}
Q = \pm  \sqrt{2N/ \pi} \int dx \;  \partial_x \Phi_c =  \pm 2,
\label{chargekinkHaldane}
\end{equation}
which correspond to the Cooperon excitations. Using the gauge redundancy (\ref{tildeznchargebose}), we find that, in sharp
contrast to the CDW and SP phases, the insulating phase when $K_c < 4/N$ is non-degenerate, its GS being:
\begin{eqnarray}
\langle \Phi_{c} \rangle &=& 0, \;  \;  \;  \;  \;  \;  \;   \; \; {\rm if}   \;  g_c < 0 \nonumber \\
\langle \Phi_{c} \rangle &=& \sqrt{\frac{\pi}{2N}}, \; {\rm if}   \;  g_c > 0 .
\label{pinningGSNeven}
\end{eqnarray}
Starting from the CDW phase of region (II), where the $\ZZ_N$ symmetry is unbroken, 
there is necessarily a quantum phase transition to the non-degenerate Mott-insulating
phase  of region (III) with broken $\ZZ_N$ symmetry.
In particular, the disorder parameter $\mu_1$ of 
Eq. (\ref{masstermrep}) cannot be compensated using
higher orders of perturbation theory as was the case for odd $N$. It means that in region (III):
$\langle {\cal O}_{\rm CDW}  \rangle = 
\langle {\cal O}_{\rm SP}  \rangle =  0$ when the Sp($2N$) and $\ZZ_N$ degrees freedom are integrated out.
It is natural to expect that the non-degenerate insulating phases, 
described by the pinning (\ref{pinningGSNeven}), signal the emergence of the HI and RS phases
that we have identified in the strong-coupling approach (\ref{HaldaneSU2Hamiltonian}).
At this point, it is worth observing that the duality symmetry $\Omega_1$ plays a subtle
role in the even $N$ case. Indeed, under the transformation (\ref{dualitytrans}), the 
cosine term of Eq. (\ref{hcNeven}) transforms as
\begin{equation}
 \cos \left(\sqrt{2 \pi N} \; \Phi_{c} \right) \to \left(-1 \right)^{N/2} \cos \left(\sqrt{2 \pi N} \; \Phi_{c} \right),
\label{dualcosNeven}
\end{equation}
so that there is room for an interesting $N/2$ even-odd effect.

\emph{\underline{$N/2$ even case}.}

Let us first consider the $N/2$ even case.  A naive estimate of the coupling
constant $g_c$ in higher orders of perturbation theory gives: $g_c \sim - g_4^{N/2}$.
The RG flow close to the SU(2)$_c$ line ($V=NU<0$) in the  $N/2$ even case is similar to the one in
Fig.~\ref{figRGN3SU2}. In this region, we have $g_4<0$ so that the non-degenerate 
gapful phase is described by the locking $\langle \Phi_{c} \rangle = 0$ of Eq. (\ref{pinningGSNeven}).
As seen in Fig.~\ref{figRGN3SU2}, this region contains the 
SU(2)$_c$ line where the strong-coupling analysis (\ref{HaldaneSU2Hamiltonian}) predicts the emergence of  
the spin-$N/2$, i.e. even spin, SU(2) Heisenberg chain. The low-lying excitation of the resulting Haldane
phase is a gapped triplet state. From the expression of the pseudo-spin operator (\ref{spinop.eq}),
 one observes that it corresponds  
 to a Cooperon excitation in full agreement with the prediction (\ref{chargekinkHaldane}).
We thus conclude that the Mott-insulating phase
in the vicinity of the SU(2)$_c$ line ($V=NU<0$), which is described by the sine-Gordon
model (\ref{hcNeven}) with $\langle \Phi_{c} \rangle = 0$,  is  the HI phase. 

The topological order of the Haldane phase with integer spin $S= N/2 >1$ has been
less understood than the $S=1$ case. This phase displays edge states 
with localized spin $N/4$ when OBC are used \cite{Ng1994}. Unfortunately, we are not able to describe 
these boundary edge excitations by means of our low-energy approach except when $N=2$
\cite{orignac,Nonne2009,nonne2010}. On top of these end-chain states, the higher integer-spin Haldane phase
should exhibit a non-local string ordering \cite{oshikawa,totsuka,hatsugai,schollwock,totsukaspin2,aschauer,qin}.
A very naive guess is to use the generalization of the string-order parameter (\ref{stringcharge})
with spin-$N/2$ operator. In the low-energy limit, we find for $N/2$ even:
\begin{align}
\lim_{|i-j| \rightarrow \infty}
&\langle 
{\cal S}^{z}_i 
e^{ i \pi \sum_{k=i+1}^{j-1} {\cal S}^{z}_k}
{\cal S}^{z}_j
\rangle  \simeq  \nonumber \\
\lim_{|x-y| \rightarrow \infty}
&\langle{\sin \left( \sqrt{N \pi/2} \; \Phi_c \left( x \right)  \right) 
 \sin \left( \sqrt{N \pi/2} \; \Phi_c \left( y \right)  \right)}\rangle \nonumber \\
 &= 0 \;,
\label{stringchargeN4}
\end{align}
since the HI phase is described by the pinning $\langle \Phi_{c} \rangle = 0$.
This result is in full agreement with what is known at the Affleck, Kennedy, Lieb, Tasaki (AKLT)
point \cite{AKLT} of the integer-spin Heisenberg chain \cite{oshikawa,totsuka},  and also from
DMRG studies of the spin-2 Heisenberg chain \cite{hatsugai,schollwock,totsukaspin2,aschauer,qin}.
A simple non-zero string order parameter in the HI phase,  that we can estimate within our low-energy approach,
is
\begin{align}
\lim_{|i-j| \rightarrow \infty}
&\langle{\cos \left(\pi \sum_{k<i} {\cal   S}^{z}_k \right)
\cos \left(\pi \sum_{k<j} {\cal   S}^{z}_k \right)}\rangle  \simeq  \nonumber \\
\lim_{|x-y| \rightarrow \infty}
&\langle{\cos \left( \sqrt{N \pi/2} \; \Phi_c \left( x \right)  \right) 
 \cos \left( \sqrt{N \pi/2} \; \Phi_c \left( y \right)  \right)}\rangle \nonumber \\
 &\ne 0 .
\label{HaldanestringN4}
\end{align}
This lattice order parameter turns out to be non-zero at the AKLT point of even-spin 
Heisenberg chains \cite{totsukacom}. 

In summary, when $N/2$ is even, the HI phase
is described at low-energy by the sine-Gordon model (\ref{hcNeven}) with $K_c < 4/N$
and a non-degenerate GS $\langle \Phi_{c} \rangle = 0$.
The quantum phase transition between the CDW and HI phases
is difficult to determine exactly. On general grounds, we expect an Ising quantum phase
transition or  a first-order one due to the difference of the GS degeneracies
between the two phases. In the CDW and HI phases, the charge bosonic field is  locked
at $\langle \Phi_{c} \rangle = 0$ so that the CDW/HI quantum phase transition 
is governed by the $\ZZ_N$ interacting field theory:
\begin{equation}
\mathcal{H}^{\ZZ_N}_{\textrm{int}} =
\lambda_2  \; \epsilon_1+   \lambda_4 \left( \mu_2 + \textrm{H.c.} \right) .
\label{effectiveHamPara}
\end{equation}
Model (\ref{effectiveHamPara}) is a deformation of the $\ZZ_N$ CFT
perturbed by two relevant operators with scaling dimensions 
$4/(N+2)$ and $2(N-2)/N(N+2)$ respectively. 
When acting separately, each perturbation yields a massive field theory, but
the interplay between them may give rise to a second-order phase transition at
intermediate coupling. In this respect, when $\lambda_2 <0$,
the first operator in Eq. (\ref{effectiveHamPara}) orders the $\ZZ_N$ degrees of freedom
while the second one wants to disorder them.  We conjecture that this competition for $\lambda_2 <0$
leads to a massless flow to a $\ZZ_2$ quantum critical point in the IR limit.
The quantum phase transition between the CDW and HI phases thus belongs
to the 2D Ising universality class with central charge $c=1/2$. In the simplest 
$N=4$ case, we can show this result explicitly by exploiting the fact that
the $\ZZ_4$ parafermionic CFT has central charge $c=1$ and
so it should be possible to realize it with a single free
Bose field. In fact, the correct identification is quite subtle and 
 the $\ZZ_4$ CFT turns out to be
equivalent to a Bose field living on the orbifold line
at radius $R=\sqrt{3/2\pi}$ \cite{yangpara}. 
However, as shown in the Appendix of Ref. ~\onlinecite{phlegogolin}, it is still possible to bosonize
some fields of the  $\ZZ_4$ CFT with a simple (periodic) Bose field $\Phi$ defined on the circle
with radius $R=\sqrt{3/2\pi}$: $\Phi \sim \Phi  + 2 \pi R$. In this respect, the two 
operators  of Eq. (\ref{effectiveHamPara}), with
scaling dimension $2/3$ and $1/6$, take the form of vertex operators.
The bosonized description of the effective field theory (\ref{effectiveHamPara}) reads:
\begin{equation}
\mathcal{H}^{\ZZ_4}_{\textrm{int}} =
\lambda_2  \; \cos \left( \sqrt{8 \pi/3} \; \Phi \right) +   \lambda_4 \sin \left( \sqrt{2 \pi/3} \; \Phi \right).
\label{2frequency}
\end{equation}
This model is the so-called two-frequency sine-Gordon model which, for instance,
governs the transition from a band insulator to a Mott insulator 
in the 1D ionic Hubbard model \cite{fgn}. When $\lambda_2 <0$ and for all signs of $ \lambda_4$,
model (\ref{2frequency})  displays a $\ZZ_2$ quantum critical point in the IR limit 
which has been analyzed non-perturbatively in Refs. \onlinecite{mussardo,fgn,bajnok}.
We thus deduce that the quantum phase transition between the CDW and HI phases for 
$N=4$ belongs to the 2D Ising universality class.

Let us now investigate the fate of the HI phase as one deviates from the SU(2)$_c$ line.
As in the $N$ odd case, there is a regime in region (III), away from the SU(2)$_c$ line, where 
the coupling $g_3$ that appears in the Luttinger parameter expression (\ref{Luttingerpara}) 
decreases and then increases as function of the interaction (see 
Figs. \ref{figRGN3SU2}, \ref{figRGN3SU2dual}). In the vicinity of the minimum of 
$g_3$, we expect the emergence of a gapless phase associated to the sine-Gordon 
model (\ref{hcNeven}) with $K_c > 4/N$. The existence of this intermediate gapless phase
will be confirmed numerically in Sec. VI by means of DMRG calculations. In this respect,
the $N=2$ case is very special since this phase shrinks to a line which marks
the phase transition between HI and RS phases (see Fig. \ref{figRGN2}).
This critical phase has only one gapless charge mode
and the singlet-pairing has the same low-energy behavior as in the $N$ odd case:
$P^{\dagger}_{00} \sim \exp \left(i\sqrt{2 \pi/N} \; \Theta_c\right)$. However, this
phase is different from the gapless BCS phase of the $N$ odd case.
Indeed, as already stressed, the disorder parameter $\mu_1$ of 
Eq. (\ref{masstermrep}) cannot be compensated using
higher orders of perturbation theory which means 
that the alternating part of the CDW operator is short-ranged.
We then deduce the following leading asymptotics of the equal-time correlation functions:
\begin{eqnarray}
\langle  P^{\dagger}_{00} \left(x\right)   P_{00} \left(0\right) \rangle
&\sim&  x^{- 1/NK_c} \nonumber \\
\langle   n \left(x\right)  n \left(0\right) \rangle &\sim&
-\frac{NK_c}{\pi^2 x^2},
\label{correlBCSNeven}
\end{eqnarray}
where in the density correlator, only the uniform part has a power-law decay.
The leading instability is the singlet-pairing one when $K_c > 4/N$.
The main difference with the gapless BCS phase of Fig.~\ref{PhDiagNodd}
stems from the fact that the alternating part of the density correlator 
(\ref{correlBCSNeven}) has now an exponential decay. The quantum phase transition between HI and BCS phases
belongs to the BKT universality class.

The last regime of region (III), which corresponds to the transition between (III) and (I)
of Fig.~\ref{figRGN3SU2dual}, can be identified by means of the duality symmetry $\Omega_1$.
Under this transformation (\ref{dualitytrans}), the CDW phase is  changed into the SP phase.
In contrast, from Eq.~(\ref{dualcosNeven}), we deduce that the sine-Gordon operator of the 
low-energy Hamiltonian (\ref{hcNeven}) remains invariant when $N/2$ is even. 
The Mott-insulating phase in the vicinity of the ${\widetilde{\text{SU}(2)}}_c$ line of Fig.~\ref{figRGN3SU2dual}
is thus described by the strong-coupling regime of the sine-Gordon model (\ref{hcNeven}) 
with $K_c < 4/N$ and the pinning $\langle \Phi_{c} \rangle = 0$. This phase is expected to be the RS phase, i.e.
the so-called large D phase of the integer spin Heisenberg chain \cite{schulz}, which appears
in the strong-coupling approach (\ref{HaldaneSU2Hamiltonian}) for a sufficiently strong positive $D$.
Interestingly enough, within our low-energy approach, this phase is described exactly in the same way as the 
HI phase. Thus, the two phases necessarily share the same order 
parameters, such as the string orders (\ref{stringchargeN4}, \ref{HaldanestringN4}) for instance. 
However, they should have different
edge states but we could not, very unfortunately, investigate these boundary end excitations
in our CFT approach. Recently, it has been argued that the edge-state structure of 
the even-spin Heisenberg chain is not protected by symmetry 
in contrast to the odd case \cite{pollman09}. In particular, the authors of Ref. ~\onlinecite{pollman09}
have conjectured that there is an adiabatic continuity between the Haldane and large D phases
in the even-spin case. The Haldane phase is thus equivalent to a topologically trivial insulating phase in this 
case. This adiabatic continuity has been shown numerically in the spin-2 XXZ Heisenberg chain with a single-ion
anisotropy by finding a path where the two phases are connected without any 
phase transition  \cite{tonegawa}. In our problem, the HI and RS phases are separated by an intermediate gapless
BCS phase. However, within our low-energy approach, the two non-degenerate Mott-insulating phases
are described in the same manner by the sine-Gordon model (\ref{hcNeven}) 
with $K_c < 4/N$ and the pinning $\langle \Phi_{c} \rangle = 0$. In this respect, our results
strongly support the conjecture put forward in Ref. ~\onlinecite{pollman09}.

\begin{figure}[t]
\centering
\includegraphics[width=0.75\columnwidth,clip]{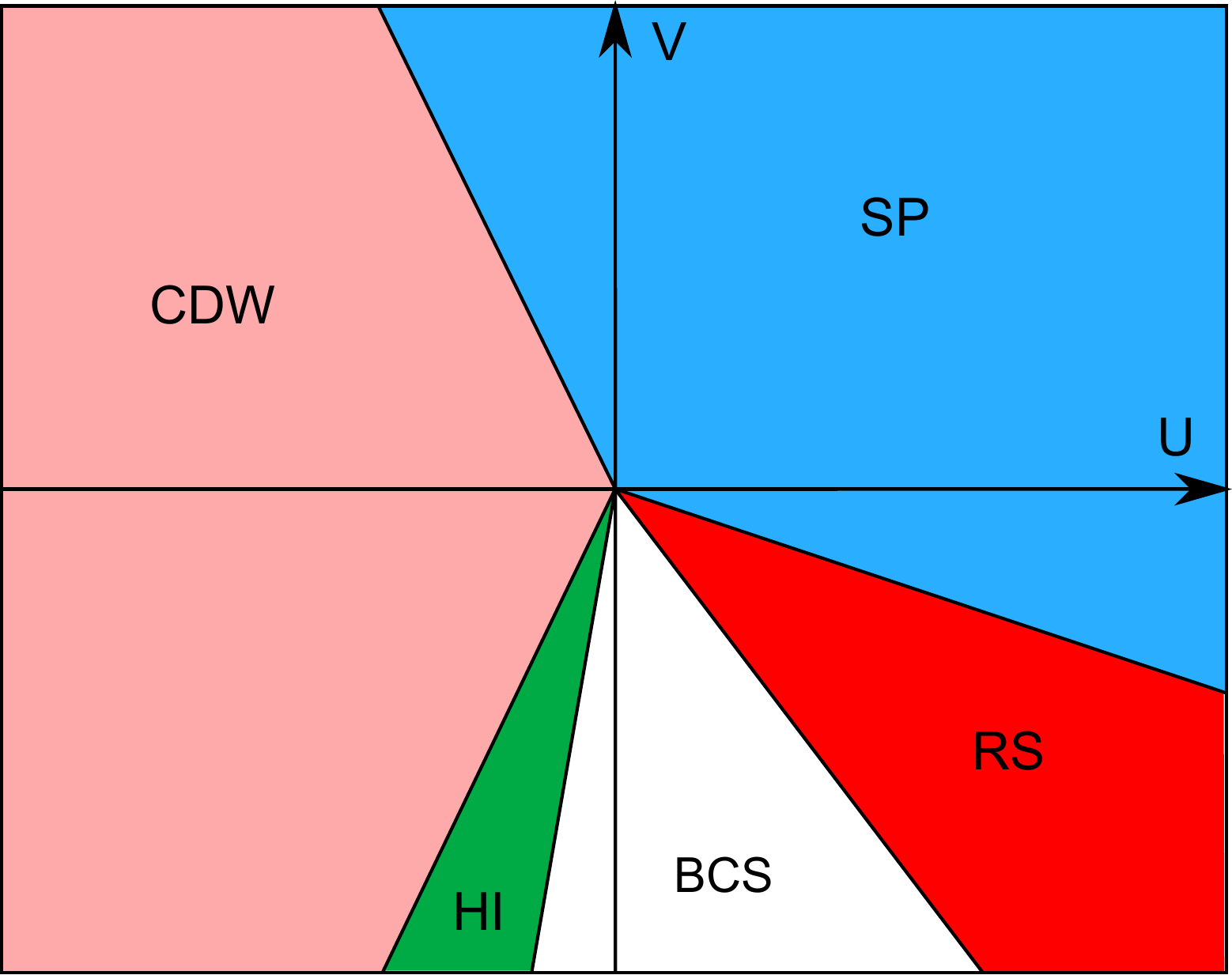}
\caption{Phase diagram obtained by the low-energy approach in the $N$ even case ($N>2$).}
\label{PhDiagNeven}
\end{figure}

Finally, the quantum phase transition between RS and SP phases is obtained 
from the effective theory (\ref{effectiveHamPara}) by the application of the duality symmetry $\Omega_1$.
The latter transformation changes the sign of the coupling constant
$\lambda_4$ of the $\mu_2$ operator.
However, this sign is irrelevant for the competition between the two relevant operators
in model (\ref{effectiveHamPara}). We thus expect that the resulting quantum phase transition
still belongs to the 2D Ising universality class.
In summary, Fig.~\ref{PhDiagNeven} presents the phase diagram, 
in terms of the lattice parameters $U,V$ in the $N/2$ even case. 

\emph{\underline{$N/2$ odd case}.}

The last case to consider is the case where $N/2$ is odd. In region (III), in the vicinity of the SU(2)$_c$ line,
the charge bosonic field $\Phi_c$ of the sine-Gordon model  (\ref{hcNeven}) with $K_c < 4/N$
is now pinned at $ \langle \Phi_{c} \rangle = \sqrt{\pi/2N}$ since  $g_c > 0$. The non-degenerate
Mott-insulating phase is the HI phase. This phase can be described by 
the generalization of the string-order parameter (\ref{stringcharge})
with spin-$N/2$ operator. Indeed, in the low-energy limit and for the $N/2$ odd case, we find:
\begin{align}
\lim_{|i-j| \rightarrow \infty}
& \langle
{\cal S}^{z}_i
e^{ i \pi \sum_{k=i+1}^{j-1}{\cal   S}^{z}_k}
{\cal   S}^{z}_j 
\rangle  \simeq  \nonumber \\
\lim_{|x-y| \rightarrow \infty}
&\langle{\sin \left( \sqrt{N \pi/2} \; \Phi_c \left( x \right)  \right) 
 \sin \left( \sqrt{N \pi/2} \; \Phi_c \left( y \right)  \right)}\rangle \nonumber \\
 &\ne 0\, ,
\label{stringchargeNodd}
\end{align}
in sharp contrast to the result (\ref{stringchargeN4}) of the $N/2$ even case.
For general odd-spin Heisenberg chain, the order parameter (\ref{stringchargeNodd}) 
is known to be non zero contrarily to the even-spin case \cite{oshikawa,totsuka}.
In this respect, there is thus a clear dichotomy in the HI phase, depending on the parity of $N/2$.
For odd-spin Heisenberg chains, the authors of Ref. ~\onlinecite{pollman09} have predicted
that the Haldane phase displays a topological order and is not equivalent to the large D phase
as in the even-spin case. 
This scenario is in perfect agreement with our low-energy approach.
Indeed, according to Eq. (\ref{dualcosNeven}), the duality symmetry $\Omega_1$ changes the sign of 
the vertex operator of model (\ref{hcNeven}) when $N/2$ is odd. The physical properties of the RS phase are thus 
governed by the sine-Gordon model (\ref{hcNeven})  with $K_c < 4/N$ and the pinning $\langle \Phi_{c} \rangle = 0$.   
In the $N/2$ odd case, the HI and RS phases are  described by two different locking of the 
charge bosonic field in sharp contrast to the $N/2$ even case. In particular, 
the RS phase is described by the string-order parameter (\ref{HaldanestringN4}) and not
(\ref{stringchargeNodd}) as the HI phase is. The HI and RS phases are thus
totally distinct phases that cannot be adiabatically connected.

Finally, as in the $N/2$ even case, the transition between these two non-degenerate phases 
is accompanied  by the formation of an intermediate gapless BCS phase with 
the properties (\ref{correlBCSNeven}). Unfortunately, in the $N/2$ odd case, 
we do not have access to a theory of the quantum phase transition between CDW (respectively 
SP) phase and the HI (respectively RS) phase.
We suspect, as in the $N/2$ even case,  that the transition belongs to the 2D Ising 
universality class but it certainly requires a proof. 
Fig.~\ref{PhDiagNeven} presents the phase diagram in the $N/2$ odd case
which, apart from the subtleties on the topological nature of the HI phases, is identical
to  the $N/2$ even case. Last, we would like to emphasize
that the $N=2$ case (see Fig.~\ref{figRGN2}) is not representative of the even family but turns out
to be special.

\section{Phase diagram of half-filled spin-3/2 fermions ($N=2$)}
\label{section2}

\begin{figure}[t]
\centering
\includegraphics[width=\columnwidth,clip]{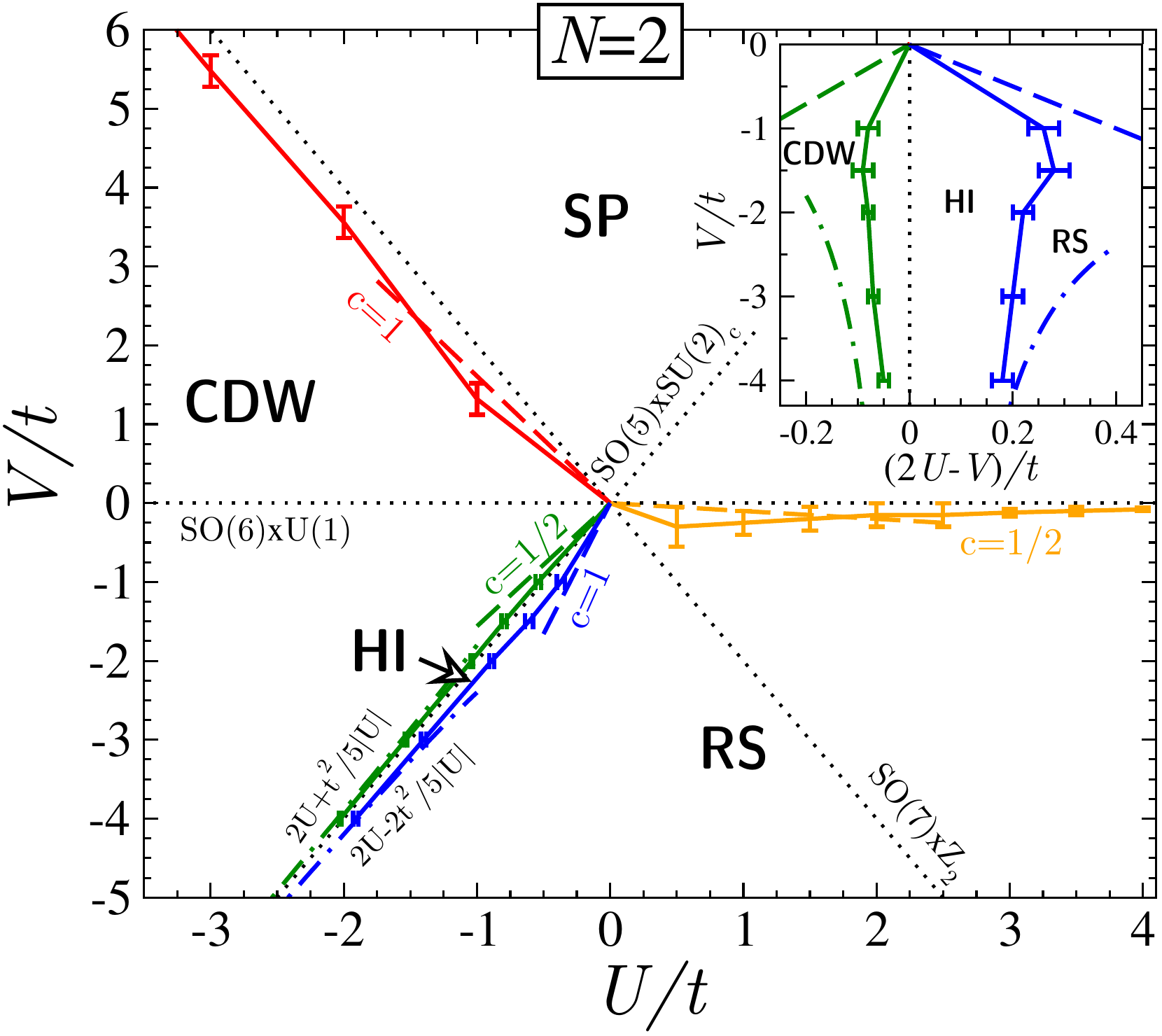}
\caption{(Color online) \textit{Phase diagram of the $N=2$ case}.
  Full lines are DMRG results (see text for discussion) and dashed
  lines are results from solving numerically the RG flow in the
  weak-coupling limit. $c$ stands for the expected central charge on
  the transition line. Lines with higher symmetries are indicated in
  black dashed lines. Boundaries of the HI phase obtained by the
  strong-coupling are shown in dotted lines. As a rule of thumb, DMRG
  results can be trusted if $|U|,|V| \gtrsim t$, while weak-coupling
  predictions are exact close to the free fermions limit at the
  origin. \emph{Inset}: Zoom on the region of the HI phase where
  the x-axis $(2U-V)/t$ is perpendicular to the SU(2)$_c$ $\times$SO(5)
  line.}
\label{fig:phasediagN2}
\end{figure}

In this section, we give the phase diagram of
model~\eqref{hubbardSgen} when $N=2$ in the $(U/t,V/t)$
plane, obtained from numerical calculations. Four phases are found and reported on
Fig.~\ref{fig:phasediagN2}: two phases which break translational
invariance, the SP and CDW
phases, and two with non-degenerate GS which can only be
distinguished through non-local string orders, the HI and RS phases. These phases are
separated by transition lines determined numerically (full lines), and compared to weak- and strong-coupling predictions
displayed in dashed lines. In addition, three particular lines are plotted where
the model has an \emph{exact} enlarged symmetry that we have discussed in Sec. II.

The numerical calculations are performed with DMRG on chains, each
site containing the 16 states of the onsite basis (for $N>2$, since the local Hilbert space is too large,
we must use other strategies as discussed below). We
fix three quantum numbers: the spin part $S^z = \frac 1 2
\sum_{\alpha,i} (-1)^{\alpha+1} n_{\alpha, i}$, $T^z = \frac 1 2
\sum_i(n_{1,i}+n_{2,i}-n_{3,i}-n_{4,i})$ as well as the total number
of particles $N_f = 2L$, i.e. the total charge. The GS lies in the $S^z = T^z = 0$
sector. The number of kept states is typically $m=2000$ and OBC
 are used if not stated otherwise. Denoting by $L$ the length of the chain, the local order
parameters are computed numerically by taking their value in the
bulk of the chain (we choose to work with an even number of sites):
\begin{eqnarray}
{\cal O}_{\text{CDW}}(L) &=& n_{L/2}-n_{L/2-1}\\
{\cal O}_{\text{SP}}(L)  &=& t_{L/2}-t_{L/2-1},
\end{eqnarray}
where $n_j = \sum_{\alpha} n_{\alpha,j}$ is the total onsite density
and $t_{j} = \sum_{\alpha} c^{\dag}_{\alpha,j+1} c_{\alpha,j} +
\mathrm{H.c.}$ the local kinetic energy on bond $(j,j+1)$.

\subsection{The HI phase}

We start a more detailed discussion of the phase diagram from the
$V=2U$ line which shows the remarkable SU(2)$_c$ symmetry,
leading to the effective spin-1 Heisenberg model (\ref{HaldaneSU2Hamiltonian})
in terms of charge degrees of freedom. We have recently
demonstrated~\cite{Nonne2009} that the gapped HI phase is
realized for a given value of $V/t$, and that its extension is rather
small.  We here refine the description of the boundaries of the
Haldane phase and discuss the nature of the transition lines to
respectively the CDW and RS phases. In order to find the transition
line from CDW to HI, we use ${\cal O}_{\text{CDW}}$ which vanishes in the
HI phase and which is straightforward to compute. The transition
from HI and RS is more difficult to determine as no local order
parameter can discriminate between the two phases. In
Ref. ~\onlinecite{Nonne2009}, we gave several signatures of the
transition which can be used to locate it: non-local charge string
order parameters and the presence of edges states, which are observed here by considering a
charge excited state with two additional fermions; for OBC, this state has a vanishing gap to the GS. The simplest way to
determine the transition with our numerical scheme is to look at the
distribution of the charge in the excited state with $N_f=2L+2$: an
excess $N_f=1$ charge will be stuck at each edge in the HI phase
(equivalent to the spin-1/2 edge state of the spin-1 Haldane phase), while an
$N_f=2$ excitation lies in the bulk of the RS phase (equivalent to the
$S=1$ magnon of the Heisenberg ladder). We thus use this change in the
density profile of the charge excited state (benchmarked with other
signals of the transition for $V=-2t$) to give the estimate of the
transition line in Fig.~\ref{fig:phasediagN2}.

In the weak-coupling regime, $|U|,|V|\lesssim t$, DMRG calculations
become hard due to the relevance of many low-energy onsite states. In the strong-coupling limit (large $U,V$), 
onsite energy scales are well-separated so that DMRG efficiently eliminates high energy
irrelevant states. The numerical predictions of the RG flow provides a
better prediction for the transition lines in this weak-coupling
regime: these estimates are $V \simeq 3.33 U$ for RS-HI and $V=1.56U$
for CDW-HI.

The two transition lines in Fig.~\ref{fig:phasediagN2} are also compared to the strong-coupling predictions
of Sec.~\ref{sec:strong-coupling}. For large $|U|/t$ and $|V|/t$, the effective Hamiltonian
around the SU(2)$_c$ line is a spin-1 model with 
antiferromagnetic coupling $J=2t^2/5|U|$ and anisotropy $D=2U-V$ (see 
Eqs. (\ref{HaldaneSU2Hamiltonian}, \ref{Haldanecouplings})).
The phase diagram of this model has been extensively studied
\cite{spinonephasediag,dennijs,schulz,Degli2003,hamer}
and shows that a Haldane-N\'eel transition (equivalent to the HI-CDW one) occurs 
for $D/J \simeq -0.5$ while a Haldane-large-$D$ transition (equivalent to the HI-RS one)
is obtained for $D/J \simeq 1$. This gives the two curves $V_{\text{HI-CDW}} = 2U+t^2/|U|$ and
$V_{\text{HI-RS}} = 2U-2t^2/|U|$ explaining both the shrinking 
and the asymmetry of the extension of the HI phase in the strong-coupling regime.

\begin{figure}[t]
\centering
\includegraphics[width=\columnwidth,clip]{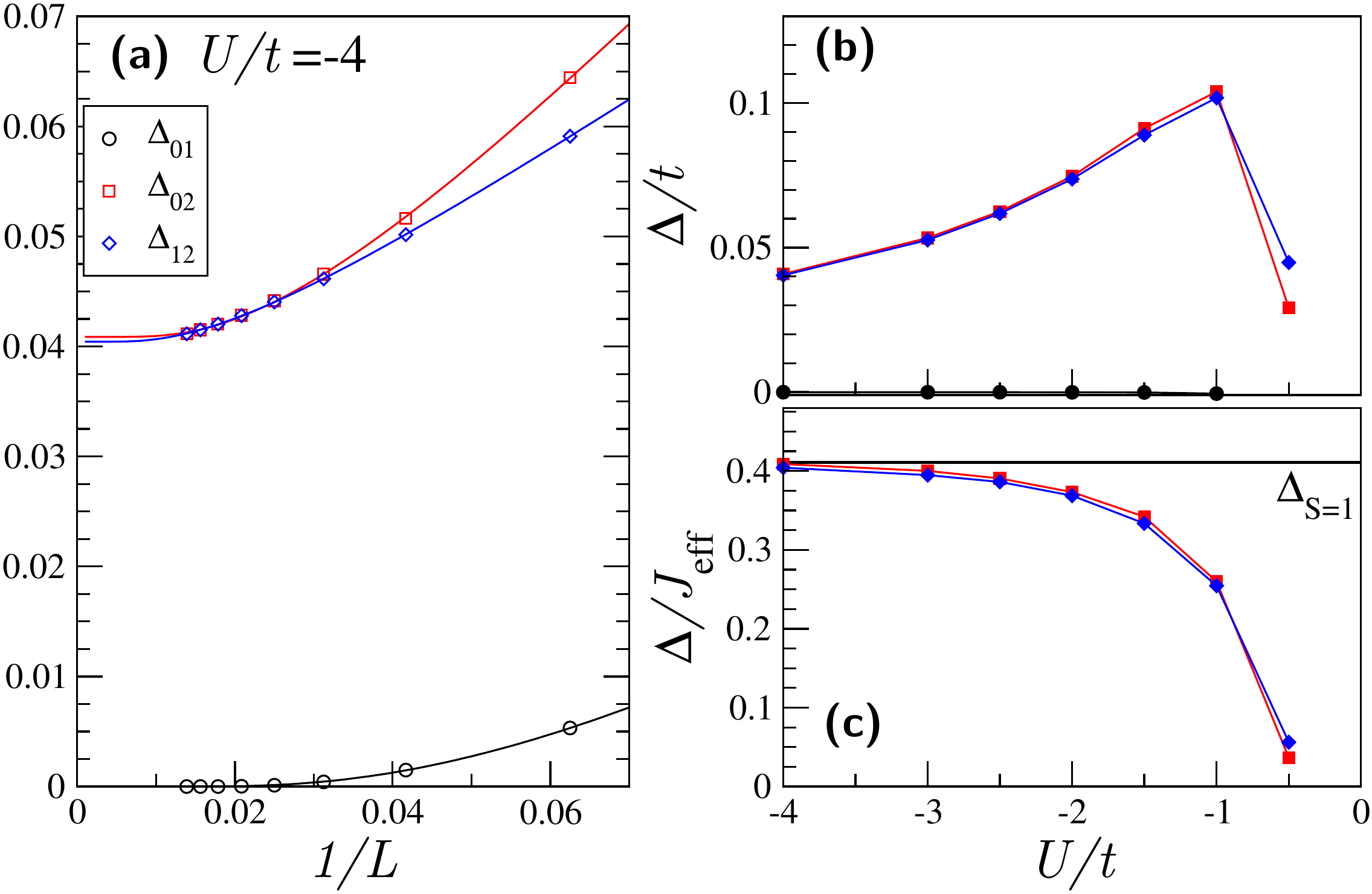}
\caption{(Color online) Behavior of the Haldane gap $\Delta$ along the SU(2)$_c$ line $V= 2U <0$ 
of Fig.~\ref{fig:phasediagN2}. (a) gaps and finite-size scalings (see text for discussion).
(b-c) the Haldane gap as a function of $U/t$ in units of respectively $t$ and the effective
antiferromagnetic coupling $J$. The $\Delta_{\text{S=1}}$ line indicates the value of the gap
known for a Heisenberg spin-one chain.}
\label{fig:HaldaneGap}
\end{figure}

Although the Haldane gap decreases in the strong-coupling regime simply because $J$ decreases,
the agreement between the fermionic spin-3/2 Hubbard model under study and the spin-1 effective
model improves as irrelevant degrees of freedom are pushed to high energies.
This can be illustrated numerically by the behavior of the Haldane gap along the SU(2)$_c$ line
as a function of $U/t$. The Haldane gap is computed using OBC
from the following gaps: 
\begin{equation}
\Delta_{ab} = E_0(2L+2b)-E_0(2L+2a)\;,
\end{equation}
where $E_0(N_f)$ stands for the GS energy with $N_f$ fermions.
As evoked previously, the presence of edge states with OBC makes 
the first excited state collapse onto the GS, so that
 $\Delta_{01}$ vanishes in the thermodynamical limit. Still, both $\Delta_{02}$
and $\Delta_{12}$ must remain finite and 
tend to the bulk Haldane gap for sufficiently large sizes. 
These behaviors, together with finite-size extrapolations
of the gaps using the ansatz
\begin{equation}
\Delta_{ab}(L) = \Delta_{ab}(\infty) + \text{const.}\,e^{-L/\xi} / L\;,
\end{equation}
are clearly shown by the numerical results of Fig.~\ref{fig:HaldaneGap}(a).
Fig.~\ref{fig:HaldaneGap}(b) and (c) display the extrapolated gaps 
as a function of $U/t$ in units of respectively $t$ and $J$. 
While the weak-coupling opening of the gap cannot be reliably studied
here, we observe that the gap passes through a maximum around $U/t\simeq -1$ 
which is close to value  $U/t\simeq -1.5$ for which the width of the HI phase is maximal.
In the strong-coupling regime, the gap in units of $t$ decreases as expected (see
Fig.~\ref{fig:HaldaneGap}(b)), while, put in units of $J$ (see Fig.~\ref{fig:HaldaneGap}(c)), 
it eventually reaches the value $\Delta_{\text{S=1}} \simeq 0.41J$ known~\cite{DMRG}
for the spin-1 Heisenberg chain: $U\simeq -4t$ is already deep in the
strong-coupling regime along this SU(2)$_c$ line.

\begin{figure}[t]
\centering
\includegraphics[width=\columnwidth,clip]{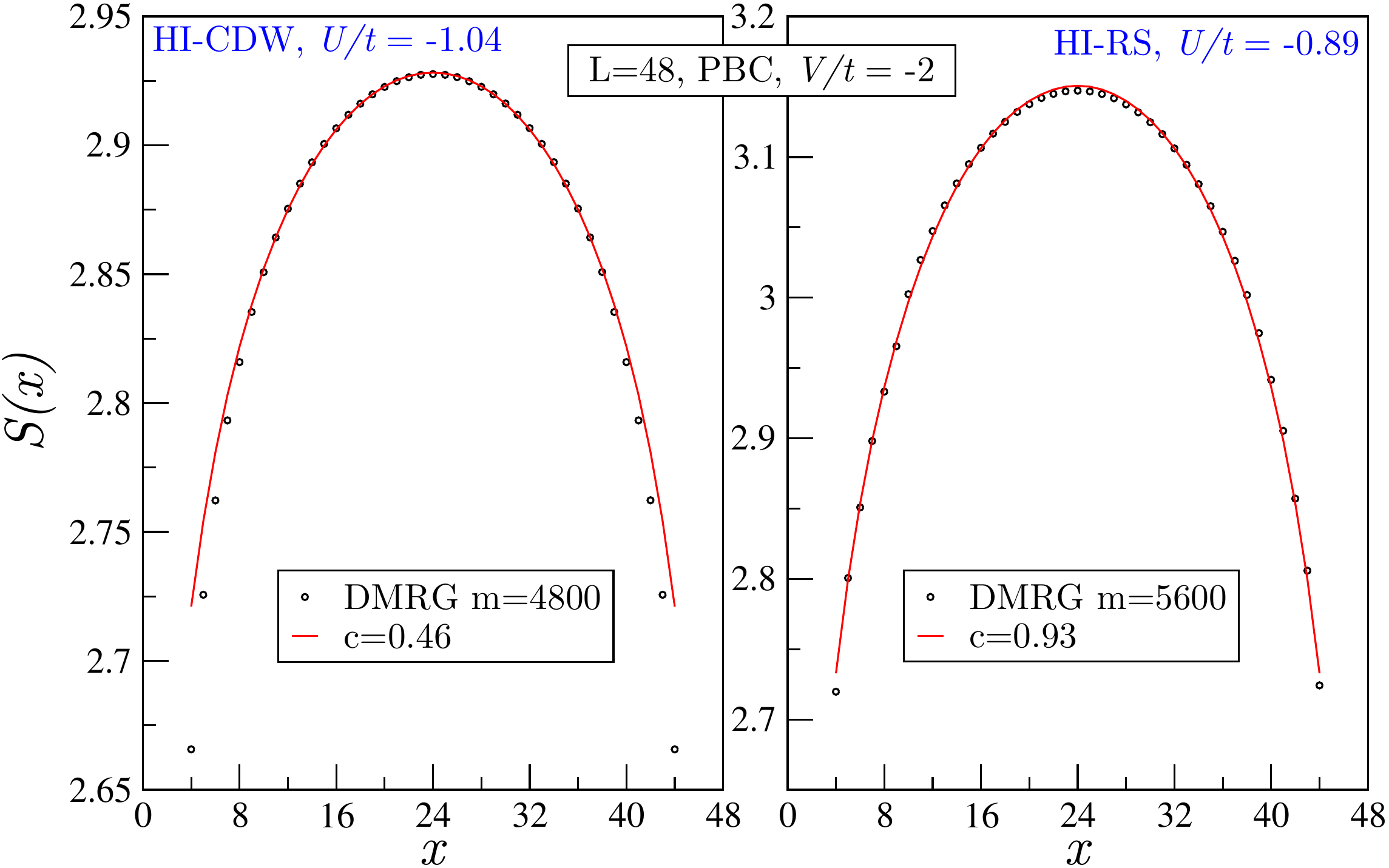}
\caption{(Color online) Fitting the entanglement entropy close to the critical lines
  surrounding the Haldane phase provides central charges $c$ close to
  the expected values $c=1/2$ (for HI-CDW) and $c=1$ (for HI-RS). The
  best agreement is found using PBC with DMRG and keeping a large
  number of kept states $m$.}
\label{fig:centralChargeHaldane}
\end{figure}

Lastly, we investigate the nature of the critical points at the two boundary lines
of the HI phase. From the low-energy results of Sec. III B,
we expect that the HI-CDW is an Ising transition with a central charge
$c=1/2$ while the HI-RS transition belongs to the BKT type, associated
with a central charge $c=1$. In the strong-coupling limit, 
this has been observed numerically for the spin-1 chain with single-ion
anisotropy~\cite{Degli2003}. To check these predictions from the 
DMRG data, we use the universal scaling of the entanglement entropy (EE)
in a critical phase, which gives a direct access to the central charge.
We obtained the most convincing results using periodic boundary conditions (PBC)
at the price of keeping a much larger number of states and using small system sizes.
Similar calculations have been performed in the context of the SU($N$) generalization
of Haldane's conjecture \cite{greiter}.
The results of the EE on a finite chain of length $L=48$ along the $V=-2t$ around the
HI phase are given in Fig.~\ref{fig:centralChargeHaldane}. The central
charge is obtained from the data using the universal formula~\cite{EE}
\begin{equation}
S(x) = \frac{c}{3}\ln d(x|L+1) + \text{const.}
\label{eq:EE}
\end{equation}
with $d(x|L) = \frac{L}{\pi}\ln\left(\frac{\pi x}{L}\right)$ the cord function
and $S(x)$ the EE of a block of size $x$ with the rest of the chain.
The values obtained for $c$ are in good agreement with the expected values considering 
the large number of local degrees of freedom. There is an uncertainty on the location
of the critical points but, on a finite system, as long as $L\ll \xi$, with $\xi$ the correlation
length associated to the closing gap, the physics will be effectively that of the critical point.

\subsection{The RS-SP transition}

We now turn to the discussion of the RS-SP transition in the right-down
quadrant of Fig.~\ref{fig:phasediagN2}.
The two phases RS and SP can be simply distinguished by the local
spin-Peierls order parameter ${\cal O}_{\rm SP}$ which is finite in
SP while it is zero in RS. The vanishing of the order as the system
size increases provides a good estimate of the transition line.

We further try to give evidence for the nature of the transition and
check whether it lies in the Ising universality class. A possible approach
is to use the EE again and look for $c=1/2$. However, the ${\cal O}_{\rm SP}$ 
order parameter appears as the leading corrections to the EE with OBC
and gives strong oscillations in the signals, particularly in the SP phase and
up to the critical point. These oscillations render the fits difficult,
and the value of $c$ is not reliably extracted for the accessible system
sizes. Using PBC improves a bit the situation, but the oscillating 
parts of the EE could not be suppressed (as expected for the GS) 
with DMRG, even by increasing the number of kept states and sweeps. 

\begin{figure}[t]
\centering
\includegraphics[width=\columnwidth,clip]{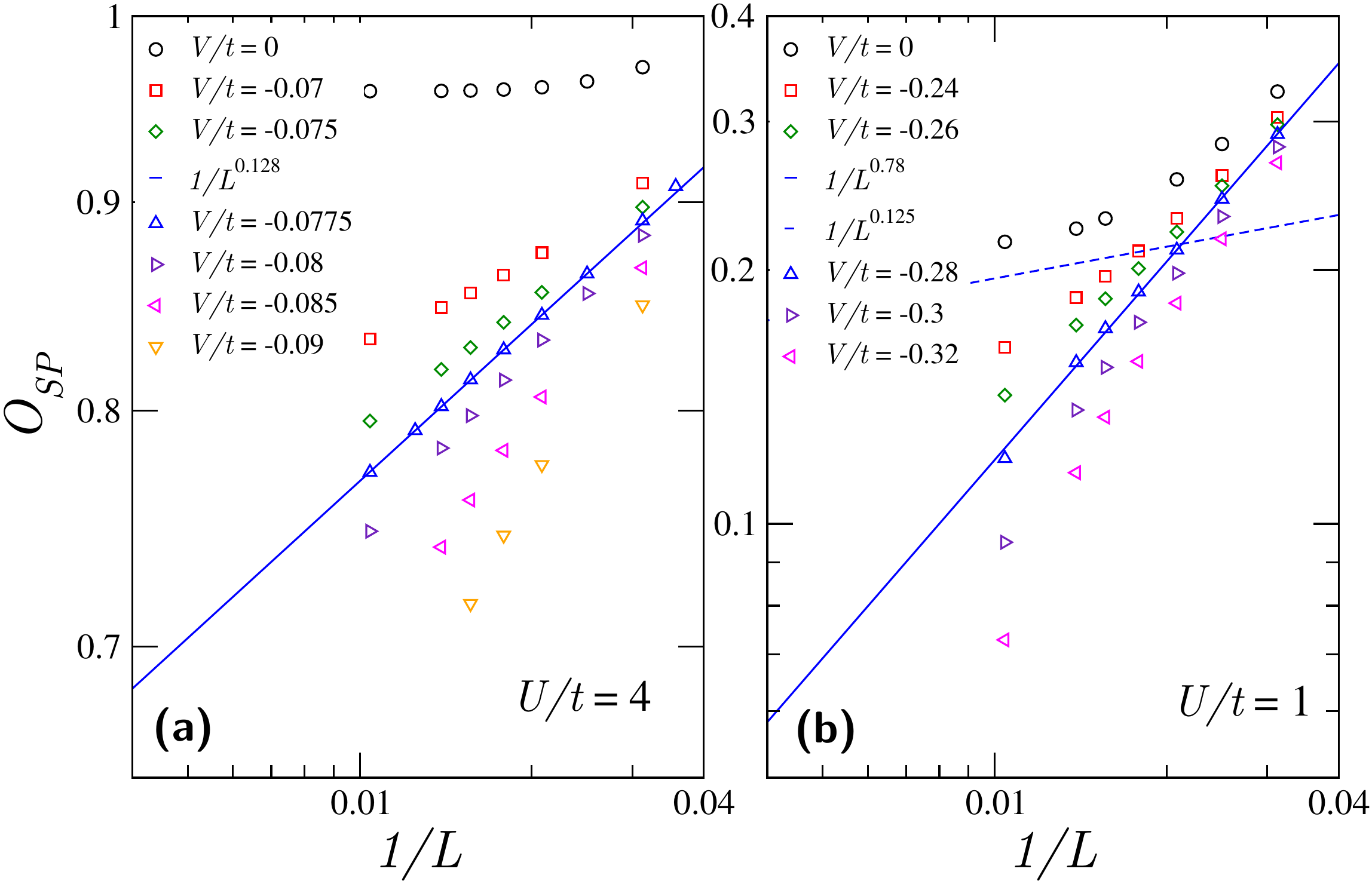}
\caption{(Color online) Scaling of the order parameter ${\cal O}_{\rm{SP}}$ at the RS-SP transition : 
(a) for $U/t=4$, the exponent is quite close to $1/8$;
(b) for $U/t=1$, it is closer to 0.8.}
\label{fig:RS-SP-Scaling}
\end{figure}

Consequently, we use another strategy to identify the Ising universality class.
We know that the correlation function of the order parameter has
a universal exponent $1/4$ at the critical point. Then, Friedel oscillations
of the order parameter gives the scaling ${\cal O}_{\rm SP}(L) \propto L^{-1/8}$
on the critical point. In the SP phase, ${\cal O}_{\rm SP}(L)$ reaches a constant in
the thermodynamic limit, while it decreases exponentially in the RS phase.
Thus, by looking at the scaling of ${\cal O}_{\rm SP}(L)$ for different
parameters, we are able to give both a precise estimate of the critical
point and to check that the exponent is indeed close to $1/8$.
The results along two cuts at $U=4t$ and $U=t$ are reported in
Fig.~\ref{fig:RS-SP-Scaling}.  In the strong-coupling regime $U=4t$,
we do observe a very good agreement with an exponent $1/8$, typical
of the Ising universality class. However, in the weak-coupling
regime, a much larger exponent of $0.78\simeq 6/8$ fits well
the scaling curves. We understand this discrepancy in the following way:
in the weak-coupling regime, the gaps to higher excited states are too
small to be thrown away in the low-energy regime of a \emph{finite} system.
In other words, the correlation lengths associated with theses gaps become
too large and we could not reach sizes sufficiently large to freeze them.
A speculative picture can account for the observed number: at weak
coupling, the transition line gets very close to the SO(6) line
which has the equivalent of six gapped Ising degrees of freedom,
but with an exponentially small gap of the order $t/U$ \cite{assaraf}. In this weak-coupling
regime, the numerics cannot resolve these gaps and the Ising degrees
of freedom appear critical, each contributing to $1/8$ in the exponent which 
then should be close to $6/8$.

This comment brings us to the discussion of effect of the proximity of the 
SO(6) line (an exact enlarged symmetry) to the RS-SP transition line.
The $V=0$ and $U>0$ line has been studied analytically and numerically
in Ref. ~\onlinecite{assaraf}: the charge and spin gaps open slowly
with $U/t$ and are numerically negligible below $U\simeq 2t$.
In the weak-coupling regime, the low-energy physics has an emerging enlarged
SO(8) symmetry. In the strong coupling regime, the spin gap decreases after
passing through a maximum around $U\simeq 6t$. The data shows that 
the RS-SP transition line has a non-monotonic behavior, first
following the weak-coupling RG predictions and then being attracted 
by the SO(6) line at large interactions (see Fig.~\ref{fig:phasediagN2}). This attraction can be
qualitatively understood by the behavior of the spin gap as $U$
increases. Considering $V$ as a perturbation which closes the spin gap $\Delta_S$,
the line should typically behave as $V_c(U) \sim -\Delta_S(U)$ which is 
non-monotonous and stick to the SO(6) line in the strong-coupling limit. 
In the weak-coupling limit $U\lesssim t$, the RG prediction $V=-0.10U$ is
more reliable than the numerics.

\subsection{The CDW-SP transition}
\label{part:CDWSPtransition}

Lastly, we briefly discuss the CDW-SP transition between these two
phases which breaks translational symmetry. Numerically, the
precise determination of the transition with ${\cal O}_{\rm SP}$ and ${\cal O}_{\rm CDW}$ using DMRG
turns out to be difficult due to formation of domains of each kind of
orders close to the transition line. Changing the number of kept
states, the number of sweeps and the size, slightly moves the transition point
determined by the order parameter at the center of the chain. This
leads to error bars in the phase diagram which are relatively small
compared to the parameter scales of Fig.~\ref{fig:RS-SP-Scaling}, 
but are too large to focus on the critical features of the transition line.
We could not check the $c=1$ expectation of this transition, due to both
the difficulty in locating the transition point, and because of
strong SP oscillations in the EE. Notice that on the critical line,
the correlations of the quartet operator $c^{\dag}_{1,i}  c^{\dag}_{2,i}  c^{\dag}_{3,i}  c^{\dag}_{4,i} $ become 
critical which is qualitatively in agreement with numerical observations.

\begin{figure}[t]
\centering
\includegraphics[width=0.8\columnwidth,clip]{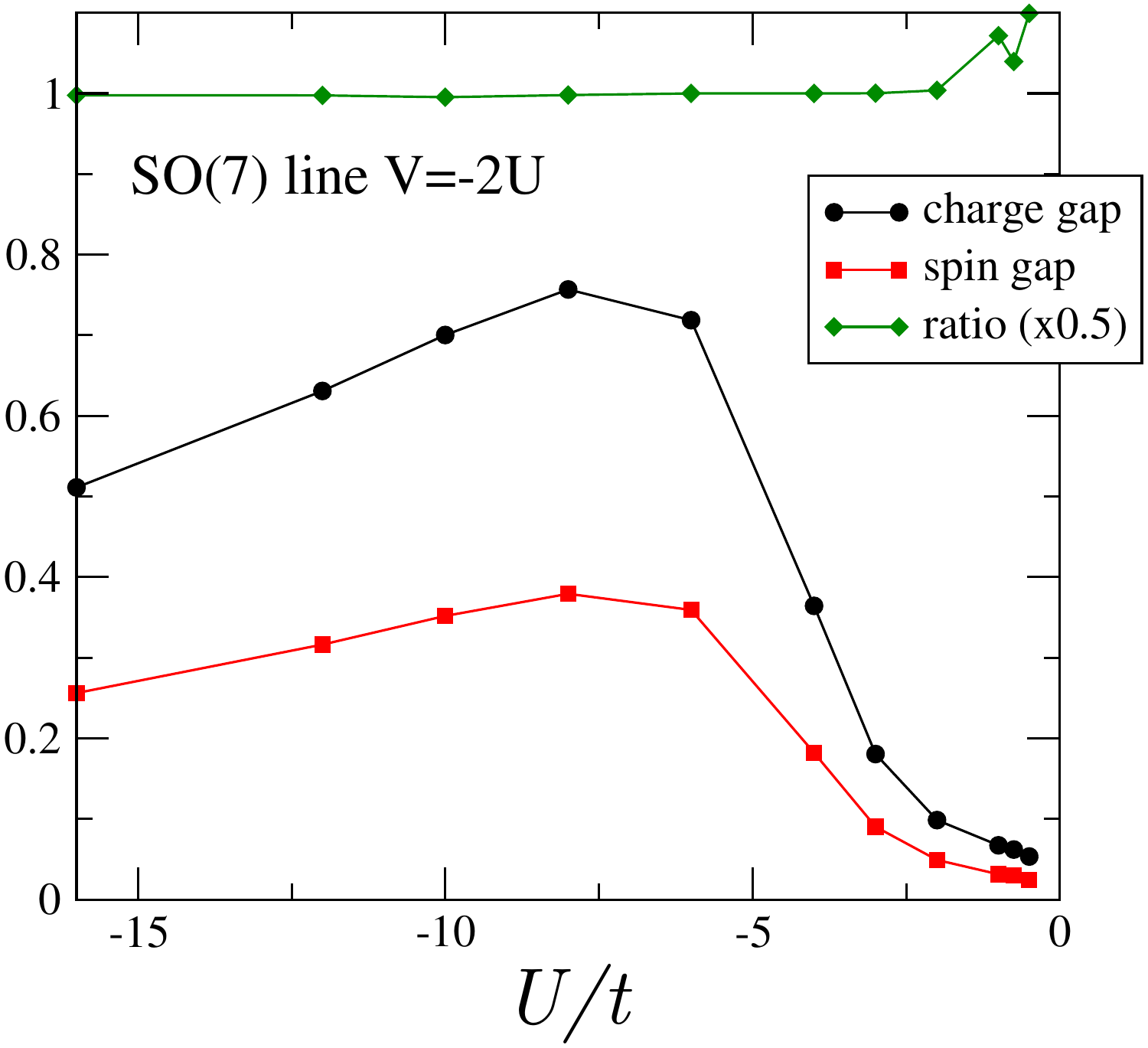}
\caption{(Color online) Charge and spin gaps and their ratio along the SO(7) line
$V=-2U$ of Fig.~\ref{fig:phasediagN2}.}
\label{fig:SO7}
\end{figure}

Here again, we see that the transition line is rather close to a high
symmetry line of the phase diagram, namely the SO(7) line $V=-2U$ \cite{zhang}. In the
weak-coupling regime, the numerical solution of the RG Eqs. (\ref{RGbetafunctionredef}) for 
$N=2$ gives $V=-1.61U$ but, for larger $|U|$, DMRG calculations indicate that the transition is attracted
to the vicinity of the SO(7) line. A argument similar to the one used for the RS-SP transition can be drawn: we see that SO(7) line is in
a SP gapped phase. The strong-coupling spin-model along this line is an SO(7) Heisenberg model
where the spins belong to the vectorial representation of SO(7) \cite{zhang} and
our analysis predicts a SP bond ordering.
Numerically, we compute the spin gap $\Delta_s$
and charge gap $\Delta_c$ defined as follow:
\begin{eqnarray*}
\Delta_s &=& E_0(N,1)+E_0(N,-1)-2E_0(N,0)\\
\Delta_c &=& E_0(N+2,0)+E_0(N-2,0)-2E_0(N,0)
\end{eqnarray*}
where $E_0(N_f,S^z)$ is the GS energy with $N_f$ fermions in the
$S^z$ sector with $T^z=0$ and $N=2L$ is the reference number of particle at half-filling.
The results extrapolated in the thermodynamic limit are given
in Fig.~\ref{fig:SO7} for a wide range of $U/t$ values. 
The gaps open slowly in the weak-coupling regime and then reach
a maximum around $U\simeq -7t$, before decreasing in the strong-coupling
regime. The ratio of the gaps $\Delta_c/\Delta_s$ is very close to two,
everywhere but in the weak-coupling limit where the numerics are challenging
for accurate predictions. 

\section{Phase diagram in the $N=3$ case}

In this section, we investigate the phase diagram of
model~\eqref{hubbardSgen} when $N=3$ and in the $(U/t,V/t)$ plane
using extensive DMRG simulations. Since the local Hilbert space on
each site contains $2^6=64$ states and is quite large, 
we have implemented the following
strategy: we use a mapping to a 3-leg Hubbard ladder where the chains
correspond to fermionic states with $S_z$ equal to $\pm 1/2$, $\pm 3/2$, and $\pm 5/2$ respectively. Then, after some algebra, we can rewrite all hoppings and interaction terms in this language, which introduce for instance rung interactions and rung pair-hopping terms. 
This mapping to a ladder allows us to
converge faster to the GS, but we have checked that the
symmetry between chains is preserved in the SU(3) case for instance. 
Typically, we keep between 1600 and 2000 states in our simulations for measuring local quantities and up to 3000 for correlations, and we use OBC. 

\begin{figure}[!ht]
\centering
\includegraphics[width=0.95\columnwidth,clip]{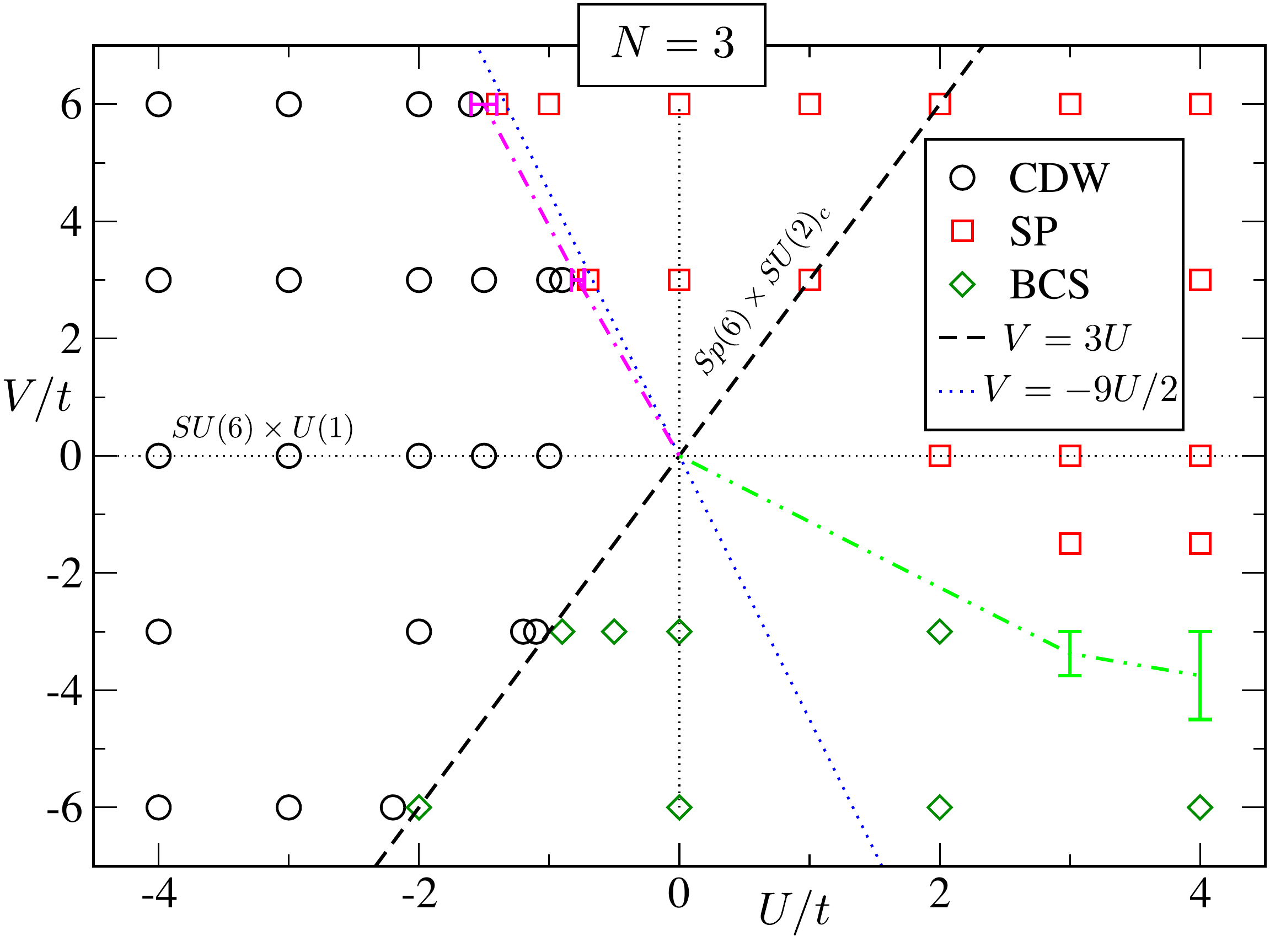}
\caption{Numerical phase diagram obtained by DMRG in the $N=3$ case.}
\label{fig:phasediag_N=3_DMRG}
\end{figure}

 Since no topological phase is
expected, we can rely on measuring local quantities such as local
density and kinetic energy, as well as density and pairing
correlations that will characterize the critical phase that has been
shown to exist along the SU(2)$_c$ line $V=3U$ in
Ref. ~\onlinecite{Nonne2009}. The following phase diagram can thus be
obtained in Fig.~\ref{fig:phasediag_N=3_DMRG} and it contains only
three phases: SP, gapless BCS and CDW.

Data points on this plot correspond to simulations done on system length $L=72$, while phase boundaries were also obtained from scaling different system sizes (see below). 

\subsection{Properties along the SU(2)$_c$ line}

\begin{figure}[t]
\centering
\includegraphics[width=\columnwidth,clip]{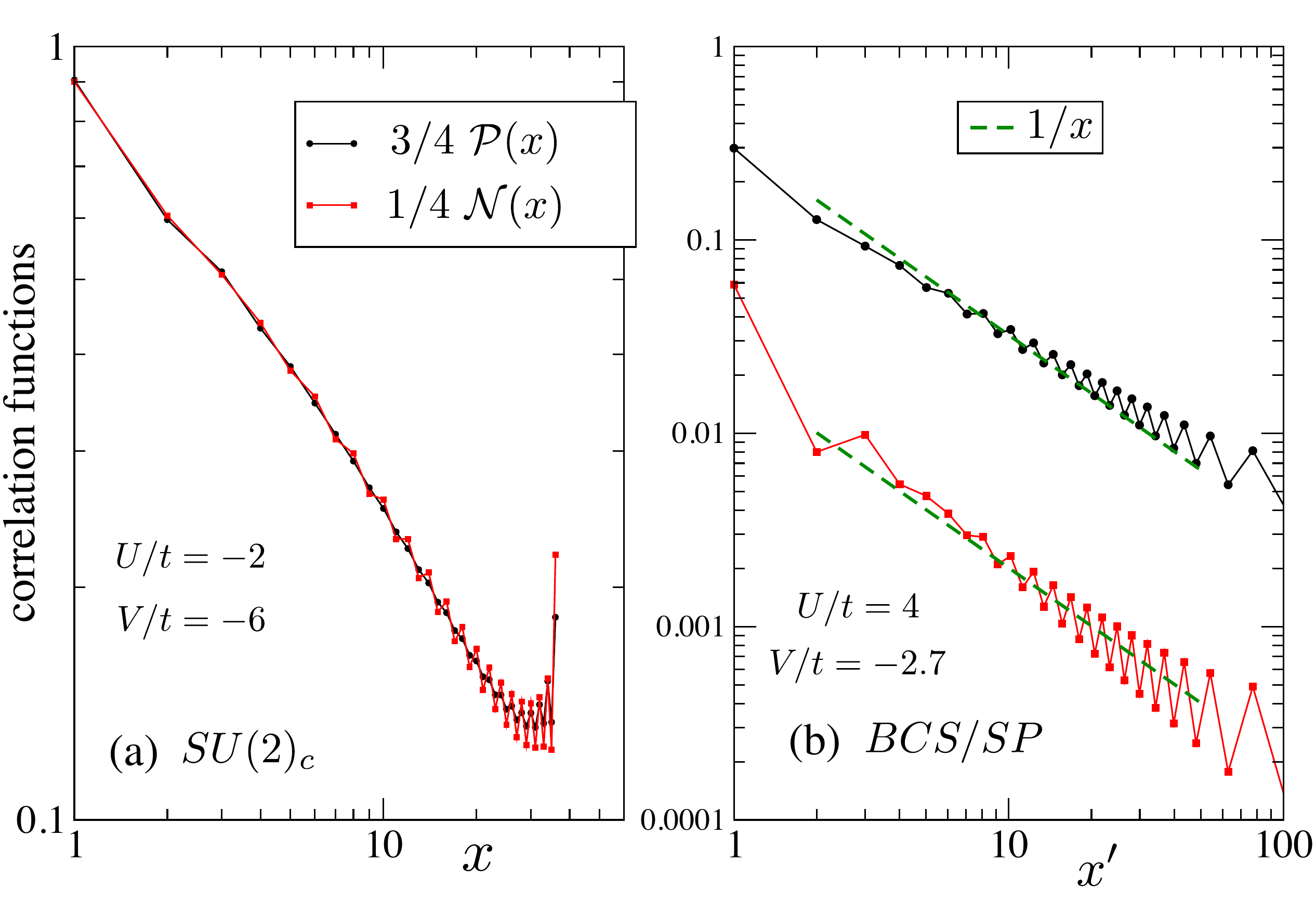}
\caption{Pairing and density correlations obtained by DMRG in the $N=3$ case for various interactions corresponding (a) to the exact SU(2)$_c$ symmetry; (b) to the emergent SU(2) symmetry 
${\widetilde{\text{SU}(2)}}_c$. Note that correlations are measured starting from the middle of the chain.}
\label{fig:corr_N=3_SU2}
\end{figure}

We start by considering the SU(2)$_c$ line $V=3U$. 
For large enough $|U|/t$, the strong coupling argument of Sec. II B tells us that the
chain will behave effectively as an antiferromagnetic Heisenberg spin-3/2 chain, which is
known to be critical.  In Fig.~\ref{fig:corr_N=3_SU2}(a), we show how pairing and
density correlations behave along this SU(2)$_c$ line. Their  long-distance form has been determined in Eq.~(\ref{correlBCS}) and, measured from the middle of the chain, reads: 
\begin{eqnarray}\label{corr.N=3.eq}
{\cal P}(x) &=& \langle P^{\dagger}_{00}(L/2+x) P_{00}(L/2) \rangle \sim \frac{A}{x^{1/(NK_c)}} \\
{\cal N}(x) &=& \langle n(L/2+x) n(L/2) \rangle - \langle n(L/2+x)\rangle \langle n(L/2)\rangle \nonumber \\
&\sim& -\frac{NK_c}{\pi^2 x^2} + \frac{(-1)^x B}{x^{NK_c}} . \nonumber
\end{eqnarray}
Using the definition of the pseudo-spin operator (\ref{spinop.eq}), we
observe that the two correlations match perfectly, as expected of
course for an exact SU(2)$_c$ symmetry. Both correlations are algebraic
and expected to decay as $\sqrt{\ln x}/x$, \cite{Afflecklog} but it is
known that checking accurately logarithmic corrections is a
challenging numerical problem~\cite{Hallberg1996} that we will not
further investigate.

\begin{figure}[!ht]
\centering
\includegraphics[width=\columnwidth,clip]{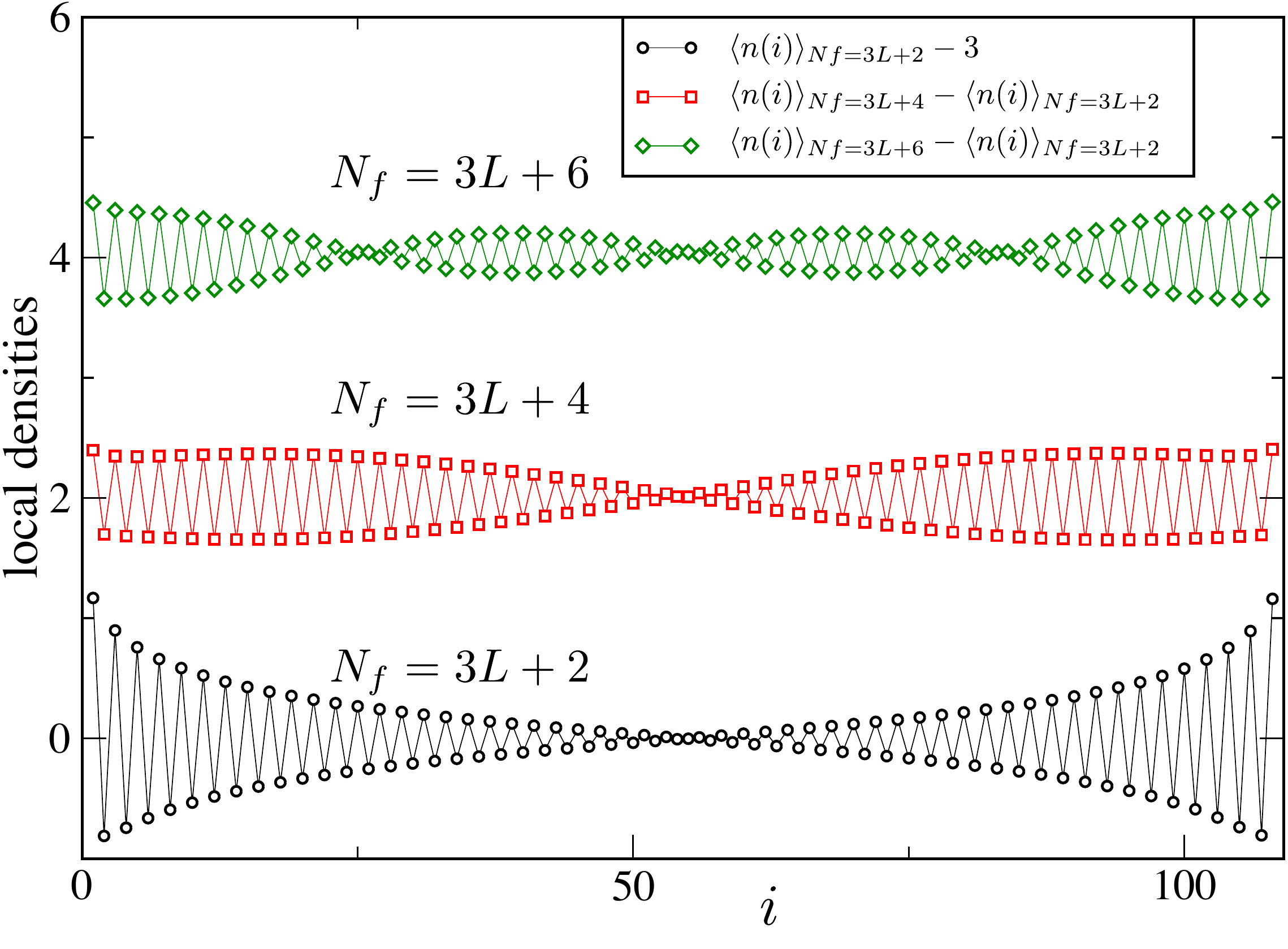}
\caption{Local densities obtained by DMRG in the $N=3$ case for $U/t=-2$, $V=3U$, and 
$L=108$. From bottom to top, data correspond to adding 2, 4 or 6 particles to the half-filled system.  Data for adding 4 and 6 particles are shifted by 2 and 4 respectively for clarity, and in these cases $\langle n(i)\rangle_{N_f=3L+2}$  has been substracted in order to get the bulk contribution.
 }
\label{fig:edgestates_N=3}
\end{figure}

Another peculiar property of spin-3/2 chain with OBC was conjectured
by Ng~\cite{Ng1994}, and confirmed later numerically: \cite{Ng1995}
even though the system is critical, one can observe ``edge states''
with OBC, in the sense that the magnetization profile will exhibit an
excess close to the edges, although there are no finite correlation
length (i.e. the magnetization profile decays algebraically away from
the edges). Here, we investigate a similar
situation, namely with a charge SU(2)$_c$ symmetry where it is the local density that 
plays the role of the magnetization for actual spin-3/2 chain. When
adding 2, 4 or 6 particles (with respect to half-filling), as shown in Fig.~\ref{fig:edgestates_N=3}, we do
observe modulations in the local densities reflecting these edge
states.  Physically, it means that the first excitation (adding 2
particles) is an edge excitation, but the next ones correspond to
making a \emph{bulk} excitation.

\subsection{The transition from critical BCS to SP}

As can be seen from the phase diagram shown in Fig.~\ref{fig:phasediag_N=3_DMRG}, the critical phase that exists
along the SU(2)$_c$ line has a rather large extension. As will be shown below, this critical phase has dominant BCS pairing correlations, thus its name. 
For fixed
negative $V$, we observe the transition to the SP phase for large enough
$U>0$. This is in agreement with the low-energy prediction and the RG phase
diagram (see Fig.~\ref{PhDiagNodd}).

\begin{figure}[!ht]
\centering
\includegraphics[width=\columnwidth,clip]{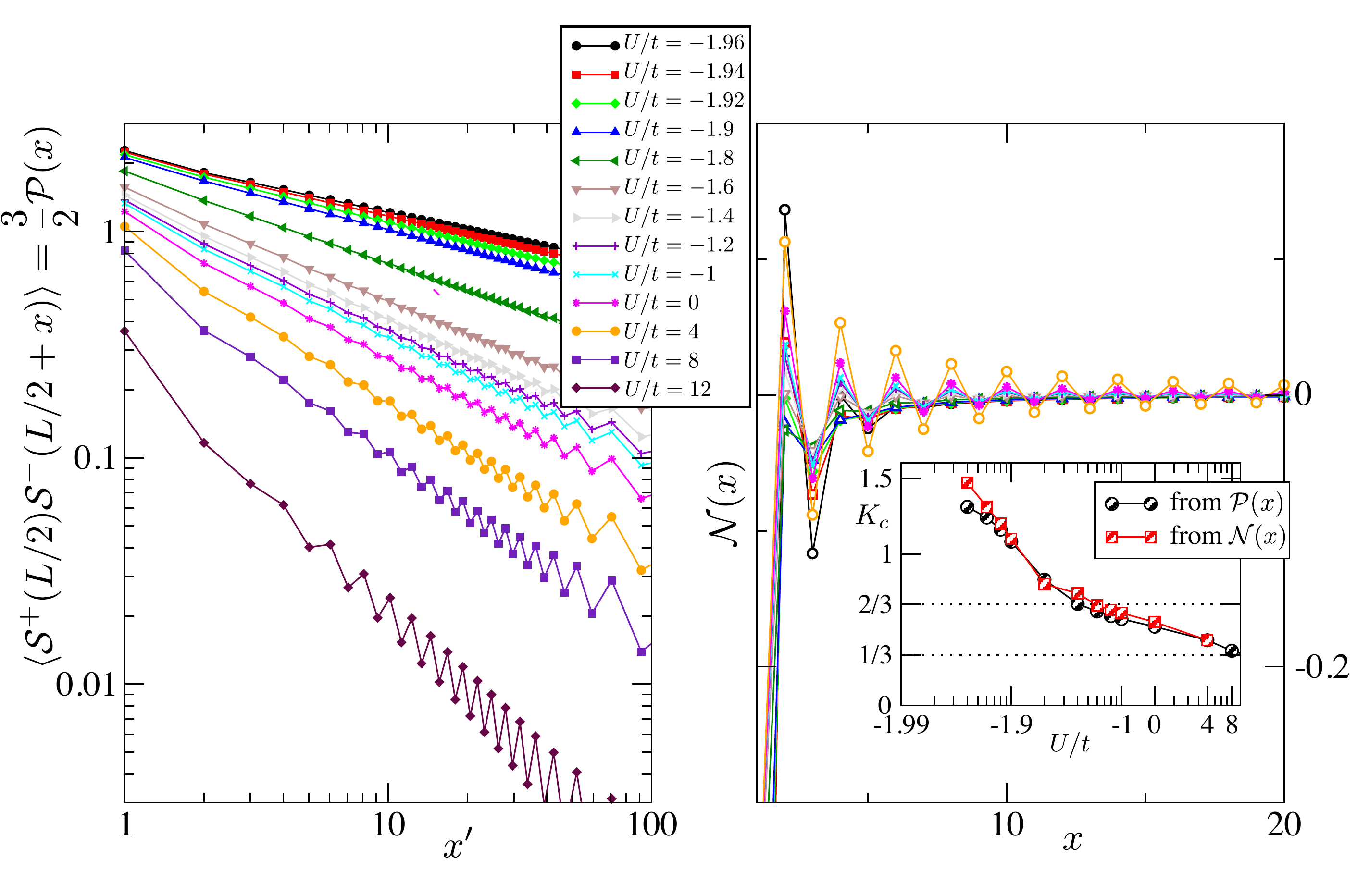}
\caption{(Color online) Pair and density correlations in the $N=3$ case 
for $V/t=-6$ and $L=72$ and
  various $U/t$. Inset: Fitting these data gives an estimate of the
  Luttinger parameter $K_c$ vs $U/t$ (using a log scale starting at $U/t=-2$). }
\label{fig:pair_N=3_V=-6}
\end{figure}

In order to characterize the critical phase, we can compute its
Luttinger parameter $K_c$ from the behavior of either pairing or 
density correlations, using Eq.~(\ref{corr.N=3.eq}). In
Fig.~\ref{fig:pair_N=3_V=-6}, we plot both correlations at $V=-6t$ and for various
values of $U$.  Let us start with the discussion of ${\cal P}(x)$ (which
corresponds up to a factor $2/3$ to the transverse pseudo-spin correlation
function). In order to be able to fit over the whole range~\cite{Cazalilla}, data are plotted vs $x'=d(x|L+1)/\sqrt{\cos(\pi x/(L+1))}$, where $d(x|L+1)$ is the cord function, defined in Eq.~(\ref{eq:EE}). We
observe a very smooth behavior, which allows to extract the behavior
of $K_c$ vs $U$ (see Inset).  Due to the logarithmic corrections which
are known to exist along the SU(2)$_c$ line, it is very hard to recover
that $K_c \rightarrow 1/3$ when $U\rightarrow -2t$ as expected from the exact SU(2)$_c$ symmetry. Moving away from
the SU(2)$_c$  line, our data indicate that $K_c$ rapidly reaches a
maximum, before going down again. The transition to SP corresponds to
pairing correlations that become exponential (not shown), and occurs 
 when $K_c=1/3$ in agreement with the low-energy approach.

Another way to compute $K_c$ consists in using Eq.~(\ref{corr.N=3.eq})
for density correlations. In principle, one can use either the uniform
or alternating part to extract it. However, in the regime where $K_c<2/3$, the alternating part is
dominant, whereas in the opposite case, the uniform part decays more
slowly. Therefore, we have fitted either the uniform part or the
staggered part to extract the value of $K_c$ shown in the Inset. 

Overall, we have an excellent agreement between the estimates of $K_c$ obtained from both correlations,  which gives confidence in the validity of the Luttinger liquid description of this critical phase.  Moreover, the behavior of $K_c$ vs $U$ is compatible with our expectation (see Sec.~\ref{part:odd_phase_diagram}): $K_c$ exceeds 1/3 in the BCS phase (giving rise to dominant BCS correlations) and the transitions to SP and CDW 
occur when $K_c$ reaches 1/3. 

In the critical phase,  the von Neumann block entropy gives access to the central charge $c$, and is consistent with a $c=1$ Luttinger liquid as expected (data not shown). 
In the gapped SP phase, the EE scaling is consistent with a saturation for large blocks.

\begin{figure}[t]
\centering
\includegraphics[width=\columnwidth,clip]{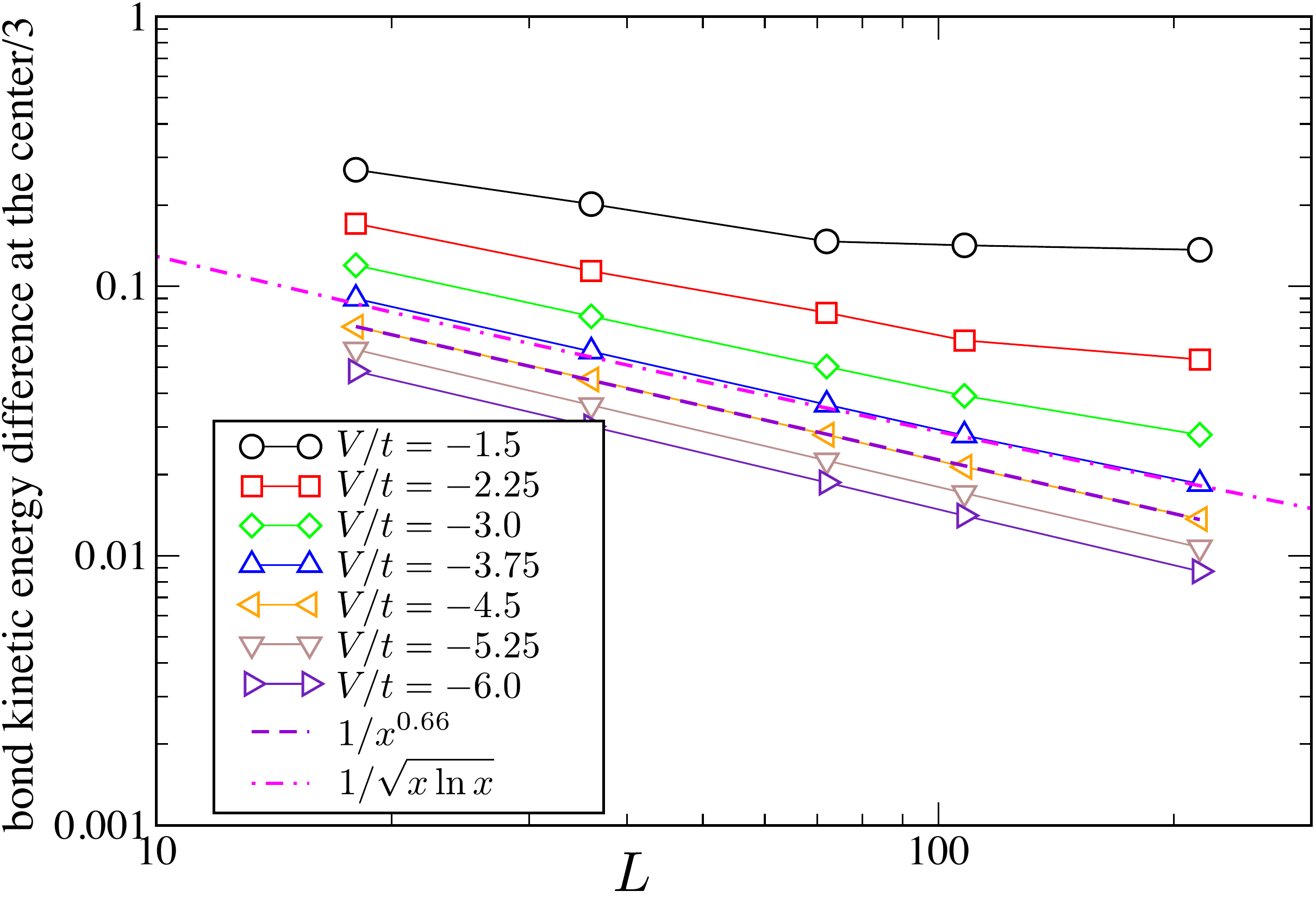}
\caption{(Color online) Bond kinetic energy modulation at the center of a chain of length $L$ for various $V$ at fixed $U/t=3$.}
\label{fig:kin_U=3_N=3}
\end{figure}

The transition from the critical BCS to the SP phase can be located when ${\cal P}(x)$
becomes exponential, or by looking at the bond kinetic modulation
scaling. In Fig.~\ref{fig:kin_U=3_N=3}, we plot the bond kinetic
energy difference at the center of the chain as a function of the
chain length $L$. We can clearly see a finite value in the SP phase
for $V/t=-1.5$ and $U/t=3$ for instance, while our data are compatible
with an algebraic power-law with exponent $0.66$ for larger
$|V|$. Locating precisely the transition is difficult since we
expect a BKT behavior at the transition, and a weakly opening gap when entering the SP phase; this
results in some uncertainty on this transition line in the phase
diagram. Using this procedure, we have determined approximately the phase transition line shown in Fig.~\ref{fig:phasediag_N=3_DMRG}. Although our data are not very accurate, our numerical findings are in agreement 
with the low-energy approach: the BCS to SP transition occurs for a \emph{finite} negative $V$ for fixed $U>0$. 

Note that the low-energy approach predicts that the transition occurs
when $K_c$ reaches 1/3. According to our fitting procedure (see
Fig.~\ref{fig:pair_N=3_V=-6}), this gives a similar estimate for its
location. According to this value, the bond kinetic energy modulation
should scale as $1/\sqrt{L}$ at the transition, while we have measured
a different exponent. In fact, it is known that logarithmic
corrections are expected at this transition, and indeed our data can
as well be fitted with a $1/\sqrt{L \log{L}}$ law.
 
Moreover, along this transition line and from the low-energy approach, we expect
an emergent SU(2) symmetry (${\widetilde{\text{SU}(2)}}_c$) that should be reflected in identical
exponents for ${\cal P}(x)$ and ${\cal N}(x)$.
Fig.~\ref{fig:corr_N=3_SU2}(b) displays our data in this region, and  we do confirm a good agreement between the two exponents (compatible with $K_c=1/3$).

Concerning the transition from the critical BCS to the CDW phase, our data are compatible with a gap opening 
as soon as $V>NU$, in perfect agreement with the low-energy prediction (see Sec. ~\ref{part:odd_phase_diagram}).
Finally, for the same reasons as in the $N=2$ case (see Sec. \ref{part:CDWSPtransition}), we could not investigate the nature of the quantum phase transition between SP and CDW phases.
We found that this transition is located in the vicinity of the $V = - 9 U/2$ line (see Fig.~\ref{fig:phasediag_N=3_DMRG}).
Unfortunately, as already stressed in Sec. II, we were not able to determine the symmetry contents of this line.

As a final remark about the BCS phase, while quasi-edge states can be
observed along the SU(2)$_c$ line or close to it (see previous section),
they no longer exist deep in the BCS phase (for instance $U/t=0$ and $V/t=-6$, data not shown). This might be understood
from the strong-coupling regime using the mapping to a spin-3/2
chain with single-ion anisotropy: for large enough $D>0$, the
relevant low-energy states consist in $S_i^z=\pm 1/2$ on each site,
thus leading to an effective spin-1/2 chain in its critical phase. In
this region, we do not expect any edge physics as is observed
numerically. We have not investigated in details the crossover between
both regimes, but it could be easily answered by studying directly a spin-3/2 anisotropic chain.

\section{Phase diagram in the $N=4$ case}

In this section, we investigate the phase diagram of
model~\eqref{hubbardSgen} when $N=4$ and in the $(U/t,V/t)$
plane. 

From a technical point of view, since the local Hilbert space is quite large, we map the one-dimensional model onto a generalized 4-leg Hubbard ladder with generalized rung interactions. Thus, we reduce the complexity of the DMRG algorithm, but we have to use a 1D path going along the ladder. We have checked that the symmetry between the chains is always restored during the simulations. 
Typically, we keep 2000 states in our simulations and use OBC.

\begin{figure}[!ht]
\centering
\includegraphics[width=0.95\columnwidth,clip]{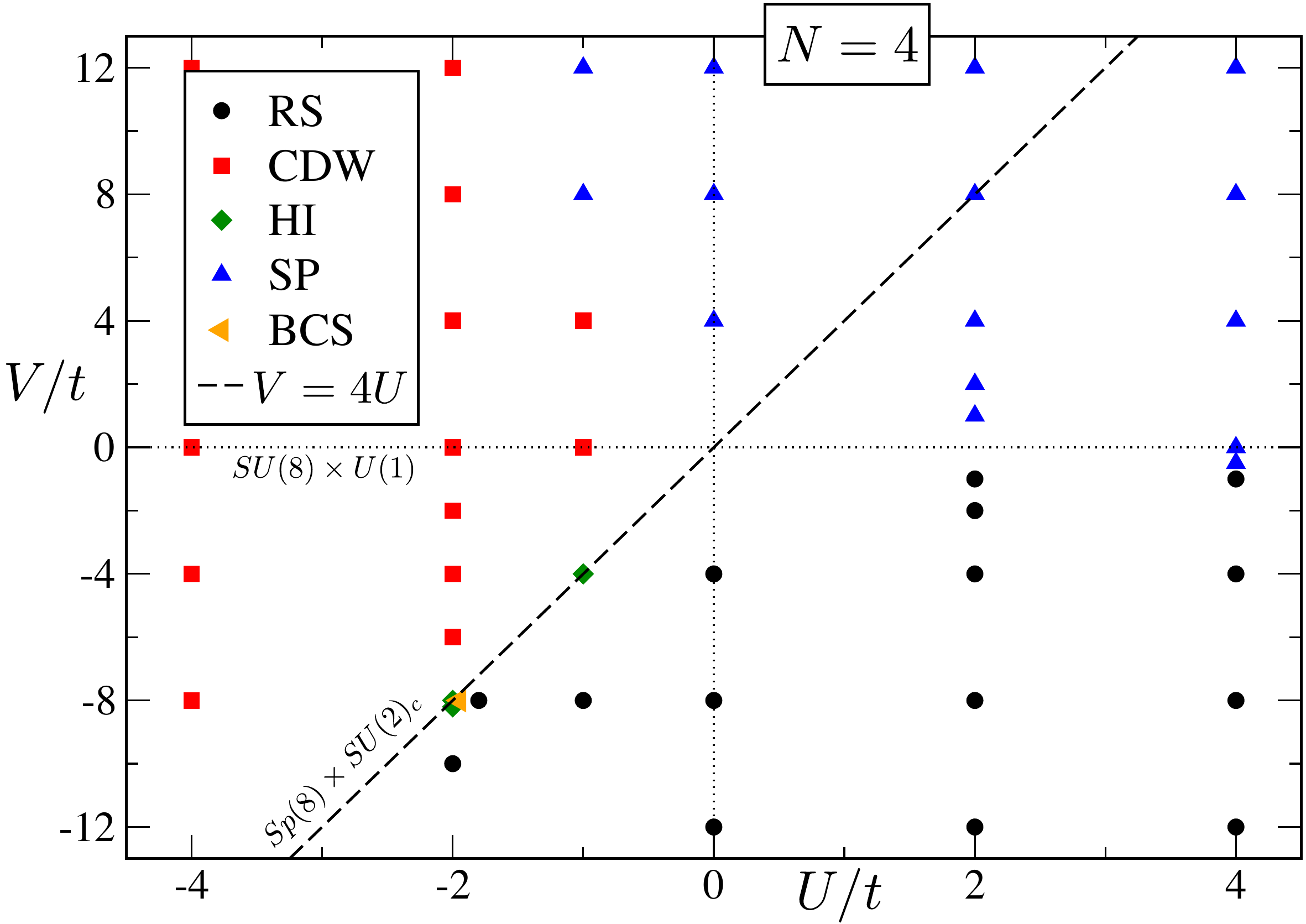}
\caption{Numerical phase diagram obtained by DMRG in the $N=4$ case
  with $L=30$. }
\label{fig:Phasediag_N=4_DMRG}
\end{figure}

Fig.~\ref{fig:Phasediag_N=4_DMRG} shows the phase diagram for $N=4$,
obtained on a $L=30$ chain.  As expected, five phases are present: on the SU(2)$_c$ line in the attractive part of the phase diagram, there is the HI phase and, close to it, the critical BCS one arises. As expected from the strong-coupling argument (see Eq.~(\ref{HaldaneSU2Hamiltonian})), it is followed by the RS phase and, on the other side of the HI phase, a CDW phase is stabilized. On the repulsive side, we detect a SP phase that was predicted in Sec. III C.
We observe a good agreement with the low-energy prediction (see
Fig.~\ref{PhDiagNeven}).

\subsection{Properties along the SU(2)$_c$ line}

We start by looking at the SU(2)$_c$ line $V=4U$ on the attractive
side. As expected from the strong-coupling argument and from the
low-energy analysis, the model should behave as an effective antiferromagnetic spin-2
Heisenberg chain, i.e. be in a Haldane phase.

We have some evidence for such a HI phase thanks to the presence of
(charge) edge states when OBC are used. Concerning the charge gap, in
order to get the bulk result (and avoid edge states effect), one needs
to compute $E_0(N_f=4L+6)-E_0(N_f=4L+4)$. Extrapolating our data on
$L=16$ and $L=32$ chains for $U=-2t$ and $V=-8t$, we obtain an
estimate of $0.0038 t$, which is extremely small. Nevertheless, using
the strong-coupling expression of the effective exchange $J_{\rm
  eff}=1/18 t$ and the known Haldane gap~\cite{Todo2001} $\Delta \simeq
0.089 J_{\rm eff} \simeq 0.0049 t$, we get a finite gap of the same
magnitude.

Moreover,  the pairing correlations (which correspond to the transverse spin correlations in the spin language) shown in Fig.~\ref{fig_corr_pair_N=4} exhibit a short-range behavior compatible with a finite correlation length and a finite gap.

However, since the correlation length of the spin-2 chain is known to be very large~\cite{Todo2001} ($\xi \sim 50$), we will not try to characterize further this HI phase (by measuring its string order for instance), but the strong-coupling argument ensures that HI phase exists in some finite region of the phase diagram around the SU(2)$_c$ line.   

\subsection{Critical BCS phase}

In Sec.~\ref{sec:strong-coupling}, we have argued why for fixed $V/t$, increasing $U/t$ gives an effective single-ion anisotropy denoted $D$.  
For the spin-2 chain, it is
known~\cite{schollwock} that such a $D$ term leads to an
\emph{extended} critical XY phase with central charge $c=1$ for $ 0.04 < D/J_{\rm eff} < 2.4$. Using our
strong-coupling estimate and assuming a fixed effective $J_{\rm eff}$, this
would predict an extended XY phase for $ -1.996 < U/t < -1.82$ for fixed
$V/t=-8$.

\begin{figure}[t]
\centering
\includegraphics[width=0.95\columnwidth,clip]{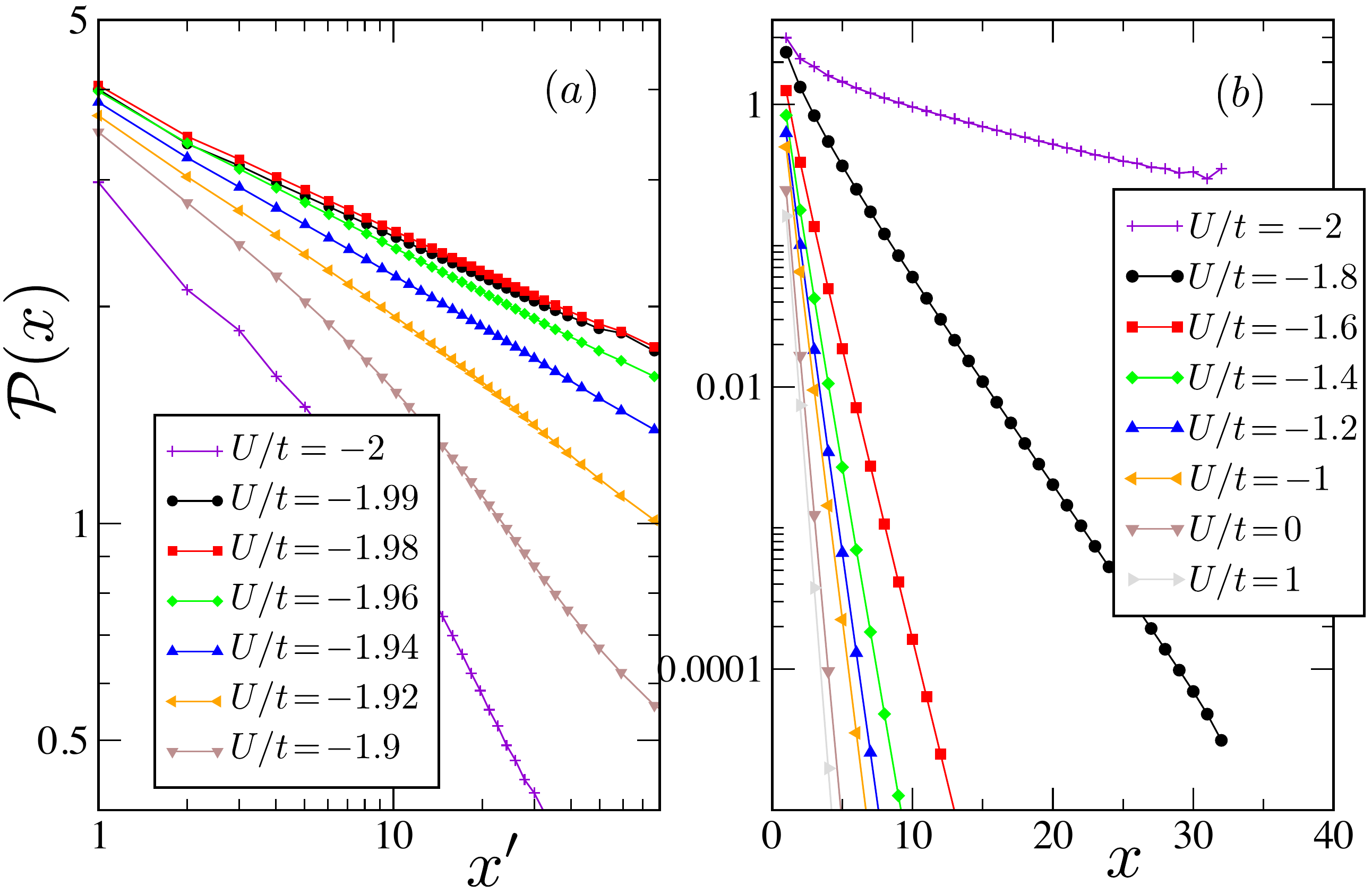}
\caption{Pairing correlations for $N=4$ at fixed $V/t=-8$ and
  $L=64$. (a) Critical correlations are observed close to the HI
  phase represented by $U/t=-2$ on the SU(2)$_c$ line. (b) Both in HI and in RS phase, the pairing correlations are short-ranged.}
\label{fig_corr_pair_N=4}
\end{figure}

Fig.~\ref{fig_corr_pair_N=4} shows numerical data for the pairing
correlations obtained for a chain with $L=64$. 
Indeed, algebraic behavior is observed close to the
Haldane phase, in a region corresponding to $ -1.99 \lessapprox U/t \lessapprox -1.9$ for $V=-8t$, in agreement with our strong-coupling estimate.

In order to be more quantitative about this Luttinger liquid phase, and make a connection with the low-energy analysis, we use the asymptotics from Eq.~(\ref{correlBCSNeven}):
\begin{eqnarray}\label{corr_Neven_DMRG.eq}
{\cal P}(x) &=& \langle  P^{\dagger}_{00}(L/2+x) P_{00}(L/2) \rangle \sim  x^{- 1/NK_c} \nonumber \\
{\cal N}(x) &=& \langle n(L/2+x) n(L/2) \rangle - \langle n(L/2+x)\rangle \langle n(L/2)\rangle \nonumber \\
 &\sim& -\frac{NK_c}{\pi^2 x^2}+(-1)^x A \exp (-x/\xi),
\end{eqnarray}
and use it  to extract the behavior of $K_c$ in the $c=1$ 
gapless phase. Note that we have measured the correlations from the center of the chain in order to minimize size effects due to OBC, and in the critical phase, we plot our data vs $x'=d(x|L+1)/\sqrt{\cos(\pi x/(L+1))}$  in order to be able to fit over the whole range~\cite{Cazalilla}.

\begin{figure}[t]
\centering
\includegraphics[width=0.95\columnwidth,clip]{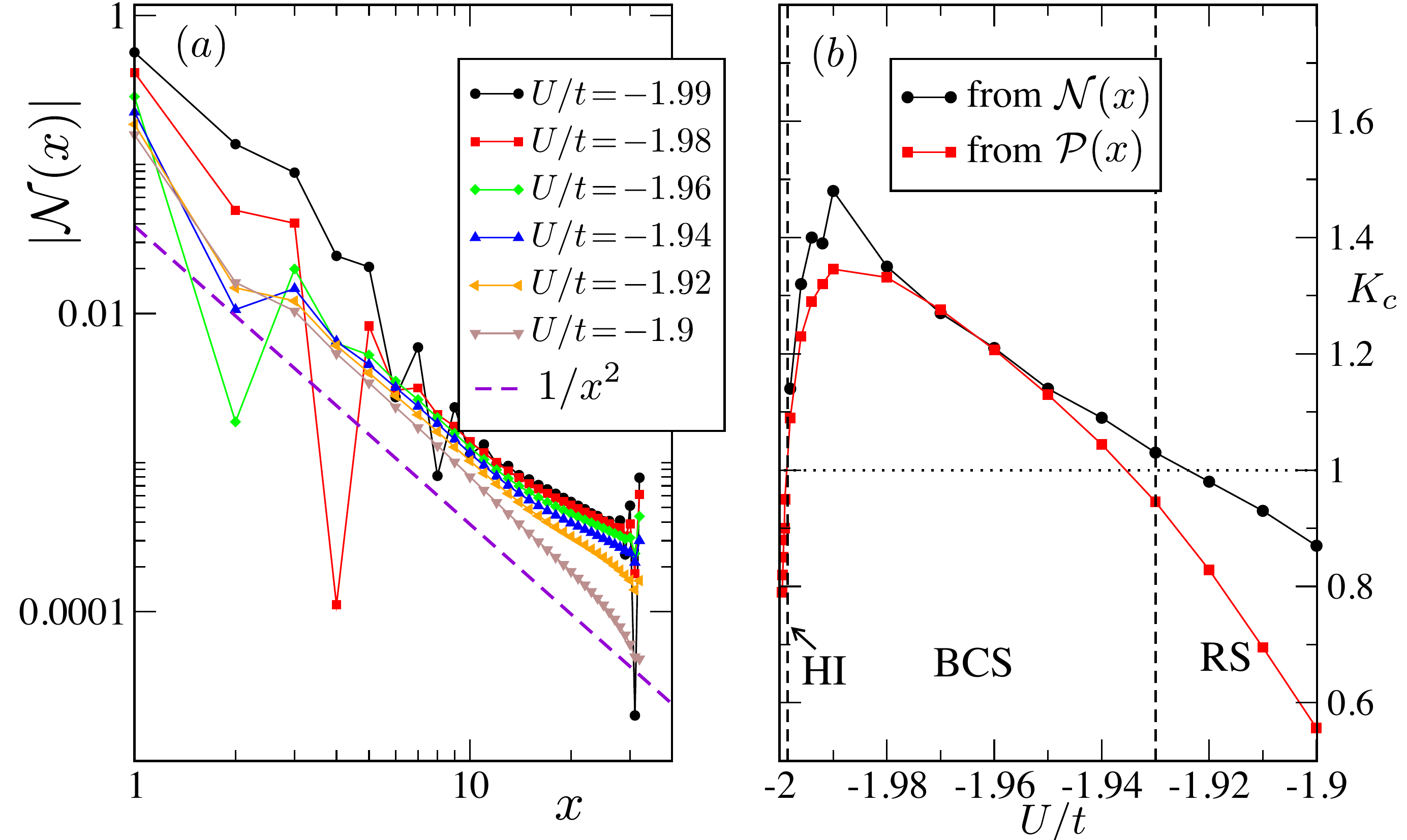}
\caption{(a) Density correlations for $N=4$ at fixed $V/t=-8$ and
  $L=64$ in the BCS phase. 
(b) Luttinger liquid parameter $K_c$ vs $U$ for fixed $V/t=-8$. BCS phase is delimited by $K_c\geq 1$.}
\label{fig_corr_CDW_N=4}
\end{figure}

Density correlations are shown in Fig.~\ref{fig_corr_CDW_N=4}(a) in the critical phase, and they exhibit an algebraic decay. The anomalies are due to the subleading short-range staggered contributions since we plot the absolute values of ${\cal N}(x)$, so that density correlations become difficult to fit close to the HI phase. We have fitted both correlations using the expressions above in Eqs.~(\ref{corr_Neven_DMRG.eq}) and the resulting Luttinger parameter $K_c$ is plotted in Fig.~\ref{fig_corr_CDW_N=4}(b). We have an overall good agreement between the two \emph{independent} fits, and we confirm the expected behavior that, starting from the HI phase, $K_c$ first increases rapidly, and then diminishes when $U/t$ increases. The gapless phase is characterized by $K_c\geq 1$, which corresponds for $V/t=-8$ to $-1.999 \leq U/t \leq -1.93$. Thus, we conclude, from Eqs.~(\ref{corr_Neven_DMRG.eq}), that the leading
instability is the BCS singlet-pairing. The extended gapless phase is thus a BCS phase which 
differs from the one in the $N=3$ case by the fact that the staggered part of the density is short-ranged.

For the choice of $V/t=-8$, we see that both pairing and density correlations become short-range when $U/t\gtrapprox -1.9$, where a RS (large-D) phase starts. As can be seen from the behavior of $K_c$, this corresponds to $K_c=1=4/N$ which is the criterion for the opening of the gap in the bosonization analysis done in Sec.~\ref{part:even_phase_diagram}.

\subsection{Quantum phase transitions}

Once we have determined the five phases in the phase diagram, we would like to clarify the nature of the quantum phase transitions. 

Starting from the SU(2)$_c$ line with $V<0$ and decreasing $U$ leads
to a CDW phase.  According to the low-energy approach, and as
confirmed numerically for $N=2$, we expect a $c=1/2$ second-order
Ising phase transition. However, due to the large correlation length
in the HI phase, we cannot get reliable results. It would be much
easier to check this criticality, as well as locate the critical
$D/J$, by studying directly the spin-2 chain with single-ion
anisotropy.

On the other side of HI phase, i.e. increasing $U$, the phase
transition to BCS critical phase was predicted to be in the BKT
universality class with $c=1$.  Scaling of EE for $L=64$ with
$V/t=-4.0$ and $U/t=-0.9$ leads to $c=0.93$ (when keeping $m=4000$ states in the DMRG simulation).  From the BCS phase and
increasing $U/t$ at fixed $V/t<0$, our correlation functions in
Figs.~\ref{fig_corr_pair_N=4}-\ref{fig_corr_CDW_N=4} are compatible
with a transition to a fully gapped RS phase when $K_c$ becomes
smaller than 1 as expected.

The transition from RS to SP is difficult to characterize due to finite-size oscillations in the quantities (including EE), but we have determined that it is located at a \emph{finite} negative $V$ for fixed $U>0$ as found in the low-energy section 
(see Fig.~\ref{PhDiagNeven}). A similar conclusion can be made for the transition from SP to CDW, which is located in the opposite quadrant as expected. 

\section{Conclusion}

We have established the zero temperature phase diagram of
multicomponent ($2N$-component) fermionic cold atoms, loaded in a 1D
optical lattice, at half filling. This entire work was done under the
hypothesis that only contact interactions matter and that the
interactions channels can be reduced to two: one singlet channel and
one non-zero spin channel. The former hypothesis is very reasonable in
the context of optical lattices, and could be relaxed without
affecting our main conclusions. The latter hypothesis requires $N-2$
independent fine-tunings, and is therefore quite restrictive for large
$N$. However, it requires no fine-tuning for $N\leq 2$, and should not
be out of reach for moderate $N\leq4$.

As soon as $N>1$, we found that the phase diagram has a rich structure
due to the degeneracy of the atomic states and the absence of
spin-charge separation at half-filling.  Several nonequivalent
Mott-insulating phases emerge. Two phases are present irrespective of
the value of $N>1$: the SP and CDW phases, that both break
translational invariance and are two-fold degenerate.  We exhibited a
hidden pseudo-spin SU(2) structure, involving spin-singlet, charged
degrees of freedom, that generalizes a structure noticed long ago for
the $N=1$ case in the context of the Hubbard model.\cite{yangzhang}
When specialized to one space dimension, this structure yields a
Haldane conjecture for attractive interactions: we show that such a
system realizes a Heisenberg antiferromagnet of magnitude $S=N/2$,
and, as a consequence, displays an alternating gapped (insulating) /
critical (BCS superfluid) behavior according to the even / odd parity
of $N$.  We have found that this parity effect has an influence on
large portions of the phase diagram, and that ultracold fermions with
$N$ even can disclose two more insulating states: the HI and RS phases
which are non-degenerate and display non-local string orderings.  The
$N=1,2$ cases turn out not to be the generic cases of the odd/even
families.  Precisely, whereas for $N=2$, only a critical quantum phase
transition occurs between the two non-degenerate insulating phases,
for even $N>2$, an intermediate gapless BCS phase arises between them.

On top of the even-odd scenario, and within the low-energy approach, we found a subtle effect
depending on the parity of $N/2$. When $N/2$ is odd, the HI and RS phases correspond to different 
phases and can be distinguished by string-order parameters. In particular, the HI phase with odd $N/2$, i.e.
odd spin,
is an exemple of a topological ordered phase with $N/4$ edge states.
In contrast, when $N/2$ is even, the HI and RS phases are related at low-energy by a duality symmetry
and share the same ground-state properties. In this respect, the HI phase with $N/2$ even, i.e.
for even spin, is not topologically protected by its edge-state structure but is equivalent
to a topological trivial insulating phase, i.e. the RS phase.
Thus, within the low-energy approach presented in this paper, our findings confirm the recent
conjecture of Ref.~\onlinecite{pollman09}. 

In the light of the recent experimental achievements where cold fermionic gases with several components could be stabilized as highly symmetry systems \cite{ytterbium}, we hope that it 
will be possible in the future to unveil part of the richness that we highlighted in this work. 
In particular, the disclosure of the HI phase would be extremely important, as it displays exotic characteristics that have attracted a lot of attention in the past years and still does nowadays.


\section*{Acknowledgements}
The authors would like to thank T. Jolicoeur, K. Totsuka, and  P. Azaria for insightful discussions. 
Numerical simulations were performed using HPC resources from GENCI-IDRIS (Grant 2009-100225) and CALMIP.
Finally, one of us (P.L.) would like to dedicate this paper to the memory of the late Sasha Gogolin.


\begin{thebibliography}{99}
\bibitem{wenbook}
X. G.~Wen, \textit{Quantum Field Theory of Many-Body Systems} 
(Oxford University Press, UK, 2004). 
\bibitem{anderson}
P. W. Anderson, \textit{Basic Notions of Condensed Matter Physics} 
(Addison-Wesley, USA, 1984). 
\bibitem{haldane}
F. D. M. Haldane, Phys. Lett. A {\bf 93}, 464 (1983);
Phys. Rev. Lett. {\bf 50}, 1153 (1983).
\bibitem{dennijs}
M. P. M. den Nijs and
K. Rommelse,
Phys. Rev. B \textbf{40},
4709 (1989).
\bibitem{kennedy}
T. Kennedy and H. Tasaki,
Phys. Rev. B \textbf{45},
304 (1992).
\bibitem{hagiwara}
M. Hagiwara, K. Katsumata, I. Affleck, B. I. Halperin, and J. P. Renard,
Phys. Rev. Lett. {\bf 65}, 3181 (1990).
\bibitem{phlereview} P. Lecheminant in {\sl Frustrated Spin Systems}, edited by H. T. Diep
(World Scientific, Singapore, 2004).
\bibitem{mikeska}
H.-J. Mikeska and A. K. Kolezhuk,
Lecture Notes in Physics vol. 645, p. 1 (2004).
\bibitem{auerbach98} A. Auerbach and  E. Altman,
Phys. Rev. Lett.  {\bf 81}, 4484 (1998).
\bibitem{cirac} J. J. Garcia-Ripoll, M. A. Martin-Delgado, and J. I. Cirac,
Phys. Rev. Lett.  {\bf 93}, 250405 (2004).
\bibitem{Berg2008} E. G. Dalla Torre, E. Berg, and E. Altman,
Phys. Rev. Lett.  {\bf 97}, 260401 (2006);
E. Berg, E. G. Dalla Torre, T. Giamarchi, and E. Altman, 
Phys. Rev. B {\bf 77}, 245119 (2008).
\bibitem{Lee2008} Y. W. Lee, Y. L. Lee, and M.-F. Yang, 
Phys. Rev. B {\bf 76}, 075117 (2007);  Y. W. Lee 
Phys. Rev. B {\bf 77}, 064514 (2008).
\bibitem{amico}
L. Amico, G. Mazzarella, S. Pasini, and F. S. Cataliotti, 
New J. Phys. {\bf 12}, 013002 (2010).
\bibitem{dalmonte}
M. Dalmonte, M. Di Dio, L. Barbiero, and F. Ortolani,
Phys. Rev. B {\bf 83}, 155110 (2011).
\bibitem{Nonne2009} H. Nonne, P. Lecheminant, S. Capponi, G. Roux, and 
E. Boulat, Phys. Rev. B {\bf 81}, 020408(R) (2010).
\bibitem{dms}
 P. Di Francesco, P. Mathieu, and D. S\'en\'echal,
\emph{Conformal Field Theory} (Springer, Berlin, 1997).
\bibitem{DMRG}
S. R. White, Phys. Rev. Lett. {\bf 69}, 2863 (1992);
Phys. Rev. B {\bf 48}, 10345 (1993);
U. Schollw{\"o}ck, Rev. Mod. Phys. {\bf 77}, 259 (2005).
\bibitem{ho}
T. L. Ho and S. Yip,
Phys. Rev. Lett. {\bf 82}, 247 (1999).
\bibitem{sachdev} 
S. Sachdev and Z. Wang,
Phys. Rev. B {\bf 43}, 10229 (1991).
\bibitem{wuzhang} 
C. J. Wu and S. C. Zhang, Phys. Rev. B {\bf 71}, 155115 (2005).
\bibitem{zhang} 
C. J. Wu, J. P. Hu,
and S. C. Zhang, Phys. Rev. Lett. {\bf 91}, 186 (2003);
C. J. Wu, Mod. Phys. Lett. B {\bf 20}, 1707 (2006).
\bibitem{Lecheminant2005}
P. Lecheminant, E. Boulat, and P. Azaria,
Phys. Rev. Lett. {\bf 95}, 240402 (2005).
\bibitem{phle}
P.~Lecheminant, P.~Azaria, and E.~Boulat,
Nucl. Phys. B {\bf 798}, 443 (2008).
\bibitem{Wu2005}
C. J. Wu, Phys. Rev. Lett. {\bf 95}, 266404 (2005).
\bibitem{sylvainMS}
S. Capponi, G. Roux, P. Azaria, E. Boulat, and P. Lecheminant,
Phys. Rev. B {\bf 75}, 100503(R) (2007);
S. Capponi, G. Roux, P. Lecheminant, P. Azaria, E. Boulat, and S. R. White, 
Phys. Rev. A {\bf 77}, 013624 (2008);
G. Roux, S. Capponi, P. Lecheminant, and P. Azaria,
Eur. Phys. J. B {\bf 68}, 293 (2009).
\bibitem{wunum} H.-H. Hung, Y. Wang, and C. Wu,
arXiv: 1103.1926.
\bibitem{yang}
C. N. Yang,
Phys. Rev. Lett. {\bf 63}, 2144 (1989).
\bibitem{yangzhang}
C. N. Yang and S. C. Zhang,
Mod. Phys.  Lett. B {\bf 4}, 759 (1990);
S. C. Zhang, Int. J. Mod. Phys. B {\bf 5}, 153 (1991).
\bibitem{bookboso}
A. O. Gogolin, A. A. Nersesyan, and A. M. Tsvelik,
{\sl Bosonization and Strongly Correlated Systems}
(Cambridge University Press, Cambridge, England, 1998).
\bibitem{giamarchi}
T. Giamarchi, \textit{Quantum Physics in One Dimension}
(Clarendon press, Oxford, UK, 2004).
\bibitem{Ng1994} T. K. Ng, Phys. Rev. B {\bf 50}, 555 (1994).
\bibitem{Ng1995} S. Qin, T. K. Ng, and Z. B. Su, Phys. Rev. B {\bf 52}, 12844 (1995);
J. Lou, S. Qin, T. K. Ng, and Z. Su, Phys. Rev. B {\bf 65}, 104401 (2002);
J. Lou, S. Qin, and C. Chen, Phys. Rev. Lett. {\bf 91}, 087204 (2003).
\bibitem{pollman09}
F. Pollmann, E. Berg, A.~M. Turner, and M. Oshikawa, arXiv: 0909.4059.
\bibitem{tonegawa}
T. Tonegawa, K. Okamoto, H. Nakano, T. Sakai, K. Nomura, and M. Kaburagi,
J. Phys. Soc. Jpn. {\bf 80}, 043001  (2011).
\bibitem{afflecksun}
I. Affleck, D. Arovas, J. B. Marston, and D. Rabson, Nucl. Phys.
B {\bf 366}, 467  (1991).
\bibitem{marston07}
A. Paramekanti and J. B. Marston, 
J. Phys. Cond. Matter {\bf 19}, 125215 (2007).
\bibitem{assaraf}
R. Assaraf, P. Azaria, E. Boulat, M. Caffarel, and P. Lecheminant,
Phys. Rev. Lett. {\bf 93}, 016407 (2004).
\bibitem{marston}
A. V. Onufriev and J. B. Marston,
Phys. Rev. B {\bf 59}, 12573 (1999).
\bibitem{nonne2010}
H. Nonne, E. Boulat, S. Capponi, and P. Lecheminant,
Phys. Rev. B {\bf 82}, 155134 (2010).
\bibitem{ueda}
J. Zhao, K. Ueda, and X. Wang,
Phys. Rev. B {\bf 74}, 233102 (2006); J. Zhao, K. Ueda,  and X. Wang, J. Phys. Soc. Jpn. {\bf 76},  114711 (2007).
\bibitem{anderson58}
P. W. Anderson, Phys. Rev. {\bf 112}, 1900 (1958).
\bibitem{wucapponi}
S. Capponi, C. J. Wu, and S. C. Zhang,
Phys. Rev. B {\bf 70}, 220505 (2004).
\bibitem{auerbach}
A. Auerbach, \textit{Interacting Electrons and Quantum Magnetism} 
(Springer-Verlag, New-York, 1994). 
\bibitem{schulz}
H. J. Schulz, Phys. Rev. B {\bf 34}, 6372 (1986).
\bibitem{cftembedding}
F. Bais and P. Bouwknegt, Nucl. Phys. B {\bf 279}, 561 (1987); A.
Schellekens and N. Warner, Phys. Rev. D {\bf 34}, 3092 (1986).
\bibitem{altschuler} D. Altschuler, Nucl. Phys. B {\bf 313}, 293 (1989).
\bibitem{knizhnik}
V. G. Knizhnik and A. B. Zamolodchikov, Nucl. Phys. B {\bf 247}, 83 (1984). 
\bibitem{boulat}
E. Boulat, P. Azaria, and P. Lecheminant, Nucl. Phys. B {\bf 822}, 367 (2009).
\bibitem{GN}
D. J. Gross and A. Neveu, Phys. Rev. D {\bf 10},  3235 (1974).
\bibitem{lin}
H.-H.~Lin, L.~Balents, and M. P. A.~Fisher,
Phys. Rev. B {\bf 58}, 1794 (1998).
\bibitem{saleur}
R. Konik, H. Saleur, and A. W. W. Ludwig,
Phys. Rev. B {\bf 66},  075105 (2002).
\bibitem{zamolo}
A. B.~Zamolodchikov and Al. B.~Zamolodchikov,
Ann. Phys. (N.Y.) {\bf 120}, 253 (1979).
\bibitem{karowski}
M. Karowski and H. J. Thun,
Nucl. Phys. B {\bf 190}, 61 (1981).
\bibitem{konik}
R. Konik and A. W. W. Ludwig,
Phys. Rev. B {\bf 64},  155112 (2001).
\bibitem{ahn90}
C. Ahn, D. Bernard,  and A. LeClair,
Nucl. Phys. B {\bf 346}, 409 (1990).
\bibitem{babichenko}
A. Babichenko, Nucl. Phys. B {\bf 697}, 481 (2004).
\bibitem{para} A. B. Zamolodchikov and V. A.  Fateev, Sov. Phys. JETP
{\bf 62}, 215 (1985).
\bibitem{gepner}
D. Gepner and  Z Qiu, Nucl. Phys. B {\bf 285}, 423 (1987).
\bibitem{affleckhaldane}
I. Affleck and F. D. M. Haldane, Phys. Rev. B {\bf 36}, 5291 (1987).
\bibitem{cabra}
D. Cabra, P. Pujol, and C. von Reichenbach, Phys. Rev. B {\bf 58}, 65 (1998).
\bibitem{fateev} V. A. Fateev, Int. J. Mod. Phys. A {\bf 6},  2109
(1991).
\bibitem{afflecksu2} I. Affleck, Nucl. Phys. B {\bf 265}, 448 (1986).
\bibitem{orignac} P. Lecheminant and E. Orignac, Phys. Rev. B {\bf 65}, 174406 (2002).
\bibitem{oshikawa} M. Oshikawa, J. Phys.: Condens. Matter {\bf 4}, 7469 (1995).
\bibitem{totsuka} K. Totsuka and M. Suzuki, J. Phys.: Condens. Matter {\bf 7}, 1639 (1995).
\bibitem{hatsugai} Y. Hatsugai, J. Phys. Soc. Jpn.  {\bf 61}, 3856 (1992). 
\bibitem{schollwock} U. Schollw\"ock and Th. Jolicoeur, Europhys. Lett. {\bf 30}, 493 (1995); U. Schollw\"ock, O. Golinelli, and Th. Jolicoeur, Phys. Rev. B {\bf 54}, 4038 (1996).
\bibitem{totsukaspin2} Y. Nishiyama, K. Totsuka, N. Hatano, and M. Suzuki, J. Phys. Soc. Jpn.  {\bf 64}, 414 (1995).
\bibitem{aschauer} H. Aschauer and U. Schollw\"ock, Phys. Rev. B {\bf 58}, 359 (1998).
\bibitem{qin} S. Qin, J. Lou, L. Sun, and C. Chen, Phys. Rev. Lett. {\bf 90}, 067202 (2003).
\bibitem{AKLT} I. Affleck, T. Kennedy, E. H. Lieb, and H. Tasaki, 
Phys. Rev. Lett. {\bf 59}, 799 (1987).
\bibitem{totsukacom} K. Totsuka, private communications.
\bibitem{yangpara} S. K. Yang, Nucl. Phys. B {\bf 285}, 183, 639 (1987).
\bibitem{phlegogolin} P. Lecheminant,  A. O. Gogolin, and A. A. Nersesyan,
Nucl. Phys. B {\bf 639}, 502 (2002).
\bibitem{fgn} M. Fabrizio, A. O. Gogolin, and A. A. Nersesyan,
Phys. Rev. Lett. {\bf 83}, 2014 (1999); M. Fabrizio, A. O. Gogolin, and A. A. Nersesyan,
Nucl. Phys. B {\bf 580}, 647 (2000).
\bibitem{mussardo} G. Delfino and G. Mussardo, Nucl. Phys. B {\bf 516},  675 (1998).
\bibitem{bajnok} Z. Bajnok, L. Palla, G. Takacs, and F. Wagner,
Nucl. Phys. B {\bf 601},  503 (2001).
\bibitem{spinonephasediag}
R. Botet, R. Jullien, and M. Kolb, Phys. Rev. B \textbf{28}, 3914 (1983);
W. Chen, K. Hida, and B.~C. Sanctuary, Phys. Rev. B \textbf{67}, 104401 (2003).
\bibitem{Degli2003}
C. Degli Esposti Boschi, E. Ercolessi, F. Ortolani, and M. Roncaglia, 
Eur. Phys. J. B \textbf{35}, 463 (2003).
\bibitem{hamer}
A. Fabricio Albuquerque, C. J. Hamer, and J. Oitmaa,
Phys. Rev. B {\bf 79}, 054412 (2009).
\bibitem{greiter} S. Rachel, R. Thomale, M. Führinger, P. Schmitteckert, and M. Greiter,
Phys. Rev. B {\bf 80}, 180420(R) (2009).
\bibitem{EE}
P. Calabrese and J. Cardy, J. Stat. Mech. (2004) P06002;
J. Phys. A \textbf{42}, 504005 (2009).
\bibitem{Afflecklog}
I. Affleck, D. Gepner, H. J. Schulz, and T. Ziman,
J. Phys. A: Math. Gen. {\bf 22},  511 (1989);
R. R. P. Singh, M. E. Fisher,  and R. Shankar,
Phys. Rev. B {\bf 39}, 2562 (1989).
\bibitem{Hallberg1996} K. Hallberg, X. Q. G. Wang, P. Horsch, and A. Moreo,  Phys. Rev. Lett. {\bf 76}, 4955 (1996).
\bibitem{Cazalilla} M.~A. Cazalilla, J. Phys. B {\bf 37}, S1 (2004).
\bibitem{Todo2001} S. Todo and K. Kato, Phys. Rev. Lett. {\bf 87}, 047203 (2001).
\bibitem{ytterbium} S. Taie, Y. Takasu, S. Sugawa, R. Yamazaki, T. Tsujimoto, 
R. Murakami, and Y. Takahashi, Phys. Rev. Lett. {\bf 105}, 190401 (2010).

\end{thebibliography}
\end{document}